\documentclass{article}[12pt]

\usepackage{amssymb}
\usepackage{amsmath}
\usepackage{enumerate}
\usepackage{latexsym}
\usepackage{mathrsfs}
\usepackage{amsthm}
\usepackage{verbatim}
\usepackage{graphicx}
\usepackage{epstopdf}
\usepackage{epsfig}
\usepackage{color}
\usepackage{float}
\usepackage{color}
\usepackage{titlesec}
\usepackage{bm}
\usepackage{tikz}
\usetikzlibrary{calc}
\usepackage{xcolor}

\usepackage[colorlinks,linkcolor=blue,anchorcolor=green,citecolor=red,hypertexnames=false]{hyperref}
\usepackage[T1]{fontenc}

\usepackage{subfig}

\usepackage{charter}
\usepackage{fancyhdr}
\usepackage{amscd}

\usepackage{caption}

\usepackage{bbm}

\makeatletter
\@addtoreset{equation}{section}
\makeatother

\makeatletter 
\@addtoreset{equation}{section}
\makeatother  
\renewcommand\theequation{\oldstylenums{\thesection}%
	.\oldstylenums{\arabic{equation}}}

\usepackage{geometry}
\usepackage{listings}
\usepackage{apptools}

\usepackage{setspace}

\DeclareMathOperator*{\argmin}{arg\,min}

\usepackage{cleveref}
\usepackage{aliascnt}

\newtheorem{theorem}{Theorem}[section]

\newaliascnt{lemma}{theorem}
\newtheorem{lemma}[lemma]{Lemma}
\aliascntresetthe{lemma}

\newaliascnt{proposition}{theorem}

\aliascntresetthe{proposition}

\newaliascnt{remark}{theorem}
\newtheorem{remark}[remark]{Remark}
\aliascntresetthe{remark}

\newaliascnt{corollary}{theorem}

\aliascntresetthe{corollary}

\newaliascnt{definition}{theorem}
\newtheorem{definition}[definition]{Definition}
\aliascntresetthe{definition}

\newcommand{\ssubset}{\subset\joinrel\subset}
\renewcommand{\phi}{\varphi}

\newcommand{\R}{\mathbb{R}}

\newcommand{\edit}[1]{{\color{black} #1}}

\begin{document}
	
	\title{The effective energy of a lattice metamaterial}
	\author{Xuenan Li \thanks{Department of Applied Physics and Applied Mathematics, Columbia University, xl3383@columbia.edu}
		\qquad Robert V. Kohn\thanks{Courant Institute of Mathematical Sciences, New York University, kohn@cims.nyu.edu}}
	\date{\today}
	
	\maketitle
	\begin{abstract}
		We study the sense in which the continuum limit of a broad class of discrete materials with periodic structures can be viewed as a nonlinear elastic material. While we are not the first
		to consider this question, our treatment is more general and more physical
		than those in the literature. Indeed, it applies to a broad class of
		systems including ones that possess mechanisms; and we discuss how
		the degeneracy that plagues prior work in this area can be avoided
		by penalizing change of orientation. A key motivation for this work
		is its relevance to mechanism-based mechanical metamaterials. Such
		systems often have ``soft modes'', achieved in typical examples by
		modulating mechanisms. Our results permit the following more general
		definition of a soft mode: it is a macroscopic deformation whose
		effective energy vanishes -- in other words, one whose spatially-averaged
		elastic energy tends to zero in the continuum limit.
	\end{abstract}
	
	\section{Introduction}\label{sec:intro}
	Homogenization was used to study large deformations of elastic composites nearly 40 years ago
	\cite{braides1985homogenization,muller1987homogenization}, and discrete-to-continuous limits of
	nonlinear elastic structures have been a focus of attention for at least 20 years
	\cite{alicandro2004general}. This paper has strong connections to both those threads, but
	its motivation comes from a much newer thread -- namely the analysis of mechanism-based
	mechanical metamaterials. As we shall explain in \cref{subsec:mechanism-based} by discussing
	some key examples, the systems we have in mind resemble porous elastic composites, but their essential
	properties can be captured by discrete lattice models. Besides their mechanisms, these systems often
	have \textit{soft modes} -- by which we mean macroscopic deformations that are not mechanisms, but that
	nevertheless have very little elastic energy. In the best-understood examples (such as
	the \edit{Rotating Squares} metamaterial \cite{czajkowski2022conformal}), the soft modes are achieved
	by modulating a mechanism. It is natural to ask for a characterization of soft modes that doesn't
	rely on a classification of the structure's mechanisms. This question is important, because there
	are interesting examples for which we have no list or classification of the mechanisms (for example
	the Kagome metamaterial, which we discuss in some detail in sections \ref{subsec:mechanism-based}
	and \ref{sec:examples}, as well as other metamaterials with many mechanisms \cite{bossart2021oligomodal}). We believe that for lattice models of mechanism-based mechanical metamaterials, the soft modes are \edit{best defined as} the macroscopic deformations \edit{for which an appropriately-defined effective energy vanishes}.
  \textit{The main goal of the present paper is to give sense to this assertion, by proving the existence of an effective elastic energy for a broad range of lattice models, including ones with mechanisms.} The assertion's consequences
	are explored and exploited in our paper \cite{liforthcoming}, which focuses particularly on some
	conformal metamaterials, including the Kagome metamaterial and the \edit{Rotating Squares} metamaterial.
	
	In the process of achieving the goal just enunciated, our paper also develops a new methodology. Indeed, we develop a new framework for the study of discrete-to-continuous limits of periodic structures, based on piecewise linear interpolation. Using our framework, arguments that are familiar for the study of continua have natural analogues for the study of discrete-to-continuous limits. While this framework is only used here to show the existence of an effective energy, we expect it to also have other applications. 
	
	Before discussing the paper's goals and accomplishments in more detail in \cref{subsec:intro-effective},
	let us elaborate further on the motivating issues discussed above.
	
	\subsection{Mechanism-based mechanical metamaterials}\label{subsec:mechanism-based}
\begin{figure}[!htb]
		\centering
		\subfloat[]{\includegraphics[width=0.3\textwidth]{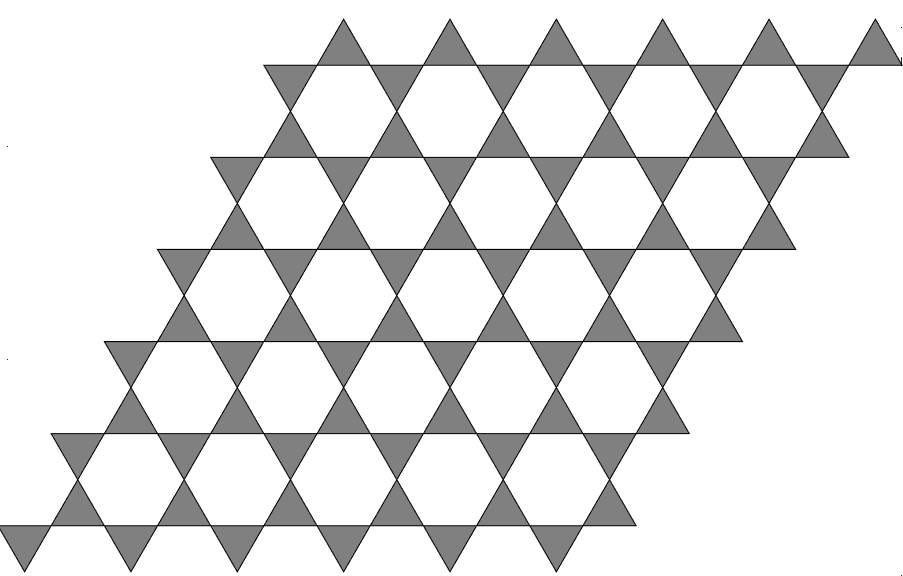}}\hfil
		\subfloat[]{\includegraphics[width=0.26\textwidth]{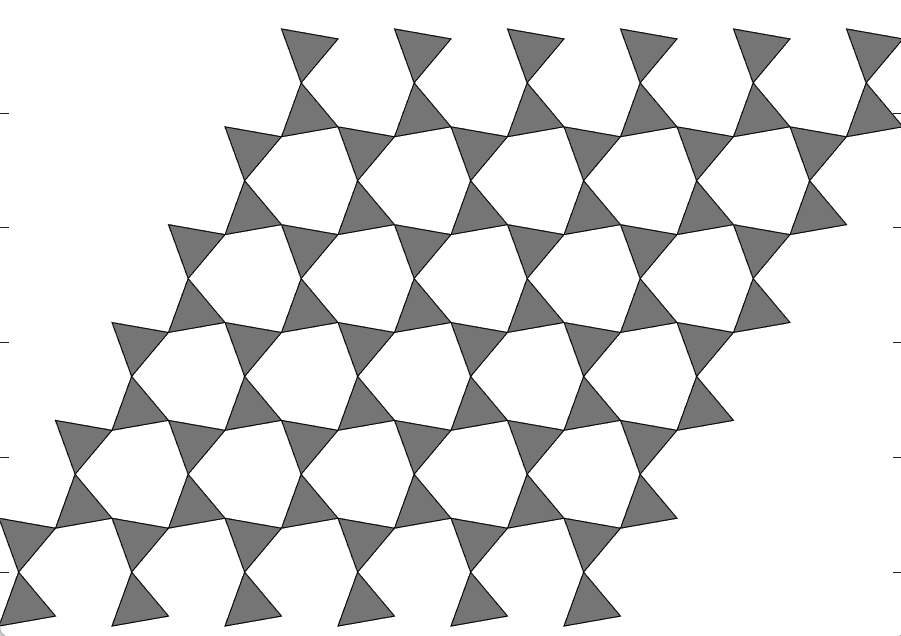}}
		\caption{One of the periodic mechanisms on the Kagome lattice: (a) the reference state; (b) the deformed state.}
		\label{fig:kagome-one-periodic}
	\end{figure}

	By definition, a mechanism of a mechanical system is a one-parameter family of deformations whose elastic energy is exactly zero, though they are not rigid motions. A broad variety of mechanism-based mechanical metamaterials have been considered in the literature
	(see, for example, \cite{bertoldi2017flexible}). To introduce the class that we shall study here it is
	convenient to begin with an example: a 2D \textit{cut-out model} of the Kagome metamaterial. It is obtained by
	tiling the plane periodically by hexagons and triangles as shown in \cref{fig:kagome-one-periodic}(a) then
	cutting out the hexagons. The triangles that are left meet just at their vertices, which we view as hinges that can rotate without costing any elastic energy. This system has a one-parameter family of mechanisms with the same periodicity as the original structure, shown
	in \cref{fig:kagome-one-periodic}(b) for a particular value of the parameter. The mechanism deforms the
	holes left by cutting out the hexagons, but moves each triangle by a rigid motion; thus its elastic energy is
	exactly zero (which is the definition of a mechanism). We note that the mechanism changes the angles
	at which the triangles meet; this costs no elastic energy since we view the nodes as hinges.
	
	This cut-out model of the Kagome metamaterial shows that our topic is closely related to the homogenization
	of nonlinear elastic composites; indeed, this model is more or less a porous elastic sheet. However, viewing the
	triangles as 2D nonlinearly elastic continua makes the model difficult to analyze. Fortunately, there is an
	alternative viewpoint which keeps the essential features of the problem and is more accessible to analysis. We
	therefore prefer the \textit{spring model} of the Kagome metamaterial, obtained by treating the edges of the triangles
	as Hookean springs. We note that the mechanisms of the spring model are the same as those of the cut-out model,
	since triangles are rigid (in other words: if the vertices are moved in a way that leaves the length of each edge
	invariant then the deformation extends to a unique rigid motion of the entire triangle). We also note that while our
	springs are Hookean, the analysis of the spring model is a nonlinear problem, since we permit large deformations (and in
	particular large rotations); correspondingly, the elastic energy of the spring connecting $x_i$ and $x_j$ is \edit{a} constant times $\big(\frac{|u(x_i) - u(x_j)|}{|x_i - x_j|} - 1\big)^2$, where $u(x_i)$ and $u(x_j)$ are the deformed
	positions of the nodes; as expected, this is not a quadratic function of the deformed positions.
	
	In summary: considering the Kagome metamaterial as a periodic lattice of springs turns its analysis into a discrete
	problem (whose degrees of freedom are the \edit{positions} of the nodes), while keeping the essential features of this system (including
	the presence of mechanisms). Another favorite example -- the \edit{Rotating Squares} metamaterial (see e.g.
	\cite{czajkowski2022conformal, deng2020characterization} and \cref{subsubsec:rotating}) -- admits a similar
	treatment: as a cut-out it is obtained by patterning the
	plane as a checkerboard then removing every white square, however its essential features are easily captured by a lattice of
	springs.\footnote{To model the \edit{Rotating Squares} metamaterial by a lattice of springs, we can start with the square lattice then
		add extra diagonal springs to make some squares rigid; see for example Figure 6 in the supplementary information of \cite{czajkowski2022conformal} and this paper's \cref{subsubsec:rotating}.}
	With these examples in mind, in the present work we shall focus on \textit{lattice metamaterials}. While this class will
	be defined in \cref{sec:setup}, we emphasize that it includes periodic lattices of springs.
	
\begin{figure}[!htb]
		\centering
		\subfloat[]{\includegraphics[width=0.48\textwidth]{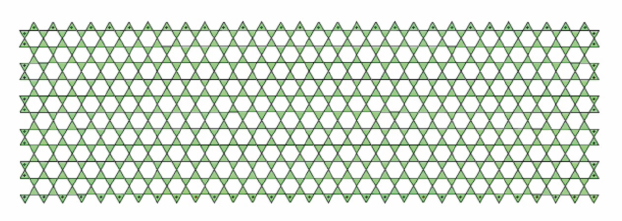}}\\
		\subfloat[]{\includegraphics[width=0.42\textwidth]{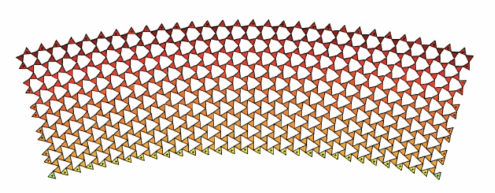}}
		\subfloat[]{\includegraphics[width=0.42\textwidth]{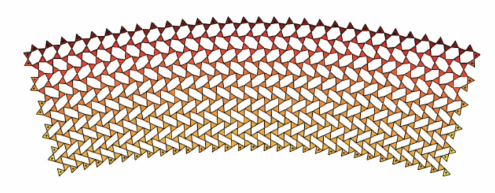}}
		\caption{Soft modes \edit{of} the Kagome lattice: (a) \edit{the reference state}; (b) a modulated version of the
		mechanism shown in \cref{fig:kagome-one-periodic}; (c) a modulated version of a \edit{different mechanism}. The color in each plot indicates the rotation angle of local equilateral triangles.}
		\label{fig:modulation}
	\end{figure}

	As already mentioned earlier, a mechanical system with a mechanism will typically also have soft modes. To explain, let us continue our focus on the
	Kagome metamaterial. It has a soft mode taking the rectangular reference domain shown in \cref{fig:modulation}(a) to the sector of an annulus shown in
	\cref{fig:modulation}(b). Microscopically, the associated deformation uses the one-parameter family of mechanisms illustrated
	in \cref{fig:kagome-one-periodic}. Since the value of the parameter varies macroscopically, the deformation shown in \cref{fig:modulation}(b)
	is not a mechanism. However, the strain in each spring tends to zero in the limit as the ratio between the microscopic and macroscopic length scales
	tends to zero. As a result, the elastic energy of the soft mode is very small as one approaches the continuum limit.
	
	It is natural to ask which macroscopic deformations can be accommodated by modulating the mechanism as in \cref{fig:modulation}(b). The answer lies beyond the
	scope of the present paper, but let us briefly discuss it anyway. 
    \edit{To be realizable this way, the macroscopic deformation $u(x)$ should be a compressive conformal map (that is, its deformation gradient should have the form $c(x) R(x)$ where $c(x)$ is scalar-valued and $R(x)$ takes values in $SO(2)$, with $0 \leq c(x) \leq 1$)}. Given such $u$, the process by which one gets an associated soft mode is discussed in \cite{czajkowski2022conformal}; an
	explanation why there are no other soft modes will be given in \cite{liforthcoming}. We note in passing that the Kagome metamaterial has \textit{many}
	periodic mechanisms \cite{kapko2009collapse} (indeed, infinitely many \cite{li2023some}), and a soft mode can be obtained by modulating any of them.
	Thus the microscopic character of a soft mode is far from unique. This is illustrated by \cref{fig:modulation}(c), which achieves the same macroscopic
	deformation as \cref{fig:modulation}(b) by modulating a different periodic mechanism.
	
	A body of literature has begun to develop concerning the mechanics of systems with a single one-parameter family of mechanisms; this includes the \edit{Rotating Squares} metamaterial (the main focus of \cite{czajkowski2022conformal}) and a related but much broader family of kirigami-based examples (the focus of
	\cite{zheng2022continuum, zheng2022modeling}). These papers identify the soft modes of the systems they study, but they do much more. Indeed, to understand
	\textit{which} soft mode will be achieved by a given loading condition, it is not enough to identify the class of all soft modes. Rather, one must minimize
	the leading-order elastic energy (plus the work done by the loads) \edit{within this class}. For the cut-out or kirigami-based examples considered by these authors, this required
	modeling quantitatively the cost of modulation and the elastic energy due to the bending of thin necks that we treat as hinges in the present work.
	To connect those studies with the present paper: we have argued that the soft modes are macroscopic deformations whose effective energy vanishes, in other
	words whose spatially-averaged energy tends to zero in the limit as the separation of scales
	$\varepsilon$ tends to zero. To predict the response of such a structure
	to loading, one should use the leading-order elastic energy (regardless of how it scales in $\varepsilon$). While the papers just discussed achieve such a goal for
	the specific systems they consider, their methods seem to require that there be a single one-parameter family of mechanisms. Thus, it remains an open question
	how something similar can be done for a system with many mechanisms like the Kagome metamaterial.
	
	Summarizing briefly the relevance of our work to mechanism-based mechanical metamaterials: while there has been impressive progress on systems
	with a single family of mechanisms, it is important to develop a framework that can also handle systems with many mechanisms (like the Kagome
	metamaterial). To get started, one requires an adequate definition of a soft mode. For periodic lattices of springs, we propose that the soft modes \edit{should be}
	the macroscopic deformations for which an appropriately-defined effective energy vanishes. Giving sense to and making use of this proposal requires
	\begin{itemize}
		\item understanding the existence and characterization of the effective energy;
		\item understanding, at least for some examples (such as the Kagome metamaterial), the macroscopic deformations \edit{whose} effective energy vanishes.
	\end{itemize}
	This paper addresses the first bullet, while our forthcoming paper \cite{liforthcoming} addresses the second one. \edit{(See, however, \cref{sec:conclusions} of this paper, which includes further information on \cite{liforthcoming} and briefly discusses the practical implications of our work.)}

    \edit{Let us also draw a connection between our work and the recent paper \cite{czajkowski2024duality}. That work discusses some mechanism-based mechanical materials that generalize the \edit{Rotating Squares} example. It argues that these structures' macroscopic behavior is described by an effective energy that vanishes at their soft modes, then studies some consequences of this assertion. By considering the existence of an effective energy for a broad range of discrete structures, our work provides a mathematical interpretation of the macroscopic energy considered in \cite{czajkowski2024duality}.}
	
	We have not attempted to review the literature on mechanism-based mechanical metamaterials. There is, of course, a large body of work
	using \emph{linear} elasticity to study lattices of springs. In that setting the discrete energy is convex, and periodic
	homogenization leads to an effective energy and even an effective Hooke's law (see
	e.g. \cite{hutchinson2006structural} and \cite{li2023some}). However, there are lattices whose macroscopic behavior
	is not correctly described by linear elasticity (see e.g. \cite{borcea2010periodic}). The Kagome metamaterial is an
	example, since its linear elastic effective Hooke's law is nondegenerate, yet it has mechanisms achieving isotropic
	compression. To model small \edit{displacements} of certain systems with mechanisms, Nassar et.al. \cite{nassar2020microtwist} have
	proposed the use of Cosserat-type models. However, this approach requires knowing what mechanism is being activated,
	so it does not fully capture the relationship between soft modes and mechanisms. Our approach is entirely different;
	in particular, it makes no use of linearization, and it does not assume the existence (let alone a classification) of
	mechanisms.
	
	\subsection{Overview of our framework and results}\label{subsec:intro-effective}
	This paper's main goals are
	\begin{enumerate}
		\item[(i)] to discuss what it should mean for a lattice metamaterial to have an effective energy, and
		\item[(ii)] to prove the existence of (and provide a characterization of) its effective energy.
	\end{enumerate}
	Our approach to point (i) is rather different from previous work on lattices of springs
	(for example \cite{alicandro2004general}). Indeed:
	\begin{itemize}
		\item our framework assumes periodicity but places virtually no other restriction on the
		geometry of the lattice; and
		
		\item it permits inclusion of a penalty for change of orientation, thereby avoiding
		the degeneracy that plagues prior treatments (without eliminating mechanisms).
	\end{itemize}
	Our approach to point (ii) is more familiar. Indeed, we adapt well-established tools from
	homogenization to the framework associated with (i), taking advantage of the analogy
	(already noted in \cref{subsec:mechanism-based}) between our problem and the analysis of
	spatially periodic nonlinearly elastic composites.
	
	This section offers a non-technical overview. The story is long, since it involves
	modeling as well as analysis; therefore, we present it in modularized form.
	\medskip
	
	\noindent {\sc The meaning of an effective energy.}
	In discussing the existence of an effective energy, we are considering a fixed macroscopic domain
	$\Omega \subset \mathbb{R}^N $ filled with an $\epsilon$-scale version of our lattice metamaterial. Informally,
	we want to know whether in the limit $\epsilon \rightarrow 0$ we can view $\Omega$ as being filled with a
	nonlinear elastic material. This question has some subtlety, since an elastic deformation of a lattice
	is determined only at its nodes, and since our (geometrically nonlinear) elastic energies are
	expected to have local minima. To deal with the first issue, our mathematical definition of a
	lattice metamaterial will include the introduction of a (periodic) triangulation of $\mathbb{R}^N$
	such that a deformation defined only at the nodes
	determines a unique piecewise linear function on all $\mathbb{R}^N$. To deal with the second issue, we use the
	framework of gamma-convergence. Informally, this means that the effective energy of a macroscopic deformation
	$u:\Omega \rightarrow \mathbb{R}^N$ is the smallest (limiting) energy achievable by \edit{microscopic} deformations $u^\epsilon$
	defined on the $\epsilon$-scale lattice, when $u^\epsilon$ approximates the desired
	macroscopic deformation (in the sense that $\lim_{\epsilon \rightarrow 0} u^\epsilon = u$).
	
	Since the effective energy captures the continuum limit of suitably-defined discrete energies, it is
	important to say a word about our discrete energies. While our framework is not limited to
	lattices of Hookean springs, it is convenient to focus for a moment on this special case.
	If $x^\epsilon_i$ and $x^\epsilon_j$ are nodes of the $\epsilon$-scale lattice that are
	connected by a spring, then a deformation $u^\epsilon$ taking
	these points to $u^\epsilon (x_i)$ and $u^\epsilon (x_j)$ gives the spring the
	strain $\frac{|u^\epsilon (x^\epsilon_i) - u^\epsilon (x^\epsilon_j)|}{|x^\epsilon_i - x^\epsilon_j|} - 1$,
	and the spring's elastic energy is a constant times the square of the strain. For a lattice of springs, the
	discrete energy $E^\epsilon(u^\epsilon,\Omega)$ of the body $\Omega$ is, roughly speaking, the volume
	of $\Omega$ times the \emph{spatial average} of the energies of all the springs in $\Omega$.
	Thus, a macroscopic deformation with effective energy zero is one which can be approximately
	achieved on the $\epsilon$-scaled lattice in such a way that the
	\emph{average energy of all the springs tends to zero as $\epsilon \rightarrow 0$}. We
	emphasize that \edit{the microscopic deformation $u^\epsilon$ need not be a mechanism, i.e. the strains of the springs in the $\epsilon$-scaled structure} need not be zero. For example, the deformations shown in
	figures \ref{fig:modulation}(b) and \ref{fig:modulation}(c) turn out to have strains of order $\epsilon$ in
	each spring; therefore the associated macroscopic deformation (which
	takes a rectangle to a sector of an annulus) has effective energy zero. However, mechanisms are
	still relevant; indeed, if there is a periodic mechanism\footnote{By definition, a periodic
	mechanism has the form \edit{$u(x) = F x + \varphi(x)$}, where $F$ is a constant matrix and $\varphi$ is a
	periodic function defined at nodes of the unit-scale lattice. In this case
	$u^\epsilon(x) = \epsilon u(x/\epsilon)$ is a mechanism of the $\epsilon$-scaled lattice, which converges
	to the macroscopic deformation \edit{$u(x) = F x$} as $\epsilon \rightarrow 0$.} with macroscopic deformation
	gradient $F$, then the effective energy must vanish \edit{for the affine deformation $u(x)=F x$}.
	\medskip
	
	\noindent {\sc Using springs alone is not sufficient.}
	It has long been understood that the effective energy of a lattice of springs can be very degenerate.
	To explain why, let us consider a simple example: the 2D lattice whose nodes are the integer points,
	with springs joining all nearest-neighbor and next-nearest-neighbor pairs.\footnote{Connecting
	only nearest neighbors would give a square lattice, which permits macroscopic shear with zero elastic energy. By introducing next-nearest-neighbor (diagonal) springs, one might expect at first to have a non-degenerate structure since the only orientation-preserving deformations with zero energy are rigid motions.} Clearly, the folding
	deformation $u(x,y) = (-x,y)$ preserves the length of every spring, so its discrete energy is zero. Similarly,
	the lattice can be folded like an accordion (using folds along lines where $x$ is an integer) to achieve any
	macroscopic compression in the horizontal direction. By symmetry, the same applies using folds where $y$ is an
	integer. Thus, the effective energy of this system vanishes at $u(x,y) = (cx,y)$ and at $u(x,y) = (x,cy)$ for any
	$0 \leq c \leq 1$.\footnote{In fact the effective energy also vanishes at $u(x,y) = (cx,dy)$ for any
	$0 \leq c \leq 1$ and $0 \leq d \leq 1$, for example by folding first along lines parallel to the $x$ axis
	then along lines parallel to the $y$ axis.} Evidently, the effective energy vanishes for deformations that \edit{we do not wish to consider soft modes}, because a discrete energy that considers only the lengths of springs does not penalize change of orientation.
	
	The resolution of this difficulty is simple: we must \emph{inform the model that change of orientation is undesirable}.
	In the preceding example, each diagonal spring breaks a square into two triangles, and for each such triangle $T$
	the nodal \edit{values of a deformation} determine an affine map $u_T$. Since \edit{$\det (\nabla u_T)$} reveals the orientation of the map $u_T$, we
	can inform the model of our preference by adding a suitable function of such determinants to our energy.
	Moreover, if we use a function that vanishes when the determinant is positive, then the energy of an orientation-preserving deformation isn't being changed at all. This idea is not new; for example, a similar penalization is used
	(for a similar purpose) in \cite{alicandro2011integral}.
	
	Should one include such a penalization for \emph{every} triangle? Not necessarily! As explained earlier, a key
	motivation for our work is the idea that the soft modes of a mechanical metamaterial are deformations for which
	a \edit{suitably-defined} effective energy vanishes. The mechanical metamaterials that we have in mind have cut-out models
	as well as spring models, and the penalization should be applied only on regions that have not been cut out.
	In considering the Kagome metamaterial, for example, we must remember that the hexagonal regions are viewed as
	``holes;'' therefore a penalization term should be included \emph{only} for the equilateral triangles in
	\cref{fig:kagome-one-periodic}(a).
	
	How large should the penalization be? This is a modeling choice, on which we need not take a definite position.
	But let us discuss what is at stake, focusing as usual on our idea that the soft modes of a lattice metamaterial
	should be the deformations \edit{whose} effective energy vanishes. Based on spring energies alone, the minimum
	energy of a triangle is zero, but this minimum is achieved \emph{both} when the triangle experiences an
	orientation-preserving rigid motion, \emph{and} when it experiences an orientation-reversing rigid motion.
	Thus, in a deformation with small spring energy, the penalization will mainly be evaluated at determinants near
	$\pm 1$. The penalization should of course be chosen so that it vanishes when the determinant is near $1$ (so it doesn't
	interfere with any mechanisms), and it should be sufficiently large when the determinant is near $-1$. We expect that
	these are the \emph{only} properties of the penalization that affect the zero-set of the effective energy.
	
	Why use penalization, rather than simply prohibiting negative determinants? The answer is technical: while our theory
	accepts penalization, the proofs break down if we try to insist that our (microscopic) deformations have pointwise positive
	determinant. Indeed, our arguments rely on piecewise affine approximation, which raises the question whether a deformation
	with \edit{$\det (\nabla u) > 0$} can be approximated by piecewise affine ones satisfying the same constraint. While an affirmative answer
	is known in two space dimensions \cite{iwaniec2011diffeomorphic, iwaniec2012hopf}, this question is open in higher dimensions.
	Another issue involves interpolation: we need at certain points to interpolate between two deformations $u$ and $v$ which are
	known to be close in a weak norm (see appendix \ref{appendix:degiorgi}). The obvious (and ultimately successful) idea is to use
	$u \phi + v (1-\phi)$ with a suitable choice of $\phi$, however it does not seem possible to assure that \edit{the deformation gradient of this}
	interpolant has pointwise nonnegative determinant. Such issues are by now familiar in nonlinear elasticity; for a relatively
	recent discussion with additional references, see \cite{conti2015theory}.

\edit{We alert the reader that while we penalize change of
orientation, we do \emph{not} penalize or prohibit interpenetration. Thus, for example, for the one-periodic
mechanism of the Kagome lattice shown in \cref{fig:kagome-one-periodic}(b), there is a value of the parameter
at which the hexagons have shrunk to slits; however our framework permits the parameter to be larger, even 
though this leads the triangles to interpenetrate. At the parameter value where the hexagons have become slits 
the macroscopic deformation gradient is $c R$ with $c=1/2$ and $R \in SO(2)$. However, there is a larger value 
of the parameter at which the triangles all lie on top of one another and the macroscopic deformation gradient is $0$.}
\medskip
	
	\noindent {\sc A geometry-independent, homogenization-like framework.}
	The systems that interest us have much in common with those studied by Alicandro and Cicalese
	in their seminal 2004 paper \cite{alicandro2004general} concerning continuum limits of systems of springs.
	However, we cannot simply rely upon that work for the existence of an effective energy,
	because the geometries considered there are not sufficiently general. Indeed, when specialized to the
	periodic setting, \cite{alicandro2004general} considers a periodic lattice of nonlinear springs
	\emph{connecting the nodes of a square lattice}. By a linear change of variables,
	the analysis also applies to springs connecting the nodes of a Bravais lattice (in which, by definition, all nodes
	are translations of a single node by period vectors of the lattice). However, our spring model of the Kagome
	metamaterial does not have this character, since its nodes do not form a Bravais lattice. (In fact, its unit cell
	contains three nodes, as shown in \cref{fig:kagome-3}.)
	
	The methods used in \cite{alicandro2004general} have been generalized to other settings, and we suppose they could be
	extended to our favorite examples (such as the Kagome lattice with a suitable penalization for change of orientation).
	In this paper, however, we pursue a different approach, which avoids any geometric hypothesis on the locations of the nodes.
	Instead, our framework emphasizes the hypothesis of periodicity and takes advantage of the analogy to homogenization
	of nonlinear elastic composites.
	
	Our approach is developed in detail in \cref{sec:setup}, but we outline it here. Since our structure is
	periodic, we consider a \emph{unit cell} $U \subset \mathbb{R}^N$ -- a \edit{parallelepiped} containing the origin whose translates by
	vectors $v_1, \ldots, v_N$ tile all $\mathbb{R}^N$. The basic object we work with is the \mbox{energy of the unit cell},
	$E(u,U)$, where $u$ is a deformation (defined at nodes of the unscaled structure). The hypothesis of periodicity is
	captured by defining the energy of a translate of $U$ by
	\begin{equation} \label{periodicity}
		E (u(x+\alpha),U + \alpha) = E(u(x), U) \, ,
	\end{equation}
	where we have introduced the convention that
	\begin{equation} \label{alpha-as-transl-unscaled}
		\alpha = \alpha_1 v_1 + \cdots + \alpha_N v_N \quad \mbox{with $\alpha_i \in \mathbb{Z}$} \, .
	\end{equation}
	The energy of the unit cell must be chosen so
	that the energy of the \emph{entire} unscaled structure is
	\begin{equation} \label{energy-of-structure}
		\sum_{\alpha_i \in \mathbb{Z}} E(u,U+\alpha) \, .
	\end{equation}
	We hasten to add: the definition of $E(u,U)$ typically involves the values of $u$ not only at the nodes in $U$,
	but also at some nodes in nearby translates of $U$; for example, in a network of springs, the springs that enter the
	definition of $E(u,U)$ may not lie entirely within $U$. In fact, the introduction of a unit cell is basically a bookkeeping
	device, which assures through \eqref{energy-of-structure} that adding the energies of $U$ and its translates gets the total
	right.
	
	To consider the effective energy, we must discuss the energy of the scaled structure. It is defined by elasticity scaling:
	if
	\begin{equation} \label{alpha-as-transl-scaled}
		\alpha = \alpha_1 v_1 + \cdots + \alpha_N v_N \quad \mbox{with $\alpha_i \in \epsilon \mathbb{Z}$}
	\end{equation}
	and $u^\epsilon(x) = \epsilon u\left(\frac{x-\alpha}{\epsilon}\right)$ then
	\begin{equation} \label{eqn:elasticity-scaling}
		E^\epsilon (u^\epsilon, \epsilon U + \alpha)= E(u,U) \, \epsilon^N \, .
	\end{equation}
	The scaled energy is again periodic, i.e. it satisfies an obvious analogue of \eqref{periodicity}. To explain the
	factor of $\epsilon^N$ on the right hand side of \eqref{eqn:elasticity-scaling}, we note that when $u^\epsilon$ is affine
	this definition makes $E^\epsilon (u^\epsilon,\epsilon U)$ proportional to the volume of $\epsilon U$.
	
	The \emph{energy of a domain $\Omega$} filled by the scaled structure is, roughly speaking, the sum of the scaled energies of all
	translates of $\epsilon U$ that lie inside $\Omega$. But we must be careful, since when $\epsilon U + \alpha$ lies near
	$\partial \Omega$ its energy could depend upon the values of $u$ outside of $\Omega$. This is a familiar issue in the area of
	discrete-to-continuum limits, and we resolve it in the usual way -- by omitting the cells so close to $\partial \Omega$ that this
	is an issue. Thus, the energy of $\Omega$ using a deformation $u^\epsilon$ (defined at the nodes of the scaled structure)
	takes the form
	\begin{align} \label{eqn:intro-energy-of-Omega}
		E^\epsilon(u^\epsilon, \Omega) &:= \sum_{\alpha \in R_\epsilon(\Omega)} E^\epsilon (u^\epsilon, \epsilon U + \alpha),
	\end{align}
	where the definition of $R_\epsilon(\Omega)$ (given in \cref{sec:setup}) is such that the sum excludes a
	boundary layer (whose width is over order $\epsilon$) near $\partial \Omega$.

\edit{We haven't yet said anything about units. When we fill a domain $\Omega$ with an $\epsilon$-scale version of the lattice, the parameter $\epsilon$ has the dimensions of length. A deformation $u^\epsilon$ of the scaled lattice is defined at its nodes $x^\epsilon$, and $u^\epsilon(x^\epsilon)$ also has the dimensions of length. 
The energy of a scaled unit cell $E^\epsilon(u^\epsilon, \epsilon U + \alpha)$ has, of course, the dimensions of elastic energy. Equation \eqref{eqn:elasticity-scaling} defines $E^\epsilon(u^\epsilon,\epsilon U + \alpha)$ in terms of the non-dimensionalized deformation $u(y) = \epsilon^{-1} u^\epsilon (\epsilon y + \alpha)$; we see from this equation that $E(u,U)$ has the dimensions of energy per unit volume (and that the unit cell $U$ is non-dimensional).}
	
	While we are mainly interested in lattices of springs (with penalization for change of orientation), our hypotheses
	upon the energy $E$ are more abstract. Besides periodicity (discussed above), they are:
	\begin{enumerate}
		\item[(a)] nonnegativity,
		\item[(b)] translation invariance, and
		\item[(c)] upper and lower bounds relating $E$ to the $L^2$ norm of $\nabla u$ on either $U$
		(for the lower bound) or the union of $U$ and finitely many nearby cells (for the upper bound).
	\end{enumerate}
	We refer to \eqref{eqn:unit-cell-eps-periodicity}--\eqref{eqn:unit-cell-lower} for precise versions of these
	hypotheses, however we offer a few comments here. Concerning (a): there would be no essential
	difference if we only assumed that the energy was bounded below, since adding a constant would then achieve
	nonnegativity. Concerning (b): it is quite natural that $u$ and $u+c$ should have the same energy when $c$ is
	a translation (i.e. it takes the same value at every node), since mechanical structures are translation-invariant.
	Concerning (c): as we shall explain in \cref{sec:setup}, we will identify a deformation $u$ (which is defined
	only at the nodes of our structure) with a piecewise linear extension, so that $\nabla u$ makes sense.
	
\edit{While the examples discussed in this paper involve lattices of springs that rotate freely at nodes, our framework also permits the use of ``torsional springs,'' for example to model resistance to bending in the 
thin necks that we have thus far treated as nodes. A typical torsional spring energy has the form $k_s (\theta - \theta_0)^2$ where $\theta_0$ is the preferred angle between two edges and $k_s$ is the torsional spring constant; thus, for the Kagome metamaterial, a typical model for the total bending energy of all the necks would be obtained by summing $k_s (\theta - 2\pi/3)^2$
over all the internal angles of all the hexagonal holes. To explain how a torsional spring law similar to
$k_s(\theta - \theta_0)^2$ can be included in our framework, let $x_1$ be a node of the lattice that belongs to the unit cell $U$, and suppose $x_1$ is connected by springs to nodes $x_2$ and $x_3$ (which may or may not belong to $U$). If we write $\ell_{ij} = u(x_j) - u(x_i)$ for the vector associated with the deformed spring from $x_i$ to $x_j$, then adding 
\begin{equation} \label{eqn:nonstandard-torsional-spring-energy}
\Big|\ell_{12} \cdot \ell_{13} - |\ell_{12}| |\ell_{13}| \cos \theta_0\Big|
\end{equation}
to the energy of the unit cell introduces a nonnegative term that vanishes
only when the cosine of the angle between $\ell_{12}$ and $\ell_{13}$ is
$\cos \theta_0$. Since this term is nonnegative with at most quadratic growth, adding it to an energy that already satisfies our lower bound will leave our framework intact. Now, the spring law 
\eqref{eqn:nonstandard-torsional-spring-energy} is a little too simple: it does \emph{not} vanish quadratically at $\theta = \theta_0$, since \eqref{eqn:nonstandard-torsional-spring-energy} is 
$|\ell_{12}| |\ell_{13}|$ times $|\cos \theta - \cos \theta_0|$. However, this is easily fixed: if $g:\R \rightarrow \R$ is nonnegative with a nondegenerate
minimum at $0$ and at most linear growth at infinity, then the spring energy 
$$
g \left(\ell_{12} \cdot \ell_{13} - |\ell_{12}| |\ell_{13}| \cos \theta_0 \right) 
$$
still fits within our framework and it vanishes quadratically at $\theta = \theta_0$ (provided the two springs are not highly deformed).}
	
	Our framework \emph{does not} require that $E(u,U)$ depend continuously on the \edit{the values of $u$ at the nodes}. Thus, for
	example, the term that penalizes change of orientation of a triangle $T$ could have the form
	$|T|$$f^\eta \left( \det (\nabla u |_T) \right)$ with
	
	\begin{equation} \label{eqn:discontinuous-penalization-term}
		f^\eta (t) = \left\{\begin{array}{cl}
			1/\eta & \mbox{if $t \leq 0$}\\ 0 & \mbox{if $t > 0$}  \end{array}  \right.
	\end{equation}
	where $\eta > 0$ is a small constant.
	\medskip
	
	\noindent {\sc Main results and methods.}
	Our main result, \cref{thm:main-theorem}, asserts that in the limit $\epsilon \rightarrow 0$, the domain $\Omega$ can indeed be
	viewed as being occupied by a nonlinear elastic solid. In more technical terms: our discrete functionals
	$E^\epsilon(u^\epsilon, \Omega)$ gamma-converge to an effective energy of the form\footnote{\edit{In \eqref{effective-energy} -- and indeed throughout this paper -- $\nabla u$ is the matrix whose $i,j$th entry is $\partial u_i/\partial x_j$. For an $N$-dimensional elastic deformation (mapping $\mathbb{R}^N$ to $\mathbb{R}^N$) this is its deformation gradient (sometimes denoted $Du$).}}
	\begin{equation} \label{effective-energy}
		E_\text{eff}(u,\Omega) = \int_\Omega \overline{W}(\nabla u) \: dx \, .
	\end{equation}
	The integrand $\overline{W}$ does not depend on $\Omega$; we view it, of course, as the hyperelastic energy density of the
	effective energy. The theorem also provides a variational characterization of $\overline{W}$, using which it is easy to see
	that $\overline{W} \geq 0$, and also that \edit{$\overline{W}$ is frame indifferent ($\overline{W}(R \lambda) = \overline{W}(\lambda)$ for all orientation-preserving rotations $R$) if the discrete energy satisfies $E(Ru,U) = E(u,U)$ for all orientation-preserving rotations $R$.}
	
	We remind the reader that to prove \edit{gamma-convergence} we must provide, for any $u$,
	\begin{enumerate}
		\item[(1)] \emph{an ansatz for the associated $u^\epsilon$} -- in other words, a family of discrete deformations
		$u^\epsilon$ (defined at the nodes of the $\epsilon$-scaled structure, and converging in a suitable sense to $u$) such that
		$E^\epsilon(u^\epsilon, \Omega) \rightarrow E_\text{eff}(u,\Omega)$; and
		
		\item[(2)] \emph{a proof that this ansatz is asymptotically energetically optimal}, by showing that if any family of discrete
		deformations $u^\epsilon$ converges to $u$, then
		$\liminf_{\epsilon \rightarrow 0} E^\epsilon(u^\epsilon,\Omega) \geq E_\text{eff}(u,\Omega)$.
	\end{enumerate}

\edit{We should, perhaps, point out a limitation of gamma-convergence. Point (1) does not place any requirement on the \emph{rate} at which $E^\epsilon(u^\epsilon, \Omega)$ converges to $E_\text{eff}(u,\Omega)$ as $\epsilon \rightarrow 0$. So when $u$ is a soft mode (in the sense that $E_\text{eff}(u,\Omega) = 0$), our gamma-convergence result assures the 
existence of $u^\epsilon$ (converging to $u$) such that $E^\epsilon(u^\epsilon, \Omega) \rightarrow 0$ as $\epsilon \rightarrow 0$;
however it does not imply, for example, that  one can achieve $E^\epsilon(u^\epsilon, \Omega) \leq C \epsilon^\alpha$ for some $\alpha > 0$ -- let alone tell us what the optimal $\alpha$ might be.}
	
	Theorem \ref{thm:main-theorem} considers our problem without boundary conditions. We also consider what happens when there is
	a Dirichlet-type boundary condition: Theorem \ref{thm:theorem-dirichlet} proves gamma-convergence in this setting as well.
	The limit is, as expected, the same effective energy \eqref{effective-energy} constrained by the Dirichlet boundary condition.
	
	Our results are not a surprise, since they were already proved in \cite{alicandro2004general} for the spring networks that satisfy
	that paper's hypotheses. However, our methods are quite different from those of \cite{alicandro2004general}. Therefore our work
	provides an alternative perspective, even for problems where the results themselves are not new.
	
	Our analysis is largely parallel to M\"{u}ller's treatment of periodic homogenization problems in nonlinear
	elasticity \cite{muller1987homogenization}. It begins by addressing assertions (1) and (2) in the special
	case when $u$ is affine; the variational characterization of $\overline{W}$ emerges from that argument. After treating
	the affine case, our analysis obtains assertion (1) for general $u$ by considering piecewise linear functions then using a density
	argument. Our proof of assertion (2) for general $u$ does not follow M\"{u}ller; instead, it uses a blowup technique
	that was first applied to periodic homogenization by Braides, Maslennikov, and Sigalotti in \cite{braides2008homogenization}.
	
	We discussed earlier our view that for mechanism-based mechanical metamaterials, the soft modes are \edit{best defined as} the \edit{macroscopic}
	deformations for which a (suitably defined) effective energy vanishes. Given the form of the effective energy, this
	means that the soft modes are deformations $u$
	such that \edit{$\overline{W}(\nabla u) = 0$} pointwise. It is therefore natural to ask: can we characterize, for specific examples,
	the zero-set of $\overline{W}$? The answer is yes: our forthcoming paper \cite{liforthcoming} shows that for our spring model
	of the Kagome metamaterial (with a suitable penalization for change of orientation), $\overline{W}(\lambda) = 0$ exactly when
	$\lambda = cR$ where $0 \leq c \leq 1$ and $R \in SO(2)$. Moreover, that paper's methods are not limited to Kagome; they also give a similar result for a spring-based model of the \edit{Rotating Squares} metamaterial, \edit{as well as for some other} conformal metamaterials. \edit{(See \cref{sec:conclusions} for additional information about \cite{liforthcoming}.)}
	\medskip
	
	\noindent {\sc Organization.}
	The paper is structured as follows. Section \ref{sec:setup} establishes our framework; this includes a careful treatment of our conditions on $E(u,U)$ and discussion about how some specific examples can be modeled this way.
	That section also gives precise statements of Theorems \ref{thm:main-theorem} and \ref{thm:theorem-dirichlet}, as well as
	several lemmas concerning useful properties of the effective energy density $\overline{W}$. Section \ref{sec:proof-thm}
	provides the proofs of these results, \edit{and} \cref{sec:examples} illustrates the scope of our framework by discussing how the
	associated energy $E(u,U)$ should be chosen in some illustrative examples.
    \edit{Finally, we close in \cref{sec:conclusions} by briefly reviewing the results obtained here and in \cite{liforthcoming}, and discussing their physical implications.}
	
	\section*{Acknowledgements}
	This research was partially supported by NSF (through grant DMS-2009746) and by the Simons Foundation (through grant 733694).
	We gratefully acknowledge comments from Gilles Francfort and Ian Tobasco concerning an earlier version of this work, which
	led to substantial improvements.

	\section{Getting started}\label{sec:setup}
	
	\subsection{Lattice nodes, some basic examples, and \texorpdfstring{$U_n$}{Lg}} \label{subsec:lattice-nodes-etc}
	To describe more precisely the $N$-dimensional lattice systems that interest us, we start by introducing some notation.
	As already indicated earlier, we start with a unit cell $U$ (a \edit{parallelepiped} containing the origin) and $N$ vectors
	$v_i$ such that the translates $U + \alpha$ tile\footnote{The translated copies of the unit cell may have overlapping
		boundaries, but their interiors remain distinct and non-intersecting.}
	all $\mathbb{R}^N$ when $\alpha = \sum_{i=1}^N \alpha_i v_i$ with $\alpha_i \in \mathbb{Z}$.
	To identify the nodes of the lattice, we fix a basic set of nodes in the unit cell, $V = \{p_1,\dots, p_{|V|}\} \subset U$;
	the full set $\mathcal{V}$ of nodes consists of all translates of elements of $V$:
	\begin{equation} \label{eqn:nodes-of-lattice-unscaled}
		\mathcal{V} = \bigcup_{\alpha_i \in \mathbb{Z}} (V + \alpha).
	\end{equation}
	We assume that no two elements of $V$ are lattice translates of one another, so each node of the lattice is \emph{uniquely} expressible as $p + \alpha$ for some $p \in V$ and
	$\alpha  = \sum \alpha_i v_i$.
	
\begin{figure}[!htb]
		\begin{minipage}{.45\linewidth}
			\centering
			\subfloat[]{\label{fig:kagome-3}\includegraphics[width=0.85\textwidth]{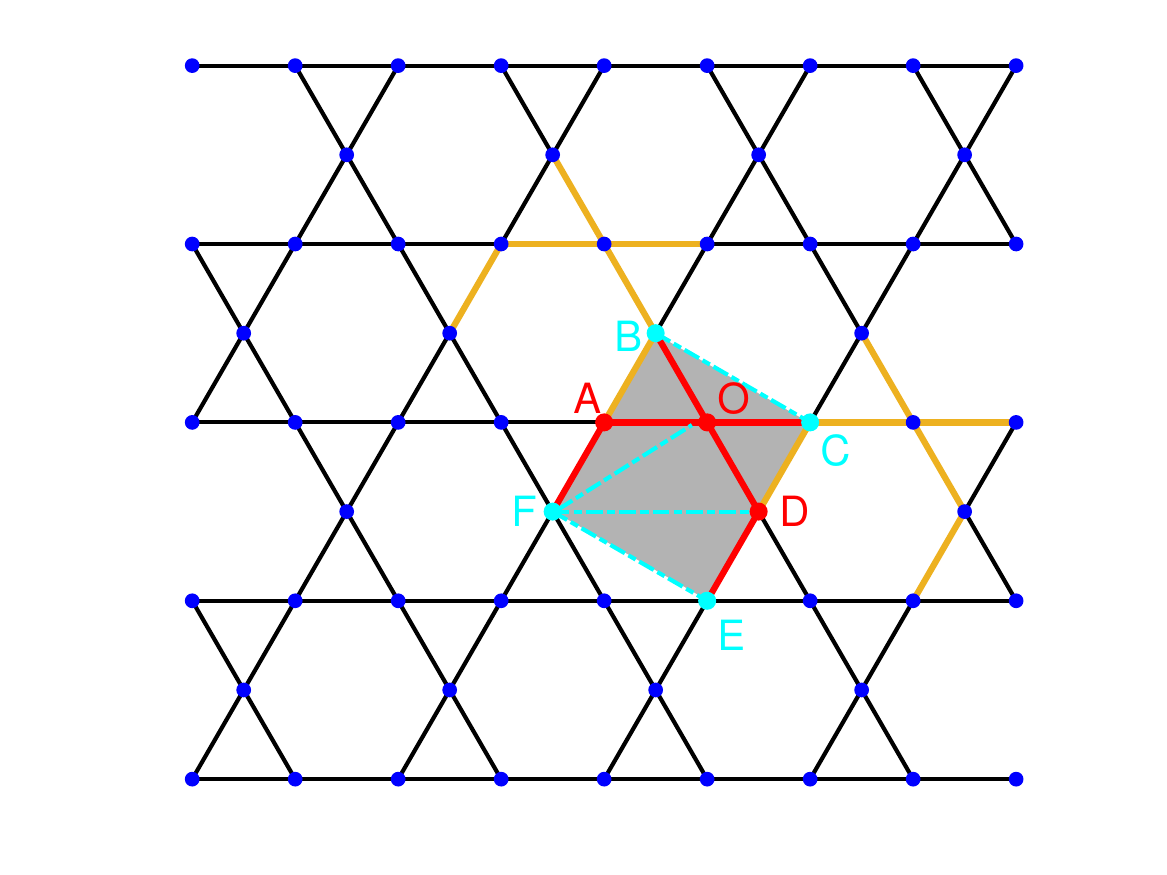}}
		\end{minipage}
		\begin{minipage}{.45\linewidth}
			\centering
			\subfloat[]{\label{fig:square-long}\includegraphics[width=0.85\textwidth]{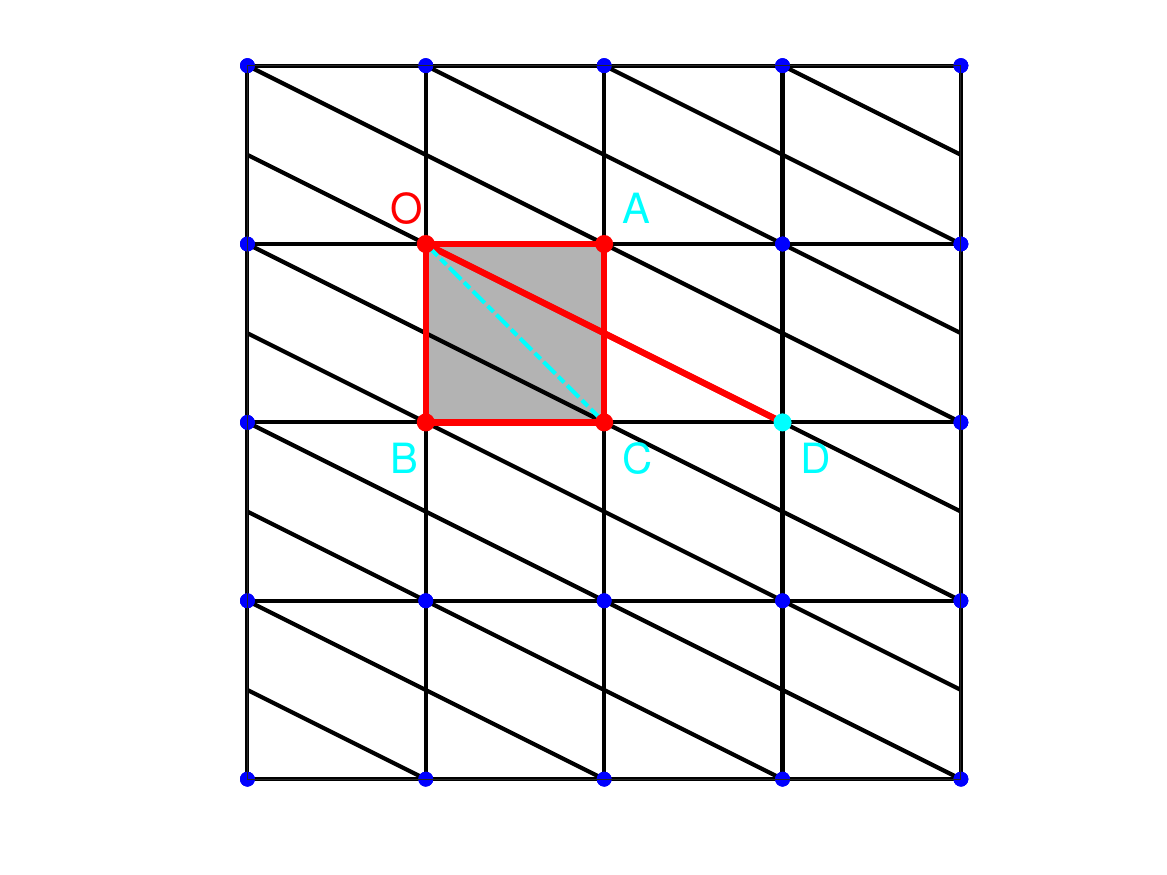}}
		\end{minipage}
		\caption{(a) The Kagome lattice: The shaded rectangle represents the unit cell $U$ for the Kagome lattice, which contains three vertices $A,O,D$ marked in red. These vertices can be translated to obtain the entire lattice. The solid red edges are those included in the energy calculation $E(u,U)$ in equation \eqref{eqn:kagome-intro-energy}. A translated copy of these edges is marked in yellow to illustrate that all edges in the Kagome lattice can be viewed as translated copies of the red solid edges. The dotted lines indicate the triangular mesh used to interpolate the admissible deformations. (b) The square lattice with long-range interactions: the same coloring scheme is used to describe the unit cell, vertices within the unit cell, edges contributing to the energy $E(u,U)$, and the triangular mesh. The details are similar to those described for the Kagome lattice and are omitted for brevity.}
	\end{figure}

	As an example, consider the 2D Kagome lattice shown in \cref{fig:kagome-3}. A convenient choice of its
	unit cell $U$ is the rectangle with vertices $B,C,E,F$, and a convenient choice of the basic set of nodes is $V = \{A,O,D\}$.
	If we choose the distance between two nearest nodes to be $1$, then the translation vectors are $v_1 = (2,0)^T$ and
	$v_2 = (1,\sqrt{3})^T$.
	
	We want to endow such a lattice with an elastic energy. To do so, it is important to be clear about what we mean by an elastic
	\edit{deformation}. We take the view that \edit{a deformation} is an $\mathbb{R}^N$-valued function defined \emph{only at nodes}.
	(Our situation is thus different from the theory of ``reticulated structures,'' discussed e.g.
	in \cite{cioranescu2012homogenization}, where the \edit{deformations} are defined on sets with nonzero volume.)
	
	As already indicated in \cref{subsec:intro-effective}, our elastic energy is determined by the energy of the unit cell $E(u,U)$,
	which we assume is nonnegative ($E \geq 0$) and translation-invariant ($E(u,U) = E(u+c,U)$ when $c$ is a translation, i.e. it
	takes the same value at every node). As an example, consider our spring model of the Kagome metamaterial, with Hookean springs
	connecting each pair of nearest-neighbor nodes. If the unit cell is chosen as shown in \cref{fig:kagome-3}, then it is convenient
	to let $E(u,U)$ be the energy of the six springs $AO,BO,CO,DO,AF,DE$, since each spring in the lattice is (uniquely) a translate of
	one of these. With this choice (and taking all the springs to be the same, and making a choice of spring constant) the energy of
	a translated unit cell $U + \alpha$ is
	\begin{align}
		E(u,U+\alpha) &= \Bigg(\Big|u(A+\alpha)-u(O+\alpha)\Big|-|A-O|\Bigg)^2 + \Bigg(\Big|u(B+\alpha)-u(O+\alpha)\Big|-|B-O| \Bigg)^2 \nonumber \\
		& \quad + \Bigg(\Big|u(C+\alpha)-u(O+\alpha)\Big|-|C-O|\Bigg)^2 + \Bigg(\Big|u(D+\alpha)-u(O+\alpha)\Big|-|D-O|\Bigg)^2 \nonumber\\
		& \quad + \Bigg(\Big|u(A+\alpha)-u(F+\alpha)\Big|-|A-F|\Bigg)^2 + \Bigg(\Big|u(D+\alpha)-u(E+\alpha)\Big|-|D-E|\Bigg)^2  .
		\label{eqn:kagome-intro-energy}
	\end{align}

\edit{A note is in order about \eqref{eqn:kagome-intro-energy}. The elastic energy of a (geometrically nonlinear) Hookean spring should be a constant times $\mbox{(strain)}^2$. The spring energy used in \eqref{eqn:kagome-intro-energy} can easily be put in this form: 
$\big( |u(A)-u(O)|-|A-O| \big)^2 = \ell^2 \left(\frac{|u(A)-u(O)|}{|A-O|} - 1 \right)^2$ with $\ell = |A-O|$. Evidently, in writing 
\eqref{eqn:kagome-intro-energy} we have made a specific choice of the spring constants; this is strictly for simplicity -- our framework would apply equally well if each spring had a different (positive) spring constant. We noted earlier that $E(u,U)$ has the dimensions of energy per unit volume, while the coordinates of $U$ and the deformation $u$ are nondimensional. Thus the entire right side of \eqref{eqn:kagome-intro-energy} should in principle be multiplied by a constant with the dimensions of energy per unit volume; for simplicity, we have set this to $1$. 
These remarks apply equally to the other examples considered in this paper.}
	
	The preceding example is somewhat unusual, because the definition of $E(u,U)$ uses only the values of $u$ at nodes that belong to
	$\overline{U}$. Our framework does not require this, and such a choice is indeed impossible for many lattices of springs. As an
	example, consider the ``square lattice with long-range interactions'' shown in \cref{fig:square-long}. The obvious unit cell $U$
	is a square with vertices $O,A,B,C$, and the natural way to define the elastic energy $E(u,U+\alpha)$ on the translated
	unit cell $U+\alpha$ is\footnote{The factor in the last row of \cref{eqn:square-long-intro-energy} is 1 because the edge $OD$ lies entirely within the unit cell $U$, whereas the other edges are equally shared with neighboring cells.}
	\begin{align}
		E(u,U+\alpha) &= \frac{1}{2} \Bigg(\Big|u(A+\alpha)-u(O+\alpha)\Big|-|A-O|\Bigg)^2 +
		\frac{1}{2} \Bigg(\Big|u(B+\alpha)-u(O+\alpha)\Big|-|B-O|\Bigg)^2 \nonumber\\
		& \quad + \frac{1}{2} \Bigg(\Big|u(B+\alpha)-u(C+\alpha)\Big|-|B-C|\Bigg)^2 +
		\frac{1}{2} \Bigg(\Big|u(A+\alpha)-u(C+\alpha)\Big|-|A-C|\Bigg)^2 \nonumber\\
		& \quad + \Bigg(\Big|u(O+\alpha)-u(D+\alpha)\Big|-|O-D|\Bigg)^2 .
		\label{eqn:square-long-intro-energy}
	\end{align}
	Evidently: the presence of long springs can require that $E(u,U)$ depend on \edit{the values of $u$ at nodes} some distance from $U$.
	
	This brings us to an important assumption which was omitted from the informal discussion in \cref{subsec:intro-effective}:
	we shall assume that for some positive integer $n$,
	\begin{align} \label{eqn:U_n} 
		\mbox{$E(u,U)$ depends only on the values of $u$ in the} & \mbox{ closure of the expanded cell} \nonumber \\
		U_n := \bigcup \limits_{\alpha_i \in [-(n-1),n-1] \cap \mathbb{Z}} & U + \alpha
	\end{align}
	with the usual convention $\alpha = \sum_{i=1}^N \alpha_i v_i $ \, .

	\subsection{The piecewise linearization of a deformation, our basic energy bounds, and \texorpdfstring{$U_m$}{Lg}} \label{subsec:the-piecewise-linearization}
	While an elastic \edit{deformation} is characterized by its values at the nodes of the lattice, we want to also view it as a piecewise linear function defined on a suitable mesh. This is convenient because deformations \edit{defined on} the scaled lattice can then be viewed as functions in a finite-dimensional subspace of $H^1$.
	
	To this end, for any given lattice we fix -- in addition to the structure introduced so far -- a
	triangulation\footnote{In dimension $3$ or more, this would be a decomposition
		of $U$ into simplices rather than triangles,
		however we shall use the term triangulation in any space dimension -- a harmless abuse of language. Our ``piecewise linear''
		functions are, of course, actually piecewise affine.}
	of the unit cell $U$.
	The vertices of the triangulation must include the nodes of the lattice that lie in $\overline{U}$. We also permit the
	triangulation to use vertices that are not lattice nodes. At any non-lattice-node vertex $y$, we choose a way of writing $y$ as a convex combination of finitely many lattice nodes (which might not belong to $\overline{U}$)
	\begin{equation} \label{eqn:piecewise-linearization-rule-a}
		y = \sum_j \theta_j z_j \quad \mbox{where each $z_j$ is a lattice node, $0 < \theta_j < 1$, and $\sum_j \theta_j = 1$}.
	\end{equation}
	These non-lattice-node vertices are known as ``ghost vertices''. At each ghost vertex $y$, we take the value of the deformation to be
	\begin{equation} \label{eqn:piecewise-linearization-rule-b}
		u(y) = \sum_j \theta_j u(z_j) \, .
	\end{equation}
	This rule has the crucial property that it \emph{preserves affine functions} -- in other words, if
	$u(x) = \lambda \cdot x + c$ at the lattice nodes for some $N \times N$ matrix $\lambda$ and some
	$c \in \mathbb{R}^N$, then $u(y) =  \lambda \cdot y + c$ at every vertex of the triangulation, so the resulting
	piecewise linear function is pointwise equal to $u(x) = \lambda \cdot x + c$.
	
	Given a piecewise linearization rule for the unit cell, we naturally obtain one for all $\mathbb{R}^N$ by
	periodic extension. It, too, preserves affine functions.
	
	The piecewise linearizations of deformations play a fundamental role in our analysis. For one thing, they make it easy to
	discuss what it means for a family of deformations $u^\epsilon$ defined on $\epsilon$-scaled versions of the lattice to converge
	as $\epsilon \rightarrow 0$ to a limit $u$, since the piecewise linearizations of $u^\epsilon$ are defined
	everywhere, not just at nodes. They also give sense to the upper and lower bounds that we require our unit cell energy $E(u,U)$
	to satisfy, namely:
	\begin{equation} \label{eqn:unscaled-upper-bound}
		E(u,U) \leq C_1 \Big( |\nabla u|^2_{L^2(U_{n})} + |U_{n}| \Big )
	\end{equation}
	and
	\begin{equation} \label{eqn:unscaled-lower-bound}
		E(u,U) \geq \max \Big\{ C_2 \Big( |\nabla u|^2_{L^2(U))} - D_2 |U| \Big) , 0 \Big\}
	\end{equation}
	for some positive constants 
    $C_1$, $C_2$, and $D_2$.\footnote{\edit{The constants $C_1$ and $C_2$ have the dimensions of energy per unit volume, while $D_2$ is dimensionless. Indeed, we observed earlier that $E(u,U)$ has dimensions of energy per unit volume, while its arguments $u$ and $U$ are dimensionless.}} 
    We discuss in \cref{sec:examples} how triangulations satisfying such
	estimates can be obtained for various 2D examples. Here, let us simply mention that for the 2D Kagome example with the unit cell
	shown in \cref{fig:kagome-3}, the triangular mesh consisting of
	$\Delta AOB, \Delta BOC$, $\Delta COD$, $\Delta AOF, \Delta DOF, \Delta DEF$ is a convenient choice.
	
	The upper bound \cref{eqn:unscaled-upper-bound} -- more precisely, the scaled version that we'll
	discuss presently -- makes it natural that our gamma-limit be defined for $u$ such that $\int_\Omega |\nabla u|^2 \, dx$ is
	finite. As the reader is by now well aware, we are especially interested
	in structures with mechanisms. The presence of a negative term $D_2 |U|$ in the lower bound \cref{eqn:unscaled-lower-bound} is
	crucial in that setting; indeed, $D_2 |U|$ must clearly be larger than the maximum of $|\nabla u|^2_{L^2(U)}$ as $u$
	ranges over mechanisms (that is, over deformations such that $E(u,U) = 0$).
	
	In using the $L^2$ norm of $\nabla u$ in the upper and lower bounds, we have made a choice. For a lattice of springs, conditions
	\eqref{eqn:unscaled-upper-bound} and \eqref{eqn:unscaled-lower-bound} place no constraint on the springs' character at small or
	even moderate strains, due to the terms involving $|U_n|$ and $|U|$ on the right hand side. However, these conditions require
	that the energies of the springs be of order $(\mbox{strain})^2$ as $\mbox{strain} \rightarrow \infty$. While our
	arguments and results have natural analogues when the $L^2$ norms in \eqref{eqn:unscaled-upper-bound} and
	\eqref{eqn:unscaled-lower-bound} are replaced by $L^p$ norms with $1 < p < \infty$, restricting our attention
	to $p=2$ simplifies the discussion. Since it includes the case of Hookean springs -- and since our motivation lies mainly
	in considering low-energy structures -- this choice also seems quite natural from a mechanical viewpoint.
	
	We note that on the right hand side of the bounds \eqref{eqn:unscaled-upper-bound} and \eqref{eqn:unscaled-lower-bound},
	the terms $|\nabla u|_{L^2(U_n)}^2$ and $|\nabla u|_{L^2(U)}^2$ refer to our piecewise linearization of $u$. If the triangulation has vertices that are not lattice
	nodes, then these terms may depend on values of the deformation at lattice nodes in nearby cells. We shall assume, however, that
	\begin{align} \label{eqn:max-dependence-of-grad-u}
		\mbox{the restriction of $\nabla u$ to $U_n$ depends only on the values} & \mbox{ of $u$ at lattice nodes in the closure of} \nonumber\\
		U_m := \bigcup \limits_{\alpha_i \in [-(m-1),m-1] \cap \mathbb{Z}} & U + \alpha \, .
	\end{align}
	(Here $n$ and $U_n$ are defined by \eqref{eqn:U_n}, and it is clear from the definition that $m \geq n$.) It will be convenient
	later to have a name for the largest distance between two points in $U_m$, so we define
	\begin{equation} \label{eqn:d_m}
		d_m = \sup_{x,y \in U_m} |x-y| \, .
	\end{equation}
	\begin{remark}
		It is worth noting that the upper bound in \cref{eqn:unscaled-upper-bound} (or the scaled version in \cref{eqn:unit-cell-upper}) for the unit cell energy depends implicitly on the value of $m$, due to the potential presence of ghost vertices. When ghost vertices are not needed to define the affine interpolation of the discrete deformations on the lattice, we have $m = n$.
	\end{remark}
	
	
	\subsection{The scaled lattice, \texorpdfstring{$E^\epsilon(u^\epsilon, \Omega)$}{Lg}, and our admissible deformations} \label{subsec:the-scaled-lattice}
	
	We have already introduced the scaled lattice and the scaled energy in \cref{subsec:intro-effective}. The nodes of the scaled
	lattice are
	\begin{equation} \label{eqn:cal-V-epsilon}
		\mathcal{V}^\epsilon := \epsilon \mathcal{V} \, ;
	\end{equation}
	the translated unit cells of this lattice are $\epsilon U + \alpha$ with $\alpha$ as in
	\eqref{alpha-as-transl-scaled}; and the scaled energy $E^\epsilon(u^\epsilon, \epsilon U + \alpha)$ was defined via elasticity
	scaling in \eqref{eqn:elasticity-scaling}. As an example: for the Kagome lattice with $U$ and $E(u,U)$ given
	by \cref{fig:kagome-3} and \cref{eqn:kagome-intro-energy}, if all the springs have length $1$ in the unscaled
	setting then
	\begin{align*}
		\edit{E^\epsilon} (u^\epsilon,\epsilon U+\alpha) &= \Bigg(\Big|u^\epsilon(\epsilon A+\alpha)-u^\epsilon(\epsilon O+\alpha)\Big|-\epsilon\Bigg)^2 + \Bigg(\Big|u^\epsilon(\epsilon B+\alpha)-u^\epsilon(\epsilon O+\alpha)\Big|-\epsilon\Bigg)^2\\
		&+ \Bigg(\Big|u^\epsilon(\epsilon C+\alpha)-u^\epsilon(\epsilon O+\alpha)\Big|-\epsilon\Bigg)^2 + \Bigg(\Big|u^\epsilon(\epsilon D+v_\alpha)-u^\epsilon(\epsilon O+\alpha)\Big|-\epsilon\Bigg)^2\\
		&+ \Bigg(\Big|u^\epsilon(\epsilon A+\alpha)-u^\epsilon(\epsilon F+\alpha)\Big|-\epsilon\Bigg)^2 + \Bigg(\Big|u^\epsilon(\epsilon D+\alpha)-u^\epsilon(\epsilon E+\alpha)\Big|-\epsilon\Bigg)^2 \, .
	\end{align*}
	Our unscaled upper and lower bounds \eqref{eqn:unscaled-upper-bound} and \eqref{eqn:unscaled-lower-bound} have scaled versions, of course.
	Their right hand sides involve the piecewise linearization of $u^\epsilon$ (determined by our unscaled piecewise linearization scheme
	and elasticity scaling, or equivalently by using \eqref{eqn:piecewise-linearization-rule-a}--\eqref{eqn:piecewise-linearization-rule-b} when $y$ is
	a vertex of the scaled triangulation and $\{ z_j \}$ are nodes of the scaled lattice).
	
	While our conditions on the energy have already been discussed, it is convenient to collect them in one place.
	Since we'll mainly be using the scaled versions, we state those here:
	
	\begin{enumerate}[(1)]
		\item the energy on the $\epsilon$-scale unit cell is periodic, i.e. we have
		\begin{align}
			E^\epsilon (u^\epsilon(x+\alpha),\epsilon U+\alpha) = E^\epsilon (u^\epsilon,\epsilon U)
			\label{eqn:unit-cell-eps-periodicity}
		\end{align}
		for any $\alpha = \sum_{i=1}^N \alpha_i v_i$ with $\alpha_i \in \epsilon \mathbb{Z}$ ;
		
		\item the energy on the $\epsilon$-scale unit cell must be translation-invariant, in the sense that for any vector
		$c \in \mathbb{R}^N$, we have
		\begin{align}
			E^\epsilon (u^\epsilon,\epsilon U+\alpha) = E^\epsilon (u^\epsilon+c,\epsilon U+\alpha) \, ;
			\label{eqn:unit-cell-translation-invariant}
		\end{align}
		
		\item an upper bound: there exists $C_1 > 0$ (independent of $\alpha$ and $\epsilon$) such that
		\begin{align}
			E^\epsilon (u^\epsilon,\epsilon U+\alpha) \leq C_1 \Big(|\nabla u^\epsilon|^2_{L^2(\epsilon U_{n}+\alpha)} +
			|\epsilon U_{n}+\alpha|\Big)
			\label{eqn:unit-cell-upper}
		\end{align}
		where $U_n$ is defined by \eqref{eqn:U_n}; and
		
		\item a lower bound: there exist $C_2 > 0$ and $D_2 \geq 0$ (independent of $\alpha$ and $\epsilon$) such that
		\begin{align}
			E^\epsilon (u^\epsilon,\epsilon U + \alpha) \geq \max \Big\{C_2 \Big(|\nabla u^\epsilon|^2_{L^2(\epsilon U+\alpha)} -
			D_2|\epsilon U + \alpha|\Big), 0 \Big\} \, .
			\label{eqn:unit-cell-lower}
		\end{align}
	\end{enumerate}
	(Note that we have included positivity in \cref{eqn:unit-cell-lower} rather than stating it as a separate condition. As
	already mentioned in \cref{subsec:intro-effective}, our energy \emph{need not} be a continuous function of the \edit{nodal values of the deformation}.)
	
	Turning now to the energy of a domain: we offered a definition of $E^\epsilon(u^\epsilon, \Omega)$ in
	\cref{eqn:intro-energy-of-Omega}, which we repeat here for the reader's convenience:
	\begin{align*}
		E^\epsilon(u^\epsilon, \Omega) &:= \sum_{\alpha \in R_\epsilon(\Omega)} E^\epsilon (u^\epsilon, \epsilon U + \alpha) \, ;
	\end{align*}
	however to make this precise we need to define the set over which the sum ranges. When considering the limiting energy of a
	fixed domain $\Omega$, it is natural to focus on deformations that are defined at lattice nodes in $\Omega$, in other
	words $u^\epsilon$ in
	\begin{align}
		\mathcal{A}_\epsilon(\Omega) &= \{u^\epsilon(x)\; |\; u^\epsilon(x) \text{ has values on } \mathcal{V}^\epsilon \cap \Omega\} \, .
		\label{eqn:admissible-eps}
	\end{align}
	Now recall from \eqref{eqn:max-dependence-of-grad-u} that to be sure right hand sides of our unscaled energy bounds
	are fully determined, we need $u$ to have values at lattice nodes in the closure of $U_m$. Scaling this statement,
	we see that the cells $\epsilon U + \alpha$ included in the definition of $E^\epsilon(u^\epsilon, \Omega)$ should have
	their closures contained in $\Omega$. To enforce this, we define
	\begin{align}
		R_\epsilon(\Omega) &:= \{\alpha = \sum_{i=1}^N \alpha_i v_i  \: : \: \alpha_i \in \epsilon \mathbb{Z} \quad \mbox{and} \quad
		\epsilon U_m + \alpha \ssubset \Omega \} \, ,
		\label{eqn:R_eps}
	\end{align}
	using the usual convention that $A \ssubset B$ means $\overline{A} \subset B$.
	
	We note that $\Omega$ need not be an open set for $E^\epsilon(u^\epsilon,\Omega)$ to be well-defined, and no regularity
	is needed for $\partial \Omega$. While our gamma-convergence results (theorems \ref{thm:main-theorem} and
	\ref{thm:theorem-dirichlet}) are restricted to Lipschitz domains, in the course of the proofs it will sometimes be
	convenient to consider the discrete energy of a domain whose boundary is not obviously Lipschitz.
	
	The upper and lower bounds \eqref{eqn:unit-cell-upper} -- \eqref{eqn:unit-cell-lower} tell us that the $\epsilon$-scale problem
	$\min_{u^\epsilon \in \mathcal{A}_\epsilon (\Omega)} E^\epsilon ( u^\epsilon, \Omega)$ is more or less a variational problem posed in
	a finite-dimensional subspace of $H^1(\Omega)$. In preparation for rigorous analysis, it is important to be clear about the class of
	admissible deformations. It is slightly different from $\mathcal{A}_\epsilon (\Omega)$, since we want to treat $u^\epsilon$
	\emph{both} as a function defined at nodes of the scaled lattice \emph{and} as a piecewise linear function, though in a neighborhood of
	$\partial \Omega$ the piecewise linearization of a deformation may depend on its values at nodes outside $\Omega$.
	While we usually use the same notation $u^\epsilon$ for both a deformation defined at lattice nodes and its
	piecewise linearization, for clarity we suspend this practice for the following definition.
	
	\begin{definition} \label{defn:admissible-deformation}
		An admissible deformation is a pair $(u^\epsilon, \tilde{u}^\epsilon)$ such that
		\begin{enumerate}
			\item[(a)] $u^\epsilon$ belongs to $\mathcal{A}_\epsilon (\Omega)$, i.e. it is a deformation defined at all nodes of the scaled
			lattice that lie in $\Omega$;
			\item[(b)] $\tilde{u}^\epsilon \in H^1(\Omega)$ is the restriction to $\Omega$ of a piecewise linear function
			obtained by applying our piecewise linearization scheme to some deformation defined at nodes of the scaled lattice;
			\item[(c)] $\tilde{u}^\epsilon = u^\epsilon$ at all nodes of the scaled lattice that lie in $\Omega$; moreover,
			$\tilde{u}^\epsilon$ agrees with our piecewise linearization of $u^\epsilon$ at all vertices of the
			triangulation where the piecewise linearization of $u^\epsilon$ is fully determined (i.e. where its value depends
			only on the deformation at nodes of the scaled lattice that lie in $\Omega$).
		\end{enumerate}
	\end{definition}
	In practice we will usually drop the tilde, writing $u^\epsilon$ instead of $\tilde{u}^\epsilon$. No confusion should result, since by
	(c) the two functions agree wherever they are both well-defined. This definition of the admissible deformations is convenient, because
	the piecewise-linear version of $u^\epsilon$ is now an element of the $\epsilon$-independent space $H^1(\Omega)$. We note that
	$E^\epsilon(u^\epsilon, \Omega)$ is finite for any admissible deformation, and summing the upper bounds
	\eqref{eqn:unit-cell-upper} for the relevant scaled cells gives
	\begin{align}
		E^\epsilon(u^\epsilon,\Omega) &\leq C_1 \Big(\sum_{\alpha \in R_\epsilon(\Omega)}|\nabla u^\epsilon|^2_{L^2(\epsilon U_{n}+\alpha)} +
		|\epsilon U_{n}+\alpha|\Big) \nonumber\\
		&\leq C_1 (2n-1)^N \big(|\nabla u^\epsilon|^2_{L^2(\Omega)} + |\Omega|\big) \, . \label{eqn:n-unit-upper}
	\end{align}
	(The second line holds since each integral $|\nabla u^\epsilon|^2_{L^2(\epsilon U+\alpha)}$ over the cell $\epsilon U + \alpha$
	can appear in the integral $|\nabla u^\epsilon|^2_{L^2(\epsilon U_{n}+\beta)}$ for some $\beta$ at most $(2n-1)^N$ times.)
	Evidently, the energy $E^\epsilon(u^\epsilon, \Omega)$ stays uniformly bounded when $|\nabla u^\epsilon|_{L^2(\Omega)}$ stays
	uniformly bounded.
	
	A similar calculation gives the following lemma, which will be used repeatedly.
	
	\begin{lemma} \label{lemma:constant-gradient-energy-bound}
		Suppose a collection of scaled unit cells $\{\epsilon U + \alpha^{(j)}\}_{j=1}^P$, a deformation $u^\epsilon$, a
		domain $\Omega$, and a constant $M$ have the properties that
		\begin{enumerate}
			\item[(a)] the piecewise linear representative of $u^\epsilon$ has $|\nabla u^\epsilon| \leq M$ on
			$\epsilon U_n + \alpha^{(j)}$ for each $j=1, \ldots, P$, and
			\item[(b)] each of the expanded cells $\epsilon U_n + \alpha^{(j)}$ is contained in $\Omega$ .
		\end{enumerate}
		Then
		$$
		\sum_{j=1}^P E^\epsilon (u^\epsilon, \epsilon U + \alpha^{(j)}) \leq C_1 (2n-1)^N (M^2 + 1) |\Omega| \, .
		$$
	\end{lemma}
	\begin{proof}
		It suffices to argue as we did for \eqref{eqn:n-unit-upper}.
	\end{proof}
	
	We also note that, as a consequence of the lower bound \eqref{eqn:unit-cell-lower}, control of
	$E^\epsilon (u^\epsilon, \Omega)$ implies control of the $L^2$ norm of $\nabla u^\epsilon$ in a
	slightly smaller domain. This too will be used repeatedly:
	
	\begin{lemma} \label{lemma:L2-bound-on-ueps}
		For any domain $\Omega$ and any admissible deformation $u^\epsilon \in \mathcal{A}_\epsilon (\Omega)$,
		\begin{equation} \label{eqn:L2-bound-on-ueps-no-bc}
			C_2 \sum_{\alpha \in R_\epsilon (\Omega)} \int_{\epsilon U + \alpha} |\nabla u^\epsilon|^2 \, dx \leq
			E^\epsilon(u^\epsilon, \Omega) + C_2 D_2 |\Omega| \, .
		\end{equation}
	\end{lemma}
	
	\begin{proof} The lower bound \eqref{eqn:unit-cell-lower} implies that
		$$
		C_2 \int_{\epsilon U + \alpha} |\nabla u^\epsilon|^2 \, dx \leq
		E^\epsilon (u^\epsilon, \epsilon U + \alpha) + C_2 D_2 |\epsilon U + \alpha| \, ,
		$$
		and our assertion follows by simply adding these inequalities over all $\alpha \in R_\epsilon(\Omega)$.
	\end{proof}

	\subsection{Boundary conditions, and gluing deformations together}
	\label{subsec:dir-bc-and-gluing}
	
	For a continous variational problem involving $\int_\Omega W(\nabla u) \, dx$, we can impose a Dirichlet
	boundary condition by specifying $u$ at $\partial \Omega$. Moreover, given a partition of $\Omega$ into
	two subdomains, we can construct a test function by specifying $u$ on each subdomain (using choices that
	agree at the partition boundary). Also, given two test functions $u_1$ and $u_2$, it can useful to
	interpolate between them by considering $\phi u_1 + (1-\phi) u_2$, where $\phi$ is smooth with $0 \leq \phi \leq 1$.
	It is well-known that things are different in the context of discrete-to-continuous limits. Focusing on
	the framework of this paper, the issues are two-fold:
	\begin{enumerate}
		\item[(a)] For any scaled cell $\epsilon U + \alpha$, the associated energy
		$E^\epsilon (u^\epsilon, \epsilon U + \alpha)$ can depend on the values of $u^\epsilon$ at nodes of the
		scaled lattice in the larger set $\epsilon U_n + \alpha$. Moreover, our basic upper bound
		\eqref{eqn:unit-cell-upper} involves the piecewise linearization of $u^\epsilon$ on $U_n$ -- which can
		depend on the values of $u^\epsilon$ in the still larger set $\epsilon U_m + \alpha$. Thus, our discrete
		problem is (a little bit) nonlocal.
		
		\item[(b)] Our use of piecewise linearization introduces an additional issue. Consider, for example, the
		construction of a test function by interpolation, whereby $u^\epsilon = \phi u^\epsilon_1 + (1-\phi) u^\epsilon_2$
		at nodes of the scaled lattice. Alas, the piecewise linearization of $u^\epsilon$ is not given by this formula.
		Therefore rather than use product rule to calculate $\nabla u^\epsilon$, we must use information from the
		piecewise linearization scheme.
	\end{enumerate}
	Point (a) is rather standard, and our way of dealing with it is rather standard as well. Point (b) is
	less standard; we shall deal with it using some basic ideas from numerical analysis.
	
	We begin with a discussion of ``Dirichlet boundary conditions.'' Given a domain $\Omega \subset \mathbb{R}^N$ and
	a Lipschitz continuous function $\psi \, : \, \partial \Omega \rightarrow \mathbb{R}^N$, how shall we
	impose in our discrete setting something similar to $u = \psi$  at
	$\partial \Omega$? Replacing $u$ by $u - \psi$, it suffices to discuss the discrete analogue of
	$u = 0$ at $\partial \Omega$. Due to the nonlocality of the discrete problem, when working at scale $\epsilon$
	we must require that $u^\epsilon$ \emph{vanish in a layer} near $\partial \Omega$,
	whose thickness is of order $\epsilon$:
	\begin{definition} \label{defn:dirichlet-condition-zero}
		For any domain $\Omega$, let
		\begin{equation}  \label{eqn:defn-omega-eps}
			\Omega_\epsilon = \Big\{x \in \Omega\: \big| \: \text{dist}(x,\partial \Omega) >  \epsilon d_m \Big\} \, ,
		\end{equation}
		where $d_m$ is defined by \eqref{eqn:d_m}. We shall say that an admissible deformation $u^\epsilon$
		``vanishes at $\partial \Omega$'' if
		it belongs to
		\begin{equation} \label{eqn:defn-A0-eps}
			\mathcal{A}_\epsilon^0(\Omega) = \Big\{ u^\epsilon \in \mathcal{A}_\epsilon(\Omega)\: \big| \:
			u^\epsilon = 0 \text{ as a piecewise linear function on } \Omega \setminus \Omega_\epsilon \Big\} \, .
		\end{equation}
	\end{definition}
	
	The logic behind this definition is captured by the following two observations.
	
	\begin{remark} \label{rmk:dirichlet-bc} If we extend $u^\epsilon \in \mathcal{A}_\epsilon^0(\Omega) $ by giving
		it the value $0$ at nodes of the scaled lattice outside $\Omega$, then the piecewise linearization of the
		extended deformation is the same as $u^\epsilon$ in the entire domain $\Omega$.
		To see why, recall that for any scaled unit cell $\epsilon U + \alpha$, the piecewise linearization of
		$u^\epsilon$ in $\epsilon U_n + \alpha$ depends only on the values of $u^\epsilon$ at scaled lattice
		nodes in $\epsilon \overline{U}_m + \alpha$. Denoting the extended deformation by $\tilde{u}^\epsilon$,
		our claim is that the piecewise linearization of $\tilde{u}^\epsilon$ is equal to the piecewise linear
		function $u^\epsilon$ in the entire domain $\Omega$. In fact:
		\begin{itemize}
			\item If $\epsilon \overline{U}_m + \alpha$ is contained in $\Omega$ then the piecewise
			linearizations of $u^\epsilon$ and $\tilde{u}^\epsilon$ in $\epsilon U_n + \alpha$ are fully
			determined by the values of $u^\epsilon$ at nodes of the scaled lattice in $\Omega$. Thus, in this case
			$\tilde{u}^\epsilon = u^\epsilon$ in $\epsilon U_n + \alpha$, which is contained in $\Omega$.
			
			\item If, on the other hand, $\epsilon \overline{U}_m + \alpha$ meets the complement of $\Omega$,
			then (using the definition of $d_m$ and $\Omega_\epsilon$), $\epsilon \overline{U}_m + \alpha$ lies in
			the exterior of $\Omega_\epsilon$. Since $u^\epsilon \in \mathcal{A}_\epsilon^0(\Omega)$,
			it vanishes at nodes of the scaled lattice that lie in $\Omega \setminus \Omega_\epsilon$. Thus, the extended
			deformation $\tilde{u}^\epsilon$ vanishes at all nodes of the scaled lattice that belong to
			$\epsilon \overline{U}_m + \alpha$. Since our piecewise linearization scheme preserves affine functions,
			the piecewise linearization of $\tilde{u}^\epsilon$ is identically zero in $\epsilon U_n + \alpha$.
			Using that the piecewise linear version of $u^\epsilon$ vanishes in $\Omega \setminus \Omega_\epsilon$,
			we conclude that $u^\epsilon$ and $\tilde{u}^\epsilon$ both vanish identically (as piecewise linear
			functions) on $(\epsilon U_n + \alpha ) \cap \Omega$.
		\end{itemize}
		As $\alpha$ varies, the sets $(\epsilon U_n + \alpha) \cap \Omega$ cover
		the entire set $\Omega$; thus $\tilde{u}^\epsilon = u^\epsilon$ in all $\Omega$, as asserted.
	\end{remark}
	
	\begin{remark}\label{rmk:bdry-layer}
		Suppose $\Omega$ is partitioned into two subdomains $\Omega_1$ and $\Omega_2$, and suppose
		$u_i^\epsilon \in \mathcal{A}^0_\epsilon(\Omega_i)$ for $i=1,2$. Then
		\begin{align*}
			u^\epsilon(x) = \begin{cases}
				u_1^\epsilon(x) & x \in \Omega_1 \\
				u_2^\epsilon(x) & x \in \Omega_2
			\end{cases}
		\end{align*}
		is an admissible deformation. When viewed as a piecewise linear function, $u^\epsilon(x) = 0$ in the region
		$\text{dist}(x,\partial \Omega_1 \cap \partial \Omega_2) \leq \epsilon d_m$; in particular, $u^\epsilon=0$
		at all nodes $x$ of the scaled lattice that lie in this region. This is consistent with our
		piecewise linearization scheme, since for any cell $\epsilon U + \alpha$ either
		\begin{enumerate}
			\item[(i)] $(\epsilon \overline{U}_m + \alpha) \cap \Omega_1 = \emptyset$, in which case the piecewise linearization of
			$u^\epsilon$ in $\epsilon U_n + \alpha$ is clearly $u_2^\epsilon$; or
			
			\item[(ii)] $(\epsilon \overline{U}_m + \alpha) \cap \Omega_2 = \emptyset$, in which case the piecewise linearization of
			$u^\epsilon$ in $\epsilon U_n + \alpha$ is clearly $u_1^\epsilon$; or
			
			\item[(iii)] $(\epsilon \overline{U}_m + \alpha)$ meets the common boundary $\partial \Omega_1 \cap \partial \Omega_2$, in which
			case $u^\epsilon = 0$ at all nodes of the scaled lattice in $\epsilon \overline{U}_m + \alpha$. Since our piecewise linearization
			scheme preserves affine functions, this is consistent (as expected) with $u^\epsilon$ being zero as a piecewise linear function
			in $\epsilon U_n + \alpha$.
		\end{enumerate}
	\end{remark}
	
	Turning now to a different issue:
	suppose $\phi$ is a continuous, piecewise linear function (with a macroscopic
	mesh that has nothing to do with our piecewise linearization scheme). What happens when we
	``discretize it'' by taking $u^\epsilon = \phi$ at nodes of the scaled lattice? The piecewise linearization
	of this $u^\epsilon$ is \emph{not} everywhere equal to $\phi$. Indeed, it agrees with $\phi$ at points which are
	far enough from a change in $\nabla \phi$; but due to the nonlocality of our piecewise linearization scheme,
	it will be different from $\phi$ in an order-$\epsilon$-thick layer around the set where $\nabla \phi$ changes. To
	show that this layer has negligible effect on the total energy, we need an upper bound on $\nabla u^\epsilon$.
	This is a typical application of the following result, whose proof is given in appendix \ref{appendix:piecewise-linearization-lemmas}:
	
	\begin{lemma} \label{lemma:interp-of-lip-fn}
		For any Lipschitz continuous function $\phi$ and any cell $\epsilon U + \alpha$ of the scaled lattice,
		suppose $u^\epsilon= \phi$ at all nodes of the scaled lattice that lie in $\overline{U}_m+ \alpha$.
		Then the piecewise linearization of $u^\epsilon$ satisfies
		\begin{align}
			| u^\epsilon |_{L^\infty(\epsilon U_n + \alpha)} & \leq
			|\phi |_{L^\infty(\epsilon U_m + \alpha)} \, , \label{eqn:interp-of-lip-fn-sup-norm} \\
			| \nabla u^\epsilon |_{L^\infty(\epsilon U_n + \alpha)} & \leq
			C |\nabla \phi |_{L^\infty(\epsilon U_m + \alpha)} \, , \mbox{ and} \label{eqn:interp-of-lip-fn} \\
			| u^\epsilon - \phi |_{L^\infty(\epsilon U_n + \alpha)} & \leq
			C' \epsilon |\nabla \phi |_{L^\infty(\epsilon U_m + \alpha)} \, . \label{eqn:interp-of-lip-fn-convergence}
		\end{align}
		The constants $C$ and $C'$ in the latter two estimates depend on the details of our piecewise linearization scheme,
		but not on $\epsilon$ or $\phi$.
	\end{lemma}
	Finally we turn to a third issue, namely: estimating the piecewise linearization of
	$\phi u_1^\epsilon + (1-\phi) u_2^\epsilon$. (This issue arises in our version of an
	argument due to De Giorgi, which is briefly discussed near the end of the proof
	of \cref{lemma:affine} then presented in full detail in appendix \ref{appendix:degiorgi}.) The required
	estimate is provided by the following result:
	
	\begin{lemma} \label{lemma:interp-of-two-defs}
		For any Lipschitz continuous function $\phi$, any cell $\epsilon U + \alpha$ of the scaled lattice, and any
		deformation $u^\epsilon$ that is defined at all nodes of the scaled lattice in $\overline{U}_m + \alpha$,
		suppose a deformation $h^\epsilon$ has
		$$
		h^\epsilon = \phi u^\epsilon \quad \mbox{at nodes of the scaled lattice in $\epsilon \overline{U}_m + \alpha$ \, .}
		$$
		Then the piecewise linearization of $h^\epsilon$ satisfies
		\begin{align}
			|\nabla h^\epsilon|^2_{L^2(\epsilon U_n + \alpha)} &\leq C \Big(
			|u^\epsilon|^2_{L^2(\epsilon U_m + \alpha)} |\nabla \varphi|^2_{L^\infty(\epsilon U_m + \alpha)} +
			|\nabla u^\epsilon|^2_{L^2(\epsilon U_n + \alpha) } |\varphi|^2_{L^\infty(\epsilon U_n + \alpha)} \Big) \ \mbox{and} \label{eqn:interpolate-scale-eps} \\
			|h^\epsilon|^2_{L^2(\epsilon U_n + \alpha)} &\leq
			C' |u^\epsilon|^2_{L^2(\epsilon U_m + \alpha)} |\varphi|^2_{L^\infty(\epsilon U_m + \alpha)} \label{eqn:interpolate-scale-eps-L2}
		\end{align}
		where the norms of $u^\epsilon$ on the right refer, as usual, to its piecewise linearization.
		The constants $C$ and $C'$ in this estimate depend only on the details of our piecewise linearization
		scheme; in particular, they do not depend on $\epsilon$, $\phi$, or $u^\epsilon$.
	\end{lemma}
	
	\noindent The proof of \cref{lemma:interp-of-two-defs} is similar to (but more complicated than) that of \cref{lemma:interp-of-lip-fn}.
	It, too, is presented in appendix \ref{appendix:piecewise-linearization-lemmas}.
	
	\subsection{Statements of our main results} \label{subsec:statements-of-main-results}
	Since our theorems use the notion of $\Gamma$-convergence, we start by defining what this means in the present context.
	Here and throughout the paper, the notation $u^\epsilon \rightharpoonup u$ means that $\{ u^\epsilon \}$ remains uniformly
	bounded in $H^1(\Omega)$ and converges \emph{weakly} to $u$.
	
	\begin{definition}[$\Gamma$-convergence] \label{defn:gamma-convergence}
		We say that the family of discrete functionals $\{E^\epsilon(u^\epsilon,\Omega)\}$ $\Gamma$-converges to a functional $E_{\text{eff}}(u,\Omega)$ (with respect to the weak topology of $H^1(\Omega)$) if
		\begin{enumerate}[(i)]
			\item for every admissible sequence $\{u^\epsilon\}_{\epsilon > 0}$ with $u^\epsilon \rightharpoonup u$ in
			$H^1(\Omega)$, we have
			\begin{align}
				\liminf_{\epsilon \rightarrow 0} E^\epsilon(u^\epsilon, \Omega)  & \geq E_{\text{eff}}(u,\Omega) \, , \quad \mbox{and} \label{eqn:liminf-domain}
			\end{align}
			\item for every $u \in H^1(\Omega)$, there is an admissible sequence $\{u^\epsilon\}_{\epsilon > 0}$ such that
			$u^\epsilon \rightharpoonup u$ in $H^1(\Omega)$ and
			\begin{align}
				\lim_{\epsilon \rightarrow 0} E^\epsilon(u^\epsilon, \Omega)  &= E_{\text{eff}}(u,\Omega) \, .
				\label{eqn:limsup-domain}
			\end{align}
		\end{enumerate}
	\end{definition}
	
	It is easy to see that when this holds, the minimizers of $E_{\text{eff}}(u,\Omega)$ are precisely the weak limits of minimizing
	sequences of $E^\epsilon(u^\epsilon, \Omega)$. Thus, for a mechanism-based mechanical metamaterial we view the deformations with
	effective energy zero as soft modes, since they capture the macroscopic behavior of sequences $u^\epsilon$ for which
	$E^\epsilon (u^\epsilon, \Omega)$ tends to zero.
	
	Our main result is the following:
	\begin{theorem}\label{thm:main-theorem}
		For any bounded, Lipschitz domain $\Omega$ the discrete energies $E^\epsilon (u^\epsilon, \Omega)$
		$\Gamma$-converge in $H^1(\Omega)$ (with respect to the weak topology of $H^1(\Omega)$) to an effective energy of the form
		\begin{align}
			E_{\text{eff}}(u, \Omega) &= \int_\Omega \overline{W}(\nabla u)\: dx \, .
			\label{eqn:effective-energy}
		\end{align}
		Moreover, the effective energy density $\overline{W}(\lambda)$ is independent of the domain $\Omega$, and it has the following variational characterization:
		\begin{align} \label{eqn:effective-energy-density}
			\overline{W}(\lambda) &= \inf_{k \in \mathbb{N}} \inf_{\psi \in \mathcal{A}^0(kU)} \frac{1}{k^N |U|} \sum_{\alpha_1, \dots, \alpha_N=0}^{k-1} E \left( \lambda x + \psi, U+ \sum_{i=1}^N \alpha_i v_i \right),
		\end{align}
		where
		\begin{equation} \label{eqn:defn-kU}
			kU = \bigcup_{\alpha_1, \dots, \alpha_N=0}^{k-1} \Big(U+\sum_{i=1}^N \alpha_i v_i \Big)
		\end{equation}
		and $\mathcal{A}^0(kU)$ is the space of (unscaled) deformations of $kU$ that ``vanish at the boundary'' in the sense of
		\cref{defn:dirichlet-condition-zero}, in other words
		\begin{align*}
			\mathcal{A}^0(kU) = \left\{
			\begin{array}{ll}
				\mbox{admissible deformations $\psi$ defined on $kU$ whose piecewise} \\
				\mbox{linear representatives vanish when dist$(x,\partial (kU)) \leq d_m$}
			\end{array}
			\right\} \, .
		\end{align*}
		(Note that this variational characterization uses the unscaled lattice, and that
		in \eqref{eqn:effective-energy-density}, the expression $E(\lambda x + \psi, U + \sum_{i=1}^N \alpha_i v_i)$ is our
		unscaled energy.)
	\end{theorem}
	
	\begin{remark} \label{rmk:word-on-defn-of-wbar}
		A word is in order about the meaning of the \eqref{eqn:effective-energy-density}, since $E(u,U + \alpha)$ depends
		on the values of $u$ in $\overline{U}_n + \alpha$, and for some of the terms
		$E(\lambda x + \psi, U + \alpha)$ in \eqref{eqn:effective-energy-density} the set $\overline{U}_n + \alpha$ extends beyond
		$kU$. Since $\psi \in \mathcal{A}^0(kU)$ and recalling \cref{rmk:dirichlet-bc}, we evaluate these terms by taking
		$\psi = 0$ outside $kU$.
	\end{remark}
	
	As the reader knows very well by now, we are especially interested in mechanical metamaterials. To explore the mechanical
	response of a metamaterial, it is natural to consider what happens when the deformation is specified at the boundary. This
	calls for an analogue of \cref{thm:main-theorem} with a Dirichlet boundary condition. Of course, for the discrete problem
	at scale $\epsilon$ the ``boundary condition'' must be imposed in an order-$\epsilon$-thickness layer near $\partial \Omega$:
	if $\psi$ is an $R^N$-valued function defined near $\partial \Omega$, we say an admissible deformation ``has boundary
	condition $\psi$'' if it belongs to
	\begin{align} \label{eqn:omega-eps-varphi}
		\mathcal{A}_\epsilon^\psi(\Omega) = \Big\{u^\epsilon \in \mathcal{A}_\epsilon(\Omega) \:\big|\:  u^\epsilon - \psi \in \mathcal{A}_\epsilon^0(\Omega)
		\Big\} \, .
	\end{align}
	This permits us to define the energy $E^\epsilon_\psi (u^\epsilon, \Omega)$ with Dirichlet boundary condition $\psi$:
	\begin{align}
		E^\epsilon_\psi (u^\epsilon, \Omega) &= \begin{cases}
			E^\epsilon(u^\epsilon, \Omega) \qquad & u^\epsilon \in \mathcal{A}_\epsilon^\psi(\Omega) \,\\
			\infty & \text{otherwise} \, .
		\end{cases}\label{eqn:discrete-energy-dirichlet}
	\end{align}
	The following result shows that the effective energy $\int_\Omega \overline{W}(\nabla u) \, dx$ introduced in
	\cref{thm:main-theorem} can also be used with a Dirichlet boundary condition.
	
	\begin{theorem}\label{thm:theorem-dirichlet}
		For any bounded, Lipschitz domain $\Omega$ and any Lipschitz continuous boundary condition
		$\psi : \partial \Omega \rightarrow \mathbb{R}^N$, the discrete energies
		$E^\epsilon_\psi (u^\epsilon, \Omega)$ $\Gamma$-converge (with respect to the weak topology of $H^1(\Omega)$)
		to the effective energy
		\begin{align} \label{eqn:effective-dirichlet-bdry}
			E_{\text{eff}}^\psi(u, \Omega) &= \begin{cases}
				\int_\Omega \overline{W}(\nabla u)\: dx \qquad & u - \psi \in H^1_0(\Omega)\\
				\infty & \text{otherwise}.
			\end{cases}
		\end{align}
	\end{theorem}
	
	Before closing this section, we present three useful properties of the effective energy density $\overline{W}(\lambda)$.
	
	\begin{lemma}\label{lemma:growth-effective-W}
		The function $\overline{W}$ defined by \eqref{eqn:effective-energy-density} satisfies a quadratic growth condition: there exist
		constants $c_1,c_2,d_1> 0$ such that
		\begin{align}
			\max\{c_1 (|\lambda|^2-d_1), 0\} \leq \overline{W}(\lambda) \leq c_2 (|\lambda|^2+1)
		\end{align}
		for all $N \times N$ matrices $\lambda$.
	\end{lemma}
	
	\begin{lemma}\label{lemma:lip-continuity-W}
		$\overline{W}$ is Lipschitz continuous; in fact, there is a constant $c_3 > 0$ such that
		\begin{align} \label{eqn:effective-continuity}
			\big|\overline{W}(\lambda) - \overline{W}(\mu) \big|& \leq c_3 (1 + |\lambda| + |\mu|) |\lambda - \mu|
		\end{align}
		for all $N \times N$ matrices $\lambda$ and $\mu$.
	\end{lemma}
	
	\begin{lemma}\label{lemma:periodic-bc}
		While the definition \eqref{eqn:effective-energy-density} of $\overline{W}(\lambda)$ uses test functions $\psi$ with a Dirichlet
		boundary condition, the effective energy density also has an alternative characterization using periodic test functions.
		Indeed, let $\mathcal{A}^{\#}(kU)$ be the set of deformations defined at all nodes of our lattice that are $k$-periodic
		(that is, deformations $\psi$ such that $\psi(x) = \psi(x+k \sum_{i=1}^N \alpha_i v_i)$ for $\alpha_1,\dots,\alpha_N \in \mathbb{Z}$);
		and let $W^{\#}$ be the function obtained by replacing $\mathcal{A}^0(kU)$ by $\mathcal{A}^{\#}(kU)$ in the definition of $\overline{W}$:
		\begin{align} \label{eqn:effective-energy-density-periodic}
			W^{\#}(\lambda) & = \inf_{k \in \mathbb{N}} \inf_{\psi \in \mathcal{A}^{\#}(kU)} \frac{1}{k^N |U|}
			\sum_{\alpha_1, \dots, \alpha_N=0}^{k-1} E(\lambda x + \psi,U+\sum_{i=1}^N \alpha_i v_i) \, .
		\end{align}
		Then in fact
		\begin{align*}
			\overline{W}(\lambda) = W^{\#}(\lambda) \, ;
		\end{align*}
		thus \eqref{eqn:effective-energy-density-periodic} gives an alternative variational characterization of $\overline{W}$.
	\end{lemma}
	
	The proofs of lemmas \ref{lemma:growth-effective-W}--\ref{lemma:periodic-bc} are presented in \cref{subsec:properties-effective}.
	
	\section{The proof of the main theorem}\label{sec:proof-thm}
	This section begins by establishing the assertions of \cref{thm:main-theorem} in the special case when the macroscopic
	deformation is affine. Then, in \cref{subsec:properties-effective}, we prove lemmas
	\ref{lemma:growth-effective-W}--\ref{lemma:periodic-bc}, which concern properties of the effective energy density
	$\overline{W}(\lambda)$. Besides being of interest in their own right, those properties are needed for the proofs of
	our theorems. With this groundwork complete, \cref{subsec:main-proof} presents the proof of our main result,
	\cref{thm:main-theorem}, establishing $\Gamma$-convergence when no boundary condition is imposed. Finally,
	\cref{subsec:dirichlet-proof} presents the proof of \cref{thm:theorem-dirichlet}, establishing $\Gamma$-convergence
	when a Dirichlet boundary condition is imposed.
	
	\subsection{The heart of the matter: affine limits}\label{subsec:affine-limits}
	The proof of \cref{thm:main-theorem} relies on the fact that every $H^1$ function is locally well-approximated by
	an affine function. Therefore, a crucial first step toward its proof lies in knowing that the theorem's assertions hold
	when the limit $u$ is affine. Using the translation invariance of our energy
	(\cref{eqn:unit-cell-translation-invariant}), it is sufficient to consider the case when $u$ is
	linear.
	\begin{lemma}\label{lemma:affine}
		For any $N \times N$ matrix $\lambda$, let $\overline{W}(\lambda)$ be defined
		by \eqref{eqn:effective-energy-density}. Then for
		any bounded, Lipschitz domain $\Omega$ we have the following results:
		
		\begin{enumerate}
			
			\item[(a)] For $u (x) =  \lambda x$, there is a sequence of admissible deformations $\{u^\epsilon\}_{\epsilon > 0}$
			such that $u^\epsilon \rightharpoonup u$ in $H^1(\Omega)$ and
			\begin{align}
				\lim_{\epsilon \rightarrow 0} E^\epsilon(u^\epsilon, \Omega) & = |\Omega| \, \overline{W}(\lambda) \, . \label{eqn:limsup}
			\end{align}
			Moreover, we can choose $u^\epsilon  - \lambda x \in \mathcal{A}_\epsilon^0(\Omega)$, the set of functions
			that ``vanish at $\partial \Omega$,'' defined by \eqref{eqn:defn-A0-eps}.
			
			\item[(b)] If $u(x) = \lambda x$, then for any sequence of admissible deformations $\{u^\epsilon\}_{\epsilon > 0}$ such that
			$u^\epsilon \rightharpoonup u$ in $H^1(\Omega)$ we have
			\begin{align}
				\liminf_{\epsilon \rightarrow 0} E^\epsilon(u^\epsilon, \Omega)  & \geq |\Omega| \, \overline{W}(\lambda) \, . \label{eqn:liminf}
			\end{align}
			
		\end{enumerate}
	\end{lemma}
	
	\begin{proof}
		Our argument follows the one M\"uller used for the continuous case in \cite{muller1987homogenization}.
		
		\textbf{Proof of part (a)} (finding a recovery sequence). We first assume that the double infimum in \eqref{eqn:effective-energy-density} is achieved for some specific $K \in \mathbb{N}$ and $\psi^K \in \mathcal{A}^0(KU)$, i.e.
		\begin{equation} \label{eqn:optimal-psi-exists}
			\overline{W}(\lambda) =
			\frac{1}{K^N |U|} \sum_{\alpha_1, \dots, \alpha_N=0}^{k-1} E(\lambda x + \psi,U+\sum_{i=1}^N \alpha_i v_i) \, .
		\end{equation}
		While $\psi^K$ is initially defined only on $KU$, it can be extended periodically to the entire lattice; the following argument
		uses this periodic extension, which (abusing notation slightly) we still call $\psi^K$. Tiling the plane with
		translates of $\epsilon K U$, we let $\widetilde{\Omega}_{K \epsilon}$ be the union of those tiles that are compactly contained
		in $\Omega_\epsilon$ (the set defined by \eqref{eqn:defn-omega-eps}). We claim that in this case
		\begin{align*}
			u^\epsilon(x) &= \begin{cases}
				\lambda x + \epsilon \psi^K(\frac{x}{\epsilon}) & \mbox{for $x \in \widetilde{\Omega}_{K \epsilon}$}\\
				\lambda x & \mbox{for $x \in \Omega \setminus \widetilde{\Omega}_{K \epsilon}$}
			\end{cases}
		\end{align*}
		has the desired properties. Indeed, by remarks \ref{rmk:dirichlet-bc} and \ref{rmk:bdry-layer} this deformation
		is well-defined; moreover, it is easy to see that $u^\epsilon - \lambda x \in \mathcal{A}_\epsilon^0(\Omega)$
		and $u^\epsilon \rightharpoonup \lambda x$ in $H^1(\Omega)$ as $\epsilon \rightarrow 0$. Our nontrivial task is to show that
		$\lim_{\epsilon \rightarrow 0} E^\epsilon(u^\epsilon, \Omega) = \overline{W}(\lambda)|\Omega|$. To this end, recall that
		$\widetilde{\Omega}_{K \epsilon} $ is a union of tiles that are translates of $\epsilon K U$. For any single tile, the sum of the energies of its cells is
		\begin{align} \label{eqn:energy-of-single-tile}
			\sum_{\alpha \, {\rm assoc \, one \, tile}} E^\epsilon(u^\epsilon, \epsilon U + \alpha) &=
			\sum_{\alpha \, {\rm assoc \, one \, tile}} E^\epsilon(\lambda x + \epsilon\psi^K (x/\epsilon), \epsilon U + \alpha) \nonumber \\
			& = \epsilon^N K^N |U| \overline{W}(\lambda)
		\end{align}
		using the definition of the scaled energy \eqref{eqn:elasticity-scaling} together with our hypothesis
		\eqref{eqn:optimal-psi-exists}. In doing this calculation, we have also used that
		$E^\epsilon(u^\epsilon, \epsilon U + \alpha)$ depends only the values of $u^\epsilon$ at nodes
		of the scaled lattice in $\epsilon \overline{U}_n + \alpha$, and that $\psi^K \in \mathcal{A}^0(KU)$; it follows
		by arguing as in \cref{rmk:dirichlet-bc} that the value of
		$E^\epsilon(\lambda x + \epsilon \psi^K (x/\epsilon), \epsilon U + \alpha)$ is oblivious
		to the fact that $\psi^K$ is periodic, and is the same as if we extended it by $0$ outside the given tile.
		Now, let us break the energy of $u^\epsilon$ in $\Omega$ into the part associated with the tiles and the rest:
		\begin{equation} \label{eqn:energy-upper-bound-two-parts}
			E^\epsilon(u^\epsilon, \Omega) =
			\sum_{\substack{\alpha \: : \: \epsilon U + \alpha \, {\rm belongs}\\ {\rm to \, a \, tile \, in \, } \widetilde{\Omega}_{K\epsilon}}} E^\epsilon (u^\epsilon, \epsilon U + \alpha) +
			\sum_{\substack{\alpha \in R_\epsilon (\Omega) \: : \: \epsilon U + \alpha\\ {\rm is \, not \, in \, any \, tile} }}
			E^\epsilon (u^\epsilon, \epsilon U + \alpha) \, .
		\end{equation}
		Applying \eqref{eqn:energy-of-single-tile} on each tile, we see that the first term equals
		$|\widetilde{\Omega}_{K\epsilon}| \overline{W}(\lambda)$. Since $\Omega$ is a Lipschitz domain, this converges to
		$ |\Omega| \overline{W}(\lambda)$ as $\epsilon \rightarrow 0$.
		
		We claim that the second term in \eqref{eqn:energy-upper-bound-two-parts} vanishes in the limit $\epsilon \rightarrow 0$.
		The key point is that for every cell $\epsilon U + \alpha$ counted in the second term we have $u^\epsilon = \lambda x$ on
		$\epsilon U_n + \alpha$ (this comes directly from the construction of $u^\epsilon$). Moreover, these cells have the property
		that $\epsilon U_n + \alpha$ is contained in an order-$\epsilon$ width layer near $\partial \Omega$. Therefore
		\cref{lemma:constant-gradient-energy-bound} shows that the second term of \eqref{eqn:energy-upper-bound-two-parts} is at most
		a constant times $\epsilon$. Thus
		$\lim_{\epsilon \rightarrow 0} E^\epsilon(u^\epsilon, \Omega) = \overline{W}(\lambda)|\Omega|$, as desired.
		
		Now we prove part (a) when the double infimum is not achieved; a diagonalization argument is needed
		in this case. We first fix $\delta > 0$ and choose $K \in \mathbb{N}$ and $\psi^\delta \in \mathcal{A}^0(KU)$ such that
		\begin{align}
			\overline{W}(\lambda) \leq \frac{1}{K^N |U|} \sum_{\alpha_1, \dots, \alpha_N=0}^{K-1} E(\lambda x + \psi^\delta, U+\sum_{i=1}^N \alpha_i v_i) \leq \overline{W}(\lambda) + \delta \, . \label{eqn:diagonal-part-a}
		\end{align}
		We obtain a lower bound on the $L^2$ norm of $\lambda + \nabla \psi^\delta$ by using the basic energy lower bound \eqref{eqn:unit-cell-lower} on each cell and adding:
		\begin{equation} \label{eqn:uniform-bound-at-scale-one}
			C_2 \int_{KU} |\lambda + \nabla \psi^\delta|^2 \, dx \leq
			(\overline{W}(\lambda) + \delta)|KU| + C_2 D_2 |KU| \, .
		\end{equation}
		Now we proceed as above with $\psi^\delta$ in place of $\psi^K$ -- setting
		$u^{\epsilon, \delta}(x) = \lambda x + \epsilon \psi^\delta\left(\frac{x}{\epsilon}\right)$ on
		$\widetilde{\Omega}_{K\epsilon}$ and $u^{\epsilon, \delta}(x) = \lambda x$ on
		$\Omega \setminus \widetilde{\Omega}_{K\epsilon}$, where $\psi^\delta$ is the periodic extension of
		$\psi^\delta \in \mathcal{A}_0(KU)$ with period $KU$. The argument used for
		\eqref{eqn:uniform-bound-at-scale-one} is applicable on each scaled copy of $KU$ in $\widetilde{\Omega}_{K\epsilon}$, and
		$\nabla u^{\epsilon,\delta} = \lambda$ in $\Omega \setminus \widetilde{\Omega}_{K\epsilon}$; these observations lead easily to
		a bound on $|\nabla u^{\epsilon,\delta}|_{L^2(\Omega)}$ that depends on $\lambda$ but is independent of $\epsilon$ and $\delta$ as they
		tend to $0$. Since the piecewise linear function $u^{\epsilon,\delta}$ equals $\lambda x$ at $\partial \Omega$, \edit{Poincar\'{e}'s}
		inequality (applied to $u^{\epsilon,\delta}(x) - \lambda x$) gives a bound on $|u^{\epsilon,\delta}|_{L^2(\Omega)}$,
		so in fact we have uniform control on $u^{\epsilon,\delta}$ in $H^1(\Omega)$.
		
		Resuming now the stream of the earlier argument, it is clear that
		$u^{\epsilon, \delta} -\lambda x\in \mathcal{A}_\epsilon^0(\Omega)$ and
		$u^{\epsilon, \delta} \rightharpoonup \lambda x$ in $H^1(\Omega)$ as $\epsilon \rightarrow 0$ with $\delta$ held fixed.
		Since bounded sets in $H^1(\Omega)$ are compact in $L^2(\Omega)$, it follows that
		$\lim_{\epsilon \rightarrow 0} |u^{\epsilon,\delta} - \lambda x|_{L^2(\Omega)} = 0$.
		We also have
		\begin{align*}
			\lim_{\epsilon \rightarrow 0} E^\epsilon(u^{\epsilon, \delta} ,\Omega) &=
			\frac{|\Omega|}{K^N |U|} \sum_{\alpha_1, \dots, \alpha_N=0}^{K-1} E(\lambda x + \psi^\delta, U+\sum_{i=1}^N \alpha_i v_i)
		\end{align*}
		by arguing as we did earlier for $\psi^K$. Using \eqref{eqn:diagonal-part-a}, we deduce that
		\begin{align*}
			|\Omega| \overline{W}(\lambda) \leq \lim_{\epsilon \rightarrow 0} E^\epsilon(u^{\epsilon, \delta} ,\Omega) \leq
			|\Omega| \big(\overline{W}(\lambda) + \delta\big) \, .
		\end{align*}
		We now use a well-known diagonalization result, which is stated at the end of this
		subsection as \cref{lemma:diagonal-argument}. Taking the function $f$ in that lemma to be
		$$
		f(\epsilon, \delta) = \big| E^\epsilon(u^{\epsilon, \delta} ,\Omega) - |\Omega| \overline{W}(\lambda) \big| +
		\int_\Omega |u^{\epsilon,\delta}(x) - \lambda x |^2  \, dx \, ,
		$$
		we obtain a sequence $u^{\epsilon, \delta(\epsilon)}$ converging strongly to $\lambda x$ in the $L^2$ norm such that
		\begin{align*}
			\lim_{\epsilon \rightarrow 0} E^\epsilon(u^{\epsilon, \delta(\epsilon)} ,\Omega) = |\Omega| \overline{W}(\lambda) \, .
		\end{align*}
		Since bounded subsets of $H^1(\Omega)$ are compact in the topology of weak convergence, we also have
		$u^{\epsilon,\delta(\epsilon)} \rightharpoonup \lambda x$, and the proof of part (a) is complete.
		\medskip
		
		\textbf{Proof of part (b)} (the lower bound). This proof has three steps. In the first two, we assume that
		$u^\epsilon -\lambda x \in \mathcal{A}_\epsilon^0(\Omega)$, i.e. that  $u^\epsilon - \lambda x$
		``vanishes at $\partial \Omega$.'' The third step removes this restriction by using an argument due to
		De Giorgi. We shall present the first two steps here. Since the arguments used for the third step are very similar to
		those used in the setting of continuous periodic homogenization, this section provides just some references and
		a brief discussion about what is different in our setting. We do, however, provide the full details of this step in  appendix \ref{appendix:degiorgi}.
		\medskip
		
		\noindent{\sc Step 1:} We consider the special case when $\Omega = \Omega_{\xi,s}$ is a scaled and translated version of $U$:
		\begin{equation} \label{eqn:defn-omega-xi-s}
			\Omega_{\xi,s} = \{x \: : \: \mbox{$x = \xi + s y$ for some $y$ in the interior of $U$}\}
		\end{equation}
		where $\xi$ is any vector in $\mathbb{R}^N$ and $s$ is any positive real number. We shall show that for any
		sequence of admissible deformations satisfying $u^\epsilon - \lambda x \in \mathcal{A}_\epsilon^0(\Omega_{\xi,s})$,
		$\liminf_{\epsilon \rightarrow 0} E^\epsilon ( u^\epsilon, \Omega_{\xi,s}) \geq |\Omega_{\xi,s}| \overline{W}(\lambda) $.
		(Weak convergence of $u^\epsilon$ to $\lambda x$ is not needed in this case.)
		
		For any $\epsilon > 0$ we can choose a positive integer $k$ and a translation
		$\alpha = \sum_{i=1}^N \alpha_i v_i$ with $\alpha_i \in \epsilon \mathbb{Z}$ such that
		$$
		\Omega_{\xi,s} \subseteq \epsilon kU + \alpha \quad 
        \text{where $kU$ is defined by \eqref{eqn:defn-kU}}
		$$
		and such that the difference $(\epsilon kU + \alpha) \setminus \Omega_{\xi,s}$ has measure of order $\epsilon$. Indeed, it suffices
		to choose the integer $k$ so that $ \epsilon < \epsilon k - s \leq 2\epsilon$. Since $\Omega_{\xi,s}$ is a \edit{parallelepiped} whose edges
		are $\{ s v_i \}_{i=1}^N$ while $\epsilon kU$ is a \edit{ parallelepiped} whose edges are $\{\epsilon k v_i \}_{i=1}^N$, the condition
		$\epsilon k - s > \epsilon$ assures enough room to find the desired translation $\alpha$ while the condition
		$\epsilon k - s \leq 2 \epsilon$ assures that the difference between the two sets has measure of order $\epsilon$.
		
		Let $\tilde{u}^\epsilon$ be the natural extension of $u^\epsilon$ to the larger domain $\alpha + \epsilon kU$:
		\begin{align*}
			\tilde{u}^\epsilon = \begin{cases}
				u^\epsilon \qquad & x \in \Omega_{\xi,s}\\
				\lambda x \qquad & x \in (\epsilon kU + \alpha) \setminus \Omega_{\xi,s} \, ,
			\end{cases}
		\end{align*}
		which clearly has the property that $\tilde{u}^\epsilon  - \lambda x  \in \mathcal{A}_\epsilon^0(\epsilon kU + \alpha)$.
		The advantage of considering this extension is that we can estimate its energy using the definition of $\overline{W}$. Indeed,
		writing $\tilde{u}(x) = \epsilon^{-1} \tilde{u}^\epsilon(\epsilon x)$ and using the elasticity-scaling-based
		definition of the effective energy \eqref{eqn:elasticity-scaling} we have
		\begin{equation} \label{eqn:elasticity-scaling-for-utilde}
			\sum_{\epsilon U + \beta \, \subseteq \, \epsilon k U + \alpha} E^\epsilon(\tilde{u}^\epsilon, \epsilon U + \beta) =
			\sum_{U + \gamma \, \subseteq \, kU + \epsilon^{-1}\alpha} \epsilon^N E(\tilde{u}, U + \gamma)
		\end{equation}
		where on the left $\beta$ is a translation preserving the scaled lattice while on the right
		$\gamma$ is a translation preserving the unscaled lattice. (A word is in order about the meaning of \eqref{eqn:elasticity-scaling-for-utilde}. In a term on the left where $\epsilon U_n + \beta$ extends
		beyond $\epsilon k U + \alpha $ or a term on the right where $U_n + \gamma$ extends beyond
		$kU + \epsilon^{-1}\alpha$, we treat $\tilde{u}^\epsilon$ and $\tilde{u}$ as being equal
		to $\lambda x$ outside their respective domains. This is appropriate
		on account of \cref{rmk:dirichlet-bc}, and it is consistent with \cref{rmk:word-on-defn-of-wbar} concerning
		the definition of $\overline{W}$.) Now using periodicity together with
		the definition of $\overline{W}$ and the fact that
		$\tilde{u} - \lambda x \in \mathcal{A}^0 (kU + \epsilon^{-1}\alpha)$, we find that
		\begin{equation} \label{eqn:use-defn-of-Wbar}
			\mbox{value of \eqref{eqn:elasticity-scaling-for-utilde}} \geq
			\epsilon^N k^N |U| \overline{W}(\lambda) = |\epsilon kU|\overline{W}(\lambda) \, .
		\end{equation}
		To obtain the desired conclusion that
		$\liminf_{\epsilon \rightarrow 0} E^\epsilon ( u^\epsilon, \Omega_{\xi,s}) \geq |\Omega_{\xi,s}| \overline{W} (\lambda) $, we
		need only verify that
		\begin{gather}
			\lim_{\epsilon \rightarrow 0} |\epsilon kU| = |\Omega_{\xi,s}| \, , \label{eqn:convergence-of-volumes}\\
			\lim_{\epsilon \rightarrow 0} \big| E^\epsilon (\tilde{u}^\epsilon, \epsilon kU + \alpha) \quad -
			\sum_{\epsilon U + \beta \subseteq \epsilon kU + \alpha} E^\epsilon(\tilde{u}^\epsilon, \epsilon U + \beta) \big| = 0 \, ,
			\quad \mbox{and} \label{eqn:sum-vs-domain} \\
			\lim_{\epsilon \rightarrow 0} \left| E^\epsilon (\tilde{u}^\epsilon, \epsilon kU + \alpha) -
			E^\epsilon (u^\epsilon, \Omega_{\xi,s}) \right| = 0 \, . \label{eqn:energy-difference-vanishes}
		\end{gather}
		The first assertion follows immediately from our choice of $\alpha$ and $k$, which were such that
		$ \Omega_{\xi,s} \subseteq \epsilon kU + \alpha$ and $(\epsilon kU + \alpha) \setminus \Omega_{\xi,s}$
		has measure of order $\epsilon$. For the second assertion, we observe that the difference being estimated
		is
		\begin{equation} \label{eqn:second-assertion-rewritten}
			\sum_{\substack{
					\epsilon U + \beta \subseteq \epsilon k U + \alpha\\
					\beta \notin R_\epsilon(\epsilon kU + \alpha)}}
			E^\epsilon(\tilde{u}^\epsilon, \epsilon U + \beta) \, .
		\end{equation}
		In each of the cells $\epsilon U + \beta$ participating in this sum, $\tilde{u}^\epsilon = \lambda x$
		throughout the expanded cell $\epsilon U_n + \beta$. Moreover, all these expanded cells lie within in an
		order-$\epsilon$ width layer near the boundary of $\epsilon k U + \alpha$, and the measure of this layer
		is of order $\epsilon$. Therefore \cref{lemma:constant-gradient-energy-bound} is applicable, and it bounds
		\eqref{eqn:second-assertion-rewritten} by a constant times $\epsilon$.
		Turning now to the third assertion: since $\Omega_{\xi,s} \subset \epsilon kU + \alpha$, we have
		$R_\epsilon (\Omega_{\xi,s}) \subset R_\epsilon (\epsilon k U + \alpha)$, and the quantity to be estimated is
		\begin{equation} \label{eqn:third-assertion-rewritten}
			\sum_{\beta \in R_\epsilon(\epsilon k U + \alpha) \setminus R_\epsilon(\Omega_{\xi,s})}
			E^\epsilon ( \tilde{u}^\epsilon, \epsilon U + \beta ) \, .
		\end{equation}
		Once again, for each cell $\epsilon U + \beta$ participating in this sum we have $\tilde{u}^\epsilon = \lambda x$
		in the expanded cell $\epsilon U_n + \beta$; moreover, the expanded cells lie within an order-$\epsilon$ width
		layer about the boundary of $\epsilon kU + \alpha$. Therefore \cref{lemma:constant-gradient-energy-bound} is
		again applicable, and it bounds \eqref{eqn:third-assertion-rewritten} by a constant times $\epsilon$. This establishes
		\eqref{eqn:energy-difference-vanishes}, completing Step 1.
		\medskip
		
		\noindent {\sc Step 2:} Our goal in this step is the analogue of Step 1 with $\Omega_{\xi,s}$ replaced by any bounded,
		Lipschitz domain. Thus, we shall show for such $\Omega$ that for any sequence of admissible deformations
		satisfying $u^\epsilon - \lambda x \in \mathcal{A}_\epsilon^0(\Omega)$,
		$\liminf_{\epsilon \rightarrow 0} E^\epsilon ( u^\epsilon, \Omega) \geq |\Omega| \overline{W}(\lambda) $.
		(As in Step 1, weak convergence of $u^\epsilon$ to $\lambda x$ is not needed for this argument.)
		
		Since $\Omega$ is bounded, we can choose $\xi$ and $s$ such that $\overline{\Omega} \subset \Omega_{\xi,s}$.
		By part (a) of this Lemma there is a sequence $v^\epsilon$ with
		$v^\epsilon - \lambda x \in \mathcal{A}^0_\epsilon (\Omega_{\xi,s} \setminus \overline{\Omega})$ such that
		$$
		\lim_{\epsilon \rightarrow 0} E^\epsilon (v^\epsilon, \Omega_{\xi,s} \! \setminus \overline{\Omega}) =
		|\Omega_{\xi,s} \! \setminus \overline{\Omega}| \, \overline{W}(\lambda) \, .
		$$
		Combining $v^\epsilon$ and $u^\epsilon$, we obtain a sequence of deformations $\tilde{u}^\epsilon$ defined on $\Omega_{\xi,s}$:
		\begin{align*}
			\tilde{u}^\epsilon(x) &= \begin{cases}
				u^\epsilon(x) \qquad & x \in  \Omega\\
				v^\epsilon(x) \qquad & x \in \Omega_{\xi,s} \! \setminus \overline{\Omega} \, .
			\end{cases}
		\end{align*}
		By \cref{rmk:bdry-layer}, we have
		\begin{equation} \label{eqn:energy-of-omega-xi-s}
			E^\epsilon(\tilde{u}^\epsilon, \Omega_{\xi,s}) = E^\epsilon (u^\epsilon, \Omega) +
			E^\epsilon(v^\epsilon, \Omega_{\xi,s} \! \setminus \overline{\Omega}) +
			\sum_{\alpha \: : \: (\epsilon \overline{U}_{m}+\alpha) \cap \partial \Omega \neq \emptyset}
			E^\epsilon(\tilde{u}^\epsilon, \epsilon U+\alpha) \, .
		\end{equation}
		The $\liminf$ of the left hand side is estimated by Step 1, and the limit of the second term on the right is known. We claim
		that the last term on the right is at most a constant times $\epsilon$. Indeed, the cells $\epsilon U + \alpha$ that
		participate in the sum have the property that $\tilde{u}^\epsilon = \lambda x$ in $\epsilon \overline{U}_n + \alpha$, and
		these expanded cells lie in an order-$\epsilon$ width layer about $\partial \Omega$. Since $\Omega$ is a bounded, Lipschitz
		domain, the volume of that layer is of order $\epsilon$. Thus \cref{lemma:constant-gradient-energy-bound} applies, and it
		estimates the last term in \eqref{eqn:energy-of-omega-xi-s} by a constant times $\epsilon$. Using these observations,
		we deduce from \eqref{eqn:energy-of-omega-xi-s} that
		$$
		|\Omega_{\xi,s}| \overline{W}(\lambda) \leq
		\liminf_{\epsilon \rightarrow 0} E^\epsilon (u^\epsilon, \Omega) +
		|\Omega_{\xi,s} \! \setminus \overline{\Omega}| \, \overline{W}(\lambda) \, ,
		$$
		which leads immediately to the desired conclusion that
		$\liminf_{\epsilon \rightarrow 0} E^\epsilon (u^\epsilon, \Omega) \geq |\Omega| \overline{W}(\lambda)$.
		\medskip
		
		\noindent {\sc Step 3:} We have thus far established the desired lower bound when ``$u^\epsilon = \lambda x$ at $\partial \Omega$,''
		even for a sequence that does not converge
		weakly to $\lambda x$. The crucial third step is to prove the lower bound
		\emph{without any boundary condition, provided that the sequence converges weakly to $\lambda x$}.
		The technique for doing this relies on an argument of De Giorgi \cite{degiorgi-1975}, and is by now well known.
		M\"{u}ller provides full details in his paper \cite{muller1987homogenization} on the homogenization of
		nonlinear variational problems with periodic microstructure, and the discussion we offer here is parallel to
		his. There is, it seems, no known alternative to this argument; for example, Braides and Defranceschi use it
		in \cite{braides2008homogenization} (stating the required result as Lemma 2.2, and pointing to
		Section 11.1 of \cite{braides-book-1998} for the proof). This technique has also been used in the discrete-to-continuous
		setting; in particular, Alicandro and Cicalese use appropriately adapted versions of it in Lemmas 3.7 and 3.8 of
		\cite{alicandro2004general}.
		
		The main idea is to use ``cutoff functions'' $\phi^\epsilon$ that are identically $1$ in most of $\Omega$
		but equal to $0$ near $\partial \Omega$, and to consider
		$w^\epsilon (x) = u^\epsilon(x) \phi^\epsilon(x) + (\lambda x) (1 - \phi^\epsilon(x))$. The
		conclusion of Step 2 applies to $w^\epsilon$; however, this is only useful if the energy of $w^\epsilon$ is asymptotically
		the same as that of $u^\epsilon$. De Giorgi's argument demonstrates the existence of such $\phi^\epsilon$.
		
		There is something different in our setting compared to that of continuous homogenization. Indeed, in the
		continuous setting one estimates $\nabla w^\epsilon $ by simply using the product rule from calculus. In our setting,
		on the other hand, the relation
		$w^\epsilon (x) = u^\epsilon(x) \phi^\epsilon(x) + (\lambda x) (1 - \phi^\epsilon(x))$ can only be imposed
		at nodes of the lattice. Since $\nabla w^\epsilon$ is the gradient of the \emph{piecewise linearization} of this
		deformation, it cannot be estimated using the product rule; rather, one must use information about the
		\edit{piecewise} linearization scheme. This is the character of \cref{lemma:interp-of-two-defs}, which was
		stated in \cref{subsec:dir-bc-and-gluing} and is proved in appendix \ref{appendix:piecewise-linearization-lemmas}.
		
		Aside from the difference just noted, the arguments for Step 3 are rather familiar; therefore rather than present
		them here we have relegated them to appendix \ref{appendix:degiorgi}. Combining those arguments with the conclusion of Step 2
		completes the proof of part (b).
	\end{proof}
	
	Before closing this section, we state the diagonalization lemma that was used above for part (a) of
	\cref{lemma:affine}.
	
	\begin{lemma}\label{lemma:diagonal-argument}
		Let $f$ be a function of two positive real numbers, taking values in the extended real line. Then there is a
		mapping $\epsilon \rightarrow \delta(\epsilon)$ such that $\epsilon \rightarrow 0$ implies $\delta(\epsilon) \rightarrow 0$ and
		\begin{align*}
			\limsup_{\epsilon \rightarrow 0} f(\epsilon, \delta(\epsilon)) \leq \limsup_{\delta \rightarrow 0} \limsup_{\epsilon \rightarrow 0} f(\epsilon, \delta) \, .
		\end{align*}
	\end{lemma}
	
	\noindent For a proof see for example Corollary 1.16 of \cite{attouch1984variational}.
	
	\begin{remark} \label{rmk:diagonalization-lemma-for-sequences}
		In our applications of \cref{lemma:diagonal-argument}, $\epsilon$ and $\delta$ will often range over sequences approaching
		$0$ rather than over all positive $\epsilon$ and $\delta$ near $0$. The lemma is still applicable, by extending the discretely-defined function $f(\epsilon_j,\delta_k)$ to
		a suitable piecewise constant function $f(\epsilon,\delta)$ defined for positive $\epsilon$ and $\delta$ near $0$.
		(Alternatively, the proof of the lemma can easily be repeated in the discrete setting.)
	\end{remark}
	
	
	\subsection{Some easy properties of \texorpdfstring{$\overline{W}(\lambda)$}{Lg}}\label{subsec:properties-effective}
	We gave three useful properties of the effective energy density $\overline{W}$ at the end of \cref{sec:setup}: (i) $\overline{W}(\lambda)$ satisfies a quadratic growth condition (\cref{lemma:growth-effective-W}); (ii) $\overline{W}(\lambda)$
	is Lipschitz continuous (\cref{lemma:lip-continuity-W}); and (iii) $\overline{W}(\lambda)$ has an alternative variational
	characterization using test functions with a periodic rather than affine boundary condition (\cref{lemma:periodic-bc}).
	We shall prove these lemmas in this subsection.
	
	\begin{proof}[Proof of \cref{lemma:growth-effective-W}]
		The upper bound for $\overline{W}(\lambda)$ is obtained by taking $k=1$ and $\psi = 0$ in
		\eqref{eqn:effective-energy-density} and using \cref{lemma:constant-gradient-energy-bound}; this gives
		$$
		\overline{W}(\lambda) \leq \frac{1}{|U|} E(\lambda x, U) \leq C_1(2n-1)^N (|\lambda|^2 + 1) \, .
		$$
		For the lower bound, it is obvious that $\overline{W}(\lambda) \geq 0$ since $E(u,U)\geq 0$ for every admissible
		deformation on the unit cell $U$. To show the other part of the lower bound, we use
		the convexity of function $\lambda \rightarrow |\lambda|^2$ to see that for every $k \in \mathbb{N}$ and
		$\psi \in \mathcal{A}_0(kU)$, the average energy of $\lambda x +  \psi$ is lower bounded by
		\begin{align*}
			\frac{1}{k^N|U|} \sum_{\alpha_1, \dots, \alpha_N=0}^{k-1}  E(\lambda x + \psi, U+\sum_{i=1}^N \alpha_i v_i)  & \geq
			\frac{C_2}{k^N|U|}\sum_{\alpha_1, \dots, \alpha_N=0}^{k-1}
			\Big(|\lambda + \nabla \psi|^2_{L^2(U+\sum_{i=1}^N \alpha_i v_i)} - D_2|U|\Big) \\
			&\geq  \frac{C_2}{k^N |U|} \int_{kU} |\lambda + \nabla \psi|^2 \: dx - C_2 D_2 \geq C_2 \Big(|\lambda|^2 - D_2\Big) \, .
		\end{align*}
		In the last line we used Jensen's inequality, noting that since $\psi \in \mathcal{A}_0 (kU)$, its
		piecewise linearization vanishes at the boundary (see \cref{rmk:dirichlet-bc}), and therefore $\nabla \psi$ has integral zero.
	\end{proof}
	
	\begin{proof}[Proof of \cref{lemma:lip-continuity-W}]
		It suffices to show that $\overline{W}$ is rank-one convex, since rank-one convexity together with the quadratic
		growth condition $|\overline{W}(\lambda)| \leq C( 1 + |\lambda|^2)$ implies the desired result \eqref{eqn:effective-continuity}.
		(Indeed, rank-one convexity implies that $\overline{W}(\lambda)$ is separately convex as a function of the $N^2$ entries of the matrix
		$\lambda$; but separate convexity and the stated quadratic growth condition imply the desired result,
		see e.g. Proposition 2.32 in \cite{dacorogna2007direct}.)
		
		The proof of rank-one convexity resembles the argument used to show that quasiconvexity implies rank-one convexity. Our goal is to \edit{show} that if $B-A$ has rank one and $0 < \theta < 1$ then
		\begin{equation} \label{eqn:rank-one-convexity}
			\overline{W}(\theta A + (1-\theta)B) \leq \theta \overline{W}(A) + (1-\theta)\overline{W}(B) \, .
		\end{equation}
		The proof is easiest to visualize when $B-A = a \otimes n$ with $n$ parallel to one of the axes of $\mathbb{R}^N$, so let us focus
		for now on this case. Working on the domain $Q = (0,1)^N$, we shall use a test function that's piecewise linear except for a boundary
		layer near $\partial Q$, whose gradient takes the values $A$ and $B$ in layers orthogonal to $n$, with gradient $A$ on approximately
		volume fraction $\theta$ and $B$ on approximately volume fraction $1-\theta$. Being more quantitative: for sufficiently small
		$\delta > 0$ our test function $\phi^\delta : Q \rightarrow \mathbb{R}^N$ should be Lipschitz continuous such that
		$$
		\begin{array}{ll}
			\phi^\delta(x) = \big( \theta A + (1-\theta)B \big) x &  \text{for $x$ in a layer near $\partial Q$ and}\\
			|\nabla \phi^\delta| \leq c &  \text{on } Q \, ,
		\end{array}
		$$
		where $c>0$ is some constant (independent of $\delta$). Moreover, $Q$ should have a partition into regions
		$\Omega_1$, $\Omega_2$, and $\Omega_3 = Q \setminus \overline{\Omega_1 \cup \Omega_2}$ such that
		$$
		\begin{array}{cl}
			\Omega_1 & \text{is a union of finitely many rectangular layers where $\nabla \phi^\delta = A$,} \\
			\Omega_2 & \text{is a union of finitely many rectangular layers where $\nabla \phi^\delta = B$, } \\
			\overline{\Omega_1 \cup \Omega_2} & \text{forms a slightly smaller cube, omitting only a thin layer near $\partial Q$,}
		\end{array}
		$$
		and
		\begin{equation} \label{eqn:proportions}
			\Big||\Omega_1| - \theta|Q|\Big| \leq \delta |Q|, \qquad \Big||\Omega_2| - (1-\theta)|Q|\Big| \leq \delta |Q|,
			\qquad |\Omega_3| \leq \delta |Q| \, .
		\end{equation}
		The existence of such $\phi^\delta$ is well-known; it is shown, for example, in Step 1 of the proof of Lemma 3.11
		in \cite{dacorogna2007direct}.
		
		We now use this framework to establish \eqref{eqn:rank-one-convexity}. Using part (a) of \cref{lemma:affine}, we choose
		a sequence of admissible deformations $u^\epsilon$ defined on $Q$ such that:
		\begin{enumerate}
			\item [(i)] on each rectangular layer $L$ in $\Omega_1$ we have
			$u^\epsilon (x) - \phi^\delta (x) \in \mathcal{A}_\epsilon^0(L)$
			and $\lim_{\epsilon \rightarrow 0} E^\epsilon(u^\epsilon,L) = |L| \, \overline{W}(A)$;
			
			\item [(ii)] on each rectangular layer $L$ in $\Omega_2$ we have
			$u^\epsilon (x) - \phi^\delta (x) \in \mathcal{A}_\epsilon^0(L)$
			and $\lim_{\epsilon \rightarrow 0} E^\epsilon(u^\epsilon,L) = |L| \, \overline{W}(B)$;
			
			\item[(iii)] at all nodes of the scaled lattice outside $\Omega_1 \cup \Omega_2$ we take $u^\epsilon = \phi^\delta$.
		\end{enumerate}
		Since $\phi^\delta$ is affine near $\partial Q$, we have (for any fixed $\delta$)
		$$
		u^\epsilon - \big( \theta A + (1-\theta) B \big) x \in \mathcal{A}_\epsilon^0 (Q)
		$$
		when $\epsilon$ is sufficiently small; therefore by Step 2 in the proof of \cref{lemma:affine} part (b), we have
		\begin{equation} \label{eqn:rank-one-convexity-a}
			|Q| \, \overline{W}(\theta A + (1-\theta) B) \leq \liminf_{\epsilon \rightarrow 0} E^\epsilon(u^\epsilon,Q) \, .
		\end{equation}
		On the other hand, we claim that
		\begin{equation} \label{eqn:rank-one-convexity-b}
			E^\epsilon( u^\epsilon, Q) - \big( E^\epsilon ( u^\epsilon, \Omega_1) + E^\epsilon ( u^\epsilon, \Omega_2) +
			E^\epsilon ( u^\epsilon, \Omega_3) \big) = O(\epsilon) \, .
		\end{equation}
		Indeed, since $\Omega_1$ is a union of finitely many disjoint layers
		$L_j$ where $\nabla \phi^\delta = A$, $\partial \Omega_1$ is the union of
		those layers' boundaries, so (using the definition \eqref{eqn:intro-energy-of-Omega}) we have
		$$
		E^\epsilon(u^\epsilon, \Omega_1) = \sum_{{\rm constituent \, layers \,} L_j}
		E^\epsilon(u^\epsilon, L_j)
		$$
		when $\epsilon$ is sufficiently small. Similarly, $E^\epsilon(u^\epsilon, \Omega_2)$ is the
		sum of the energies of its constituent layers (where $\nabla \phi^\delta = B$). Therefore
		for sufficiently small $\epsilon$ the left hand side of \eqref{eqn:rank-one-convexity-b} is
		the sum of $E(u^\epsilon, \epsilon U + \alpha)$ as $\alpha$ ranges over
		$R_\epsilon (Q) \setminus \big( R_\epsilon(\Omega_1) \cup R_\epsilon(\Omega_2) \cup R_\epsilon(\Omega_3) \big)$. These scaled
		cells have the property that $\epsilon \overline{U}_m + \alpha$ meets $\partial \Omega_i$ for some $i$. For every such cell, we have
		$u^\epsilon = \phi^\delta$ at the lattice nodes in $\epsilon U_m + \alpha$, so \cref{lemma:interp-of-lip-fn} gives a
		uniform bound for $|\nabla u^\epsilon|_{L^\infty(\epsilon U_n + \alpha)}$. Moreover, for every such
		$\alpha$, $\epsilon U_n + \alpha$ lies within an order-$\epsilon$ width layer near the boundary
		of $\Omega_i$ for some $i=1,2,3$.
		Therefore \cref{lemma:constant-gradient-energy-bound} shows that the cumulative energy of all these cells is of order
		$\epsilon$ (with an implicit constant that depends on $\delta$, since the number of layers depends on $\delta$).
		Combining \eqref{eqn:rank-one-convexity-b} with properties (i) and (ii) of $u^\epsilon$, we conclude that
		\begin{equation} \label{eqn:rank-one-convexity-c}
			\liminf_{\epsilon \rightarrow 0} E^\epsilon ( u^\epsilon, Q)  \leq
			| \Omega_1 | \, \overline{W}(A) + | \Omega_2 | \, \overline{W}(B) +
			\limsup_{\epsilon \rightarrow 0} E^\epsilon(u^\epsilon, \Omega_3) \, .
		\end{equation}
		The last term on the right is at most a constant times $|\Omega_3|$, by another application of lemmas
		\ref{lemma:constant-gradient-energy-bound} and \ref{lemma:interp-of-lip-fn}. Therefore \eqref{eqn:rank-one-convexity-c}
		combines with \eqref{eqn:proportions} to give
		\begin{equation} \label{eqn:rank-one-convexity-d}
			\liminf_{\epsilon \rightarrow 0} E^\epsilon ( u^\epsilon, Q)  \leq | Q | \,
			\big( \theta \overline{W}(A) + (1-\theta) \overline{W}(B) + O(\delta) \big) \, .
		\end{equation}
		The desired conclusion \eqref{eqn:rank-one-convexity} now follows by combining \eqref{eqn:rank-one-convexity-a} with
		\eqref{eqn:rank-one-convexity-d} then taking the limit $\delta \rightarrow 0$.
		
		In the preceding argument, we restricted our attention to the case when $B-A = a \otimes n$ with $n$ parallel to one of the
		coordinate axes, since in this case the construction of $\phi^\delta$ is relatively simple and easily visualized. The general
		case is, however, almost the same: if $B-A = a \otimes n$ for any $a, n \in \mathbb{R}^N$, then an essentially identical
		argument can be used by taking $Q$ to be a cube with sides parallel and perpendicular to $n$. Thus \eqref{eqn:rank-one-convexity}
		holds whenever $B-A$ has rank one, and the proof is complete.
	\end{proof}
	
	\begin{proof}[Proof of \cref{lemma:periodic-bc} (An equivalent variational form)]
		It is obvious that $\overline{W}(\lambda) \geq W^{\#}(\lambda)$, since any $\psi \in \mathcal{A}^0(kU)$ has a natural
		periodic extension with period $kU$. We will use part (b) of \cref{lemma:affine} to prove the opposite inequality. For any
		periodic $\psi$ with periodicity $kU$, we consider a sequence of deformations of the form
		$v^\epsilon(x) = \lambda x + \epsilon \psi(\frac{x}{\epsilon})$ with $\epsilon \rightarrow 0$ chosen such that
		$1/(k\epsilon) \in \mathbb{N}$. Using the periodicity of $\psi$, we have
		$$
		E^\epsilon(v^\epsilon, U) =
		\frac{1}{k^N}  \sum_{\alpha_1, \dots, \alpha_N=0}^{k-1}  E(\lambda x + \psi, U+\sum_{i=1}^N \alpha_i v_i) \, .
		$$
		But since $\psi$ is periodic, $v^\epsilon$ converges weakly to $\lambda x$ in $H^1(\Omega)$, so we know from
		\cref{lemma:affine} part (b) that
		$$
		\liminf_{\epsilon \rightarrow 0} E^\epsilon(v^\epsilon, U) \geq |U|\, \overline{W}(\lambda) \, .
		$$
		Combining these results gives
		$$
		\frac{1}{k^N |U|} \sum_{\alpha_1, \dots, \alpha_N=0}^{k-1}  E(\lambda x + \psi, U+\sum_{i=1}^N \alpha_i v_i) \geq
		\overline{W}(\lambda)
		$$
		for any $k \in \mathbb{N}$ and any $\psi \in \mathcal{A}^{\#}(kU)$. We deduce the
		desired conclusion that $W^{\#}(\lambda) \geq \overline{W}(\lambda)$ by minimizing over $k$ and $\psi$.
	\end{proof}
	
	
	\subsection{The proof of Theorem \ref{thm:main-theorem}}\label{subsec:main-proof}
	We recall from \cref{defn:gamma-convergence} that proving $\Gamma$-convergence of $E^\epsilon(u^\epsilon, \Omega)$ to
	$\int_\Omega \overline{W}(\nabla u) \, dx$ requires showing two rather distinct results: the lower bound \eqref{eqn:liminf-domain},
	and the existence of a ``recovery sequence'' \eqref{eqn:limsup-domain}. So far we have proved these assertions
	when $u$ is affine. We turn now to the general case, when the limit can be any
	$u \in H^1(\Omega)$. As already mentioned in \cref{subsec:intro-effective}, our methods are familiar from the literature on
	continuous homogenization problems: the recovery sequence is obtained using piecewise affine approximation, while the
	lower bound is proved by adapting the blowup argument of \cite{braides2008homogenization} to our discrete setting.
	Throughout this subsection $\Omega$ is assumed to be a bounded Lipschitz domain, since this is among the hypotheses of \cref{thm:main-theorem}.
	
	\begin{proof}[Proof of \cref{thm:main-theorem}]
		We prefer to start with the recovery sequence, since the methods used for this are perhaps more familiar.
		To be clear: our goal in this part of the proof is to find, for any $u \in H^1(\Omega)$, a sequence
		$u^\epsilon \in \mathcal{A}^\epsilon(\Omega)$ such that $u^\epsilon \rightharpoonup u$ and
		$\lim_{\epsilon \rightarrow 0} E^\epsilon(u^\epsilon,\Omega) = E_{\text{eff}}(u,\Omega) =
		\int_\Omega \overline{W}(\nabla u) \, dx$.
		
		Since $\Omega$ is a bounded, Lipschitz domain, the function $u$ can be extended to a compactly
		supported $H^1$ function $\tilde{u}$ defined in all $\mathbb{R}^N$ with
		$|\tilde{u}|_{H^1(\mathbb{R}^N)} \leq C_\Omega |u|_{H^1(\Omega)}$. The extension
		$\tilde{u}$ can be approximated by a smooth function $u^\eta$ using mollification, and $u^ \eta$
		can be approximated by a piecewise linear function $u_\delta$ using a mesh of order $\delta$. (The mesh
		used to define $u_\delta$ has nothing to do with our lattice, nor with the scheme discussed in \cref{subsec:the-piecewise-linearization} for determining the piecewise linearization of a deformation.
		In 2D, for example, the vertices of the mesh for $u_\delta$ could be the nodes of the square lattice with
		side length $\delta$, if we triangulate each resulting square by introducing a diagonal edge.)
		We show in appendix \ref{appen:approximation} that by choosing the scale $\eta$ of the mollification
		to depend appropriately on $\delta$, we can arrange that
		\begin{gather}
			|\tilde{u} - u_\delta|_{H^1(\R^N)} \rightarrow 0
			\mbox{ as $\delta \rightarrow 0$ \, and} \label{eqn:udelta-approaches-utilde} \\
			|u_\delta|_{L^\infty(\R^N)} +
			|\nabla u_\delta|_{L^\infty(\R^N)} \leq c_u \delta^{-a} \label{eqn:uniform-bounds-for-udelta}
		\end{gather}
		where $c_u$ is a constant (depending on $|u|_{H^1(\Omega)}$) and $a$ is a positive constant depending only on
		the spatial dimension $N$. We note that the functions $u_\delta$ are uniformly bounded (independent of $\delta$)
		in $H^1(\R^N)$, since $|u_\delta|_{H^1} \leq |u_\delta - \tilde{u}|_{H^1} + |\tilde{u}|_{H^1}$.
		
		We shall obtain the recovery sequence by approximating $u_\delta$ with a suitable sequence of
		deformations $v^{\epsilon,\delta}$ defined on the $\epsilon$-scaled lattice, then applying
		the diagonalization \cref{lemma:diagonal-argument}. The argument shares many features with our proof of
		\cref{lemma:lip-continuity-W}. Some details are different, however, due to the negative exponent of $\delta$ in
		\eqref{eqn:uniform-bounds-for-udelta}. Fortunately, that estimate will be needed only
		in an order-$\epsilon$ width boundary layer near the faces of the triangulation, so it leads to a
		term of order $\epsilon$ times a negative power of $\delta$. Since our diagonalization lemma takes the limit
		$\epsilon \rightarrow 0$ before sending $\delta$ to $0$, a term of this type is not problematic.
		
		To define $v^{\epsilon,\delta}$ we apply part (a) of \cref{lemma:affine} (combined with the translation invariance of
		our energy) to each of the simplices $T$ in or near $\Omega$ on which $u_\delta$ is affine:
		$u_\delta |_T(x) = \lambda^\delta_T \cdot x + c^\delta_T $. The resulting $v_T^{\epsilon,\delta}$
		has the following properties:
		\begin{equation} \label{eqn:strong-L2-convergence-on-T}
			\lim_{\epsilon \rightarrow 0} \int_T |v_T^{\epsilon,\delta} - u_\delta|^2 \, dx = 0 \, ,
		\end{equation}
		\begin{equation} \label{eqn:affine-boundary-layer-on-T}
			v_T^{\epsilon,\delta} (x) - (\lambda^\delta_T \cdot x + c^\delta_T) \in \mathcal{A}_\epsilon^0 (T) \, , \quad \mbox{and} \quad
			\lim_{\epsilon \rightarrow 0} E^\epsilon(v_T^{\epsilon,\delta}, T) = |T| \overline{W}(\lambda^\delta_T) \, .
		\end{equation}
		Piecing these functions together, we define $v^{\epsilon,\delta}$ on the union of the simplices $T$ that lie in or near
		$\Omega$:
		\begin{equation} \label{eqn:recovery-sequence-limsup}
			v^{\epsilon,\delta} = v_T^{\epsilon,\delta} \ \mbox{at nodes of the $\epsilon$-scaled lattice that lie in
				$\overline{T}$, if dist$(T,\Omega) \leq 1$} \, .
		\end{equation}
		This is well-defined, even when a node of the scaled lattice belongs to two or more simplices; to explain why,
		we observe that $v_T^{\epsilon,\delta} = u_\delta$ at $\partial T$, by the first part of
		\eqref{eqn:affine-boundary-layer-on-T} combined with \cref{rmk:dirichlet-bc}. Our recovery sequence will
		be obtained by taking $\delta$ to be a suitable function of $\epsilon$ and restricting
		$v^{\epsilon,\delta(\epsilon)}$ to $\Omega$.
		
		The obvious idea is to apply our diagonalization \cref{lemma:diagonal-argument} with
		\begin{equation} \label{eqn:obvious-choice-of-f}
			f(\epsilon,\delta) = \big| E^\epsilon (v^{\epsilon,\delta},\Omega) - \int_\Omega \overline{W}(\nabla u) \big|
			+ \int_\Omega |v^{\epsilon,\delta} - u|^2 \, dx \, .
		\end{equation}
		In the end we will make a slightly different choice -- see \eqref{eqn:revised-choice-of-f} -- but to explain the main
		ideas it is convenient to focus on \eqref{eqn:obvious-choice-of-f}. For the diagonalization lemma to be applicable, we
		need to show that
		\begin{equation} \label{eqn:convergence-of-energy}
			\lim_{\delta \rightarrow 0} \, \lim_{\epsilon \rightarrow 0}
			\big| E^\epsilon (v^{\epsilon,\delta},\Omega) - \int_\Omega \overline{W}(\nabla u) \big| = 0
		\end{equation}
		and
		\begin{equation} \label{eqn:convergence-in-L2}
			\lim_{\delta \rightarrow 0} \, \lim_{\epsilon \rightarrow 0} \int_\Omega |v^{\epsilon,\delta} - u|^2 \, dx = 0 \, .
		\end{equation}
		Since $v^{\epsilon,\delta}$ has been defined simplex-by-simplex, it is convenient to work with inner and outer
		approximations of $\Omega$ that are unions of simplices on which $u_\delta$ is affine.
		For the inner approximation we choose
		\begin{equation} \label{eqn:inner-approximation}
			\Omega^\delta_{\rm in} = \ \mbox{union of all simplices $T$ such that $\overline{T} \subset \Omega$},
		\end{equation}
		while for the outer approximation we choose $\Omega^\delta_{\rm out}$ such that
		\begin{equation} \label{eqn:outer-approximation}
			\mbox{$\Omega^\delta_{\rm out}$ is a union of simplicies, it contains a $\delta$-neighborhood of $\Omega$, and
				$\left| \Omega^\delta_{\rm out} - \Omega \right| \rightarrow 0$ as $\delta \rightarrow 0$.}
		\end{equation}
		Since our simplices have diameter of order $\delta$, it is obvious that $|\Omega - \Omega^\delta_{\rm in}| \rightarrow 0$.
		So the volume of $\Omega^\delta_{\rm out} \setminus \Omega^\delta_{\rm in}$ tends to $0$ as $\delta \rightarrow 0$, and
		therefore
		\begin{equation} \label{eqn:diff-of-approxns}
			\int_{\Omega^\delta_{\rm out} \setminus \Omega^\delta_{\rm in}} (1 + |\nabla u_\delta|^2) \, dx \leq
			\int_{\Omega^\delta_{\rm out} \setminus \Omega^\delta_{\rm in}}
			(1 + 2|\nabla (u_\delta - \tilde{u})|^2 + 2|\nabla \tilde{u}|^2) \, dx
			\rightarrow 0 \quad \mbox{as $\delta \rightarrow 0$.}
		\end{equation}
		
		To justify \eqref{eqn:convergence-of-energy}, our main task is to show that
		\begin{equation} \label{eqn:limit-of-energies-out-and-in}
			\lim_{\epsilon \rightarrow 0} E^\epsilon (v^{\epsilon,\delta},\Omega^\delta_{\rm out})  =
			\int_{\Omega^\delta_{\rm out}} \overline{W}(\nabla u_\delta) \, dx \quad \mbox{and} \quad
			\lim_{\epsilon \rightarrow 0} E^\epsilon (v^{\epsilon,\delta},\Omega^\delta_{\rm in})  =
			\int_{\Omega^\delta_{\rm in}} \overline{W}(\nabla u_\delta) \, dx \, .
		\end{equation}
		To explain why this implies \eqref{eqn:convergence-of-energy}, we first observe that since
		$\Omega^\delta_{\rm in} \subset \Omega \subset \Omega^\delta_{\rm out}$, the nonnegativity of our
		energy gives
		\begin{equation} \label{eqn:sandwich-argument-a}
			E^\epsilon (v^{\epsilon,\delta},\Omega^\delta_{\rm in}) \leq
			E^\epsilon (v^{\epsilon,\delta},\Omega )\leq
			E^\epsilon (v^{\epsilon,\delta},\Omega^\delta_{\rm out})
		\end{equation}
		and the nonnegativity of $\overline{W}$ gives
		\begin{equation} \label{eqn:sandwich-argument-b}
			\int_{\Omega^\delta_{in}} \overline{W} ( \nabla u_\delta ) \, dx \leq
			\int_\Omega \overline{W} ( \nabla u_\delta ) \, dx \leq
			\int_{\Omega^\delta_{out}} \overline{W} ( \nabla u_\delta ) \, dx \, .
		\end{equation}
		On the other hand, we have $0 \leq \overline{W}(\nabla u_\delta) \leq c_2 (1 + |\nabla u_\delta|^2)$ from
		\cref{lemma:growth-effective-W}, so \eqref{eqn:diff-of-approxns} implies that
		\begin{equation} \label{eqn:sandwich-argument-c}
			\int_{\Omega^\delta_{out}} \overline{W} ( \nabla u_\delta ) \, dx -
			\int_{\Omega^\delta_{in}} \overline{W} ( \nabla u_\delta ) \, dx \rightarrow 0
		\end{equation}
		as $\delta \rightarrow 0$. Finally, the Lipschitz property of $\overline{W}$ (\cref{lemma:lip-continuity-W}) gives
		\begin{equation} \label{eqn:sandwich-argument-d}
			\left| \int_\Omega \overline{W}(\nabla u_\delta) - \overline{W}(\nabla u) \, dx \right| \leq
			c_3 \int_\Omega (1 + |\nabla u| + |\nabla u_\delta| ) |\nabla u - \nabla u_\delta| \, dx \,
		\end{equation}
		which tends to $0$ as $\delta \rightarrow 0$ by H\"{o}lder's inequality. The desired conclusion \eqref{eqn:convergence-of-energy} follows easily from \eqref{eqn:limit-of-energies-out-and-in}
		combined with \eqref{eqn:sandwich-argument-a}--\eqref{eqn:sandwich-argument-d}.
		
		We turn now to the proof of \eqref{eqn:limit-of-energies-out-and-in}. It suffices to discuss the first assertion
		(concerning $\Omega^\delta_{\rm out}$) since the justification of the second assertion
		(concerning $\Omega^\delta_{\rm in}$) is entirely parallel.  We recall from \eqref{eqn:affine-boundary-layer-on-T}
		that for each simplex $T \subset \Omega^\delta_{\rm out}$ we have
		$$
		\lim_{\epsilon \rightarrow 0} E^\epsilon(v^{\epsilon,\delta}_T,T) = \int_T \overline{W}(\nabla u_\delta) \, dx \, ,
		$$
		so we need to show that
		\begin{equation} \label{eqn:energy-out-vs-sum-over-T}
			E^\epsilon(v^{\epsilon,\delta}, \Omega^\delta_{\rm out}) -
			\sum_{T \subset \Omega^\delta_{\rm out}} E^\epsilon(v^{\epsilon,\delta}_T,T) \rightarrow 0
			\quad \mbox{as $\epsilon \rightarrow 0$}.
		\end{equation}
		Using the definition of the energy -- \eqref{eqn:intro-energy-of-Omega} and \eqref{eqn:R_eps} -- the difference
		\eqref{eqn:energy-out-vs-sum-over-T} is precisely
		$$
		\sum_{\epsilon \overline{U}_m + \alpha \subset \Omega^\delta_{\rm out}}
		E^\epsilon(v^{\epsilon,\delta}, \epsilon U + \alpha) \quad -
		\sum_{\substack{\epsilon \overline{U}_m + \alpha \subset T \, {\rm for}\\
				{\rm some \, simplex \, } T \subset \Omega^\delta_{\rm out}}} E^\epsilon(v^{\epsilon,\delta}_T, \epsilon U + \alpha) \, .
		$$
		The sum on the right is not changed if we replace $v^{\epsilon,\delta}_T$ by $v^{\epsilon,\delta}$. Indeed,
		$E^\epsilon(v^{\epsilon,\delta}_T, \epsilon U + \alpha)$ depends only on the values of $v^{\epsilon,\delta}_T$ at
		nodes of the scaled lattice in $\epsilon \overline{U}_n + \alpha$, by \eqref{eqn:U_n}; and for the $\alpha$ that
		enter the sum, $\epsilon \overline{U}_n + \alpha \subset T$ (using that $m \geq n$). Therefore the difference \eqref{eqn:energy-out-vs-sum-over-T} is equal to
		\begin{equation} \label{eqn:discrepancy-as-a-sum}
			\sum_{\substack{\epsilon \overline{U}_m + \alpha \, {\rm meets} \, \partial T \, {\rm for}\\
					{\rm some \, simplex \, } T \subset \Omega^\delta_{\rm out}}} E^\epsilon(v^{\epsilon,\delta}, \epsilon U + \alpha) \, .
		\end{equation}
		We come now to a key point: since $v_T^{\epsilon,\delta} (x) - (\lambda^\delta_T x - c^\delta_T) \in \mathcal{A}^\epsilon_0 (T)$
		by \eqref{eqn:affine-boundary-layer-on-T}, for each $\alpha$ that participates in the preceding sum we have
		$v^{\epsilon,\delta}(x)= u_{\delta}(x)$ at all nodes of the scaled lattice in $\epsilon \overline{U}_m + \alpha$. Therefore
		\cref{lemma:interp-of-lip-fn} combines with \eqref{eqn:uniform-bounds-for-udelta} to show that
		$$
		\text{for each term in \eqref{eqn:discrepancy-as-a-sum}, $|\nabla v^{\epsilon,\delta}| \leq C \delta^{-a}$
			in $\epsilon U_n + \alpha$.}
		$$
		Since each scaled unit cell that participates in \eqref{eqn:discrepancy-as-a-sum} lies in an
		order-$\epsilon$ neighborhood of $\partial T$ for some simplex $T \subset \Omega^\delta_{\rm out}$ -- and
		since the number of simplices is of order $|\Omega| \delta^{-N}$ -- we conclude from \cref{lemma:constant-gradient-energy-bound} and the nonnegativity of
		our energy that
		$$
		\text{the value of \eqref{eqn:discrepancy-as-a-sum} is nonnegative, and bounded above by $C \epsilon $}
		$$
		with a constant $C$ that depends on $\delta$, $\Omega$, and $u$ (but not $\epsilon$). Taking the
		limit $\epsilon \rightarrow 0$, we obtain the desired conclusion \eqref{eqn:energy-out-vs-sum-over-T}.
		
		We turn now to \eqref{eqn:convergence-in-L2}. Since $u_\delta$ approaches $\tilde{u}$ in $L^2$ as $\delta \rightarrow 0$ and
		$\Omega \subset \Omega^\delta_{\rm out}$, it suffices to show that
		\begin{equation} \label{eqn:v-epsilon-delta-limit-L2}
			\lim_{\epsilon \rightarrow 0} \int_{\Omega^\delta_{\rm out}} |v^{\epsilon,\delta} - u_\delta|^2 \, dx = 0 \, .
		\end{equation}
		The argument is similar to the proof of \eqref{eqn:limit-of-energies-out-and-in}. We start with the
		obvious fact that
		\begin{equation} \label{eqn:T-minus-Tepsilon-and-Tepsilon}
			\int_{\Omega^\delta_{\rm out}} |v^{\epsilon,\delta} - u_\delta|^2 \, dx =
			\sum_{T \subset \Omega^\delta_{\rm out}}
			\int_{T_\epsilon} |v^{\epsilon,\delta} - u_\delta|^2 \, dx +
			\sum_{T \subset \Omega^\delta_{\rm out}}
			\int_{T \setminus T_\epsilon} |v^{\epsilon,\delta} - u_\delta|^2 \, dx
		\end{equation}
		where each sum is over all simplices $T \subset \Omega^\delta_{\rm out}$, and (consistent with \eqref{eqn:defn-A0-eps})
		$$
		T_\epsilon = \Big\{x \in T \: \big| \: \text{dist}(x,\partial T) >  \epsilon d_m \Big\} \, .
		$$
		Observing that $v^{\epsilon,\delta}(x) = v_T^{\epsilon,\delta}(x)$ for $x \in T_\epsilon$, we have
		$$
		\int_{T_\epsilon} |v^{\epsilon,\delta} - u_\delta|^2 \, dx = \int_{T_\epsilon} |v_T^{\epsilon,\delta} - u_\delta|^2 \, dx
		\rightarrow 0 \quad \text{as $\epsilon \rightarrow 0$}
		$$
		using \eqref{eqn:strong-L2-convergence-on-T}. It follows that the first term on the right side
		of \eqref{eqn:T-minus-Tepsilon-and-Tepsilon} tends to $0$ as $\epsilon \rightarrow 0$.
		
		Preparing to estimate the other term, we claim that
		\begin{equation} \label{eqn:uniform-estimate-near-bdry-of-T}
			|v^{\epsilon,\delta}(x)| \leq |u_\delta|_{L^\infty} \quad \mbox{when $x \in T \setminus T_\epsilon$ \, .}
		\end{equation}
		Indeed, let $\epsilon U + \alpha$ be the scaled and translated unit cell that contains $x$.
		If $\epsilon \overline{U}_m + \alpha$ meets $\partial T$, then $\epsilon \overline{U}_m + \alpha$ cannot
		meet $T'_{\epsilon}$ for \emph{any} simplex $T'$, so we know from the first part of \eqref{eqn:affine-boundary-layer-on-T}
		that $v^{\epsilon,\delta} = u_\delta$ throughout $\epsilon \overline{U}_m + \alpha$, and \cref{lemma:interp-of-lip-fn}
		provides the estimate \eqref{eqn:uniform-estimate-near-bdry-of-T} at $x$.
		If, on the other hand, $\epsilon \overline{U}_m + \alpha \subset T$ then
		$v^{\epsilon,\delta} = v^{\epsilon,\delta}_T$ in $\epsilon U + \alpha$; in particular, these two functions are equal at $x$. Since $x \in T \setminus T_\epsilon$ we know that $v^{\epsilon,\delta}_T (x) = u_\delta(x)$ from the first part
		of \eqref{eqn:affine-boundary-layer-on-T}; therefore the estimate \eqref{eqn:uniform-estimate-near-bdry-of-T} is
		also valid in this case.
		
		An estimate for the second term on the right side of \eqref{eqn:T-minus-Tepsilon-and-Tepsilon}
		follows easily from \eqref{eqn:uniform-estimate-near-bdry-of-T}
		combined with our uniform bound \eqref{eqn:uniform-bounds-for-udelta} on $u_\delta$. Remembering that
		$T \setminus T_\epsilon$ is an order-$\epsilon$ thick neighborhood of $\partial T$ and that the total
		number of simplices is of order $|\Omega| \delta^{-N}$, we get that
		$$
		\text{the second term on the right side of \eqref{eqn:T-minus-Tepsilon-and-Tepsilon} is at most $C \epsilon$}
		$$
		with a constant $C$ that depends on $\delta$, $u$, and $\Omega$ (but not $\epsilon$). This converges to $0$
		as $\epsilon \rightarrow 0$, so the proof of \eqref{eqn:v-epsilon-delta-limit-L2} is complete.
		
		We are still lacking one element. The preceding results let us conclude, using \cref{lemma:diagonal-argument}, existence of
		$v^{\epsilon,\delta(\epsilon)}$ for which
		$E^\epsilon(v^{\epsilon,\delta(\epsilon)},\Omega) \rightarrow \int_\Omega \overline{W}(\nabla u) \, dx$ and
		$\int_{\Omega} |v^{\epsilon,\delta(\epsilon)} - u|^2 \, dx \rightarrow 0$
		as $\epsilon \rightarrow 0$. However, to know that $v^{\epsilon,\delta(\epsilon)} \rightharpoonup u$ in the weak topology on
		$H^1(\Omega)$ we need to know that $\nabla v^{\epsilon,\delta(\epsilon)}$ stays uniformly bounded in $L^2(\Omega)$ as
		$\epsilon \rightarrow 0$. The natural tool for proving this is our lower bound on the discrete energy, \eqref{eqn:unit-cell-lower},
		which implies that
		\begin{equation} \label{eqn:lower-bound-restated}
			C_2 \int_{\epsilon U + \alpha} |\nabla v^{\epsilon,\delta(\epsilon)}|^2 \, dx \leq
			E^\epsilon( v^{\epsilon,\delta(\epsilon)}, \epsilon U + \alpha) + D_2 |\epsilon U + \alpha| \, .
		\end{equation}
		Summing these inequalities over all $\alpha \in R_\epsilon (\Omega)$ gives an upper bound for
		\begin{equation} \label{eqn:L2-bound-misses-bdry-layer}
			\int_{\bigcup_{\alpha \in R_\epsilon(\Omega)} (\epsilon U + \alpha)} |\nabla v^{\epsilon,\delta(\epsilon)}|^2 \, dx \, ,
		\end{equation}
		which misses an order-$\epsilon$ width layer near $\partial \Omega$.
		
		We can fix this problem by changing the choice of $f(\epsilon,\delta)$ to which the diagonalization lemma is applied.
		Rather than the function $f_0(\epsilon,\delta)$ defined by \eqref{eqn:obvious-choice-of-f}, let us use
		\begin{equation} \label{eqn:revised-choice-of-f}
			f_1(\epsilon,\delta) = f_0(\epsilon,\delta) + \frac{\epsilon}{\delta} +
			\big| E^\epsilon (v^{\epsilon,\delta},\Omega^\delta_{\rm out}) - \int_{\Omega^\delta_{\rm out}}\overline{W}(\nabla u_\delta) \big|
			\, .
		\end{equation}
		The lemma is applicable, since we know using \eqref{eqn:limit-of-energies-out-and-in} that
		$\lim_{\delta\rightarrow 0} \lim_{\epsilon \rightarrow 0} f_1 (\epsilon,\delta) = 0$.
		The resulting $v^{\epsilon,\delta(\epsilon)}$ has the property that $\epsilon/\delta(\epsilon) \rightarrow 0$ as
		$\epsilon \rightarrow 0$. Adding the estimates \eqref{eqn:lower-bound-restated} over all
		$\alpha \in R_\epsilon \big( \Omega^{\delta(\epsilon)}_{\rm out} \big)$ and writing
		$$
		S_{\epsilon} = \bigcup_{\alpha \in R_\epsilon \big( \Omega^{\delta(\epsilon)}_{\rm out} \big)} (\epsilon U + \alpha)
		$$
		we get
		\begin{equation} \label{eqn:after-adding-lower-bounds}
			C_2 \int_{S_\epsilon} |\nabla v^{\epsilon,\delta(\epsilon)}|^2 \, dx \leq
			E^\epsilon(v^{\epsilon,\delta(\epsilon)},\Omega^{\delta(\epsilon)}_{\rm out}) +
			D_2 |\Omega^{\delta(\epsilon)}_{\rm out}| \, .
		\end{equation}
		To see that the left hand side of \eqref{eqn:after-adding-lower-bounds} controls
		$\int_\Omega |\nabla v^{\epsilon,\delta(\epsilon)}|^2 \, dx$ when $\epsilon$ is sufficiently small, we recall that
		$\Omega^\delta_{\rm out}$ contains a $\delta$-neighborhood of $\Omega$ by \eqref{eqn:outer-approximation}; it follows
		that $\Omega \subset S_\epsilon$ when $\epsilon/\delta(\epsilon)$ is sufficiently small. To see that the
		right hand side of \eqref{eqn:after-adding-lower-bounds} stays bounded we observe that
		$$
		\big| E^\epsilon(v^{\epsilon,\delta(\epsilon)},\Omega^{\delta(\epsilon)}_{\rm out}) -
		\int_{\Omega^{\delta(\epsilon)}_{\rm out}} \overline{W}(\nabla u_{\delta(\epsilon)}) \, dx \big| \rightarrow 0
		$$
		since $f_1(\epsilon, \delta(\epsilon)) \rightarrow 0$; moreover $u_{\delta(\epsilon)}$ stays uniformly bounded in
		$H^1$ while we know from \cref{lemma:growth-effective-W} that $\overline{W}$ has quadratic growth. Thus the
		sequence
		$$
		u^\epsilon = \text{restriction to $\Omega$ of $v^{\epsilon,\delta(\epsilon)}$}
		$$
		converges weakly to $u$ in $H^1(\Omega)$ and has
		$E^\epsilon (u^\epsilon,\Omega) \rightarrow \int_\Omega \overline{W}(\nabla u) \, dx $, as desired.
		
		We turn now to the lower bound. Our task is to show that if $u^\epsilon \rightharpoonup u$ in $H^1(\Omega)$ then
		\begin{equation} \label{eqn:lower-bound-general-u}
			\liminf_{\epsilon \rightarrow 0} E^\epsilon(u^\epsilon, \Omega) \geq \int_\Omega \overline{W}(\nabla u) \, dx \, .
		\end{equation}
		Our proof relies on the fact that this has already been established when $u$ is affine. We will localize
		the assertion \eqref{eqn:lower-bound-general-u} using a blow-up procedure. Since the blow-up of $u$ at $x_0$
		is its affine approximation, this procedure will permit us to
		deduce the desired result for any $u \in H^1(\Omega)$ from the one for affine limits. Since the argument is
		fairly long, we present it in several steps.
		\medskip
		
		\noindent {\sc Step 1: Setting up the localization.} We may (and do) focus on a subsequence
		$\epsilon_j \rightarrow 0$ such that
		\begin{equation} \label{eqn:energy-has-limit}
			\lim_{j \rightarrow \infty} E^{\epsilon_j}(u^{\epsilon_j}, \Omega) =
			\liminf_{\epsilon \rightarrow 0} E^\epsilon(u^\epsilon, \Omega) \, .
		\end{equation}
		We associate with this sequence a family of discrete nonnegative measures $\mu_j$ supported in $\Omega$, by taking
		$\mu_j$ to have a point mass at each $\alpha \in R_\epsilon(\Omega)$ with weight
		$E^{\epsilon_j}(u^{\epsilon_j},\epsilon_j U + \alpha)$; in other words
		\begin{align*}
			\mu_j(A) = \sum_{\alpha \in R_\epsilon(\Omega)} E^{\epsilon_j}(u^{\epsilon_j},\epsilon_j U + \alpha) \, \delta_\alpha(A) \, , \qquad \delta_\alpha(A) =
			\begin{cases}
				1 & \alpha \in A \\
				0 & \alpha \notin A \, .\\
			\end{cases}
		\end{align*}
		Since $E^\epsilon(u^\epsilon, \Omega) = \sum_{\alpha \in R_\epsilon(\Omega)} E(u^\epsilon, \epsilon U + \alpha)$
		we have
		\begin{equation} \label{eqn:point-mass-measure-omega}
			\mu_j(\Omega) = E^{\epsilon_j}(u^{\epsilon_j},\Omega) \, .
		\end{equation}
		Moreover, for any subset $A$ of $\Omega$ we have
		\begin{equation} \label{eqn:point-mass-inequality}
			\mu_j(A) \geq E^{\epsilon_j}(u^{\epsilon_j},A)
		\end{equation}
		since $\epsilon_j \overline{U}_m  + \alpha \subset A$ implies $\epsilon_j \overline{U}_m  + \alpha \subset \Omega$.
		(In our applications of this inequality, the set $A$ will be a small cube.) Passing to a further subsequence
		if necessary, we may suppose that the measures $\mu_j$ converge weakly to a limit $\mu$. The weak limit is
		clearly nonnegative (since each $\mu_j$ is nonnegative) and it is supported on $\overline{\Omega}$, with
		\begin{equation} \label{eqn:mu-limit-1}
			\mu(\overline{\Omega}) = \lim_{j \rightarrow \infty} \mu_j(\Omega) =
			\lim_{j \rightarrow \infty} E^{\epsilon_j}(u^{\epsilon_j}, \Omega) \, .
		\end{equation}
		Taking the Radon-Nikodym decomposition of $\mu$ with respect to Lebesgue measure on $\mathbb{R}^N$, we have
		\begin{equation} \label{eqn:mu-limit-2}
			\mu = \frac{d\mu}{d x} \mathcal{L}^N + \mu^s
		\end{equation}
		where $\mathcal{L}^N$ is Lebesgue measure and $\mu^s \perp \mathcal{L}^N$. The singular part is nonnegative
		($\mu^s \geq 0$) since $\mu(A) \geq 0$ for any measurable set $A$. Combining
		\eqref{eqn:mu-limit-1}-\eqref{eqn:mu-limit-2} and using the nonnegativity of $\mu^s$ we obtain
		\begin{equation} \label{eqn:key-result-proof}
			\lim_{j \rightarrow \infty} E^{\epsilon_j}(u^{\epsilon_j}, \Omega) = \mu(\overline{\Omega}) \geq
			\mu(\Omega) \geq \int_\Omega \frac{d\mu}{dx} \, dx \, .
		\end{equation}
		This framework reduces our task to proving that
		\begin{equation} \label{eqn:key-result}
			\frac{d \mu}{dx}(x) \geq \overline{W}(\nabla u(x)) \quad \mbox{for Lebesgue-a.e. $x \in \Omega$} \, ,
		\end{equation}
		since the lower bound \eqref{eqn:lower-bound-general-u} then follows immediately using
		\eqref{eqn:energy-has-limit} and \eqref{eqn:key-result-proof}.
		
		The rest of the proof is devoted to establishing \eqref{eqn:key-result}. We shall prove it at $x=x_0$ when
		\begin{enumerate}
			\item[(i)] $x_0$ is a Lebesque point of $\mu$, in other words
			\begin{equation} \label{eqn:lebesgue-pt-of-mu}
				\frac{d \mu}{dx}(x_0) = \lim_{\rho \rightarrow 0} \frac{\mu \big( Q_\rho(x_0) \big)}{\rho^N}
			\end{equation}
			where $Q_\rho(x_0)$ is an open cube centered at $x_0$ with side length $\rho$; and
			\item[(ii)] $x_0$ is a Lebesgue point for $u$ and $\nabla u$, and moreover $u$ is well-approximated near
			$x_0$ by its linear approximation in the sense that
			\begin{equation} \label{eqn:good-pt-for-u}
				\lim_{\rho \rightarrow 0} \frac{1}{\rho^2} \Big(\frac{1}{\rho^N}
				\int_{Q_\rho(x_0)} |u(x)-u(x_0)- \nabla u(x_0) \cdot (x-x_0)|^2 \: dx\Big) = 0 \, .
			\end{equation}
		\end{enumerate}
		This suffices, since \eqref{eqn:lebesgue-pt-of-mu} holds Lebesgue-a.e. by a standard result from measure theory, and
		\eqref{eqn:good-pt-for-u} holds Lebesgue-a.e. for any $u \in H^1(\Omega)$ (as a consequence, for example, of
		Theorem 3.4.2 in \cite{ziemer2012weakly}).
		\medskip
		
		\noindent {\sc Step 2: Blowing up the discrete deformations.} The deformation $u^{\epsilon_j}$ is defined at nodes
		of the $\epsilon_j$-scaled lattice in $\Omega$. Given $x_0 \in \Omega$ and sufficiently small $\rho > 0$, we
		want to consider the restriction of $u^{\epsilon_j}$ to a cube of size $\rho$ around $x_0$, and to rescale
		it to a deformation $w^\rho_j$ defined on the unit cube centered at $0$ (which we denote by $Q_1$).
		By defining the rescaling appropriately, we will arrange that $w^\rho_j$ be defined at the nodes of
		our $\frac{\epsilon_j}{\rho}$-scaled lattice that lie in $Q_1$.
		
		Given $x_0$ and $\epsilon_j$, there is a unique translation of the $\epsilon_j$-scaled lattice that takes $x_0$ to
		the scaled unit cell $\epsilon_j U$:
		\begin{equation} \label{eqn:x0-determines-xi0}
			\mbox{$x_0 = \epsilon_j (\xi_j + \alpha^j)$ where $\xi_j \in U$ and $\alpha^j = \sum_{i=1}^N \alpha^j_i v_i $
				with $\alpha^j_i \in \mathbb{Z}$ for each $i \, .$}
		\end{equation}
		(Note that, contrary to our usual convention, $\alpha^j$ is a translation of the \emph{unscaled} lattice
		rather than the scaled one. This is convenient because the following discussion involves two distinct scalings.)
		Our rescaled deformation is then
		\begin{equation} \label{eqn:rescaled-deformation}
			w_j^\rho(x) = \frac{u^{\epsilon_j}(x_0 - \epsilon_j \xi_j + \rho x)-u(x_0)}{\rho} \, .
		\end{equation}
		This deformation is in $\mathcal{A}^{\epsilon_j/\rho}(Q_1)$ provided that $Q_\rho(x_0) - \epsilon_j \xi_j \subset \Omega$.
		Indeed, if $x$ is a node of the $\epsilon_j/\rho$-scaled lattice, say
		$$
		\text{$x = \frac{\epsilon_j}{\rho} (p_k + \beta)$ where $p_k \in V$ and $\beta = \sum_{i=1}^N \beta_i v_i $
			with $\beta_i \in \mathbb{Z}$ for each $i \, ,$}
		$$
		then $u^{\epsilon_j}$ is evaluated in \eqref{eqn:rescaled-deformation} at
		$$
		x_0 - \epsilon_j \xi_j + \epsilon_j (p_k + \beta) =  \epsilon_j \big( p_k + [\alpha^j + \beta] \big) \, ,
		$$
		which is a node of the $\epsilon_j$-scaled lattice. A similar calculation reveals that the map
		$x \rightarrow x_0 - \epsilon_j \xi_j + \rho x$ takes the cell
		$\frac{\epsilon_j}{\rho} (U + \beta)$ of the $\frac{\epsilon_j}{\rho}$-scaled lattice to the cell
		$\epsilon_j \big( U + [\alpha^j + \beta] \big)$ of the $\epsilon_j$-scaled lattice, and
		$\frac{\epsilon_j}{\rho} (\overline{U}_m + \beta) \subset Q_1$ in $x$-space if and only if
		$\epsilon_j \big( \overline{U}_m + [\alpha^j + \beta] \big) \subset Q_\rho(x_0) - \epsilon_j \xi_j$ in the image space.
		It follows from the definition \eqref{eqn:elasticity-scaling} of our scaled energy (together with its
		translation invariance \eqref{alpha-as-transl-scaled}) that
		\begin{equation} \label{eqn:correspondence-of-energies}
			E^{\epsilon_j/\rho} (w^\rho_j, Q_1) = \rho^{-N} E^{\epsilon_j}(u^{\epsilon_j}, Q_\rho(x_0) - \epsilon_j \xi_j )
		\end{equation}
		
		While the blown-up deformation $w^{\rho}_j$ puts a spotlight on the behavior of $u^{\epsilon_j}$ near $x_0$, its
		relationship to the affine approximation of $u$ is not obvious. To make that relationship more evident, it is
		convenient to define
		\begin{equation} \label{eqn:w0-defn}
			w_0(x) = \nabla u (x_0) \cdot x
		\end{equation}
		and to observe that \eqref{eqn:rescaled-deformation} can be rewritten as
		\begin{equation} \label{eqn:w_j}
			w_j^\rho(x) =
			\frac{u^{\epsilon_j}(x_0 - \epsilon_j \xi_j +\rho x) - u(x_0) - \nabla u(x_0) \cdot (\rho x) }{\rho} + w_0(x) \, .
		\end{equation}
		Notice that the numerator of the first term on the right becomes the affine approximation of $u$ at $x_0$ if we
		ignore the small translation $\epsilon_j \xi_j$ and replace $u^{\epsilon_j}$ by $u$.
		\medskip
		
		\noindent {\sc Step 3: Taking the limit $\epsilon_j \rightarrow 0$.} In Step 4 we will apply the diagonalization lemma
		to get a sequence $\rho_j \rightarrow 0$ with the following properties:
		\begin{align}
			& \lim_{j \rightarrow \infty} \int_{Q_1} |w_{j}^{\rho_j}(x) - w_0(x)|^2 \, dx = 0 \, , \label{eqn:rho_j}\\
			& \lim_{j \rightarrow \infty} \rho_j^{-N} \mu_{j} \big( Q_{\rho_j}(x_0) - \epsilon_j \xi_j \big) =
			\frac{d \mu}{dx}(x_0) \, , \label{eqn:rho_j-2}\\
			& \lim_{j \rightarrow \infty} (2\rho_j)^{-N} \mu_{j} \big( Q_{2\rho_j}(x_0) \big)=
			\frac{d \mu}{dx}(x_0) \, , \quad \text{and} \label{eqn:rho_j-3} \\
			& \lim_{j \rightarrow \infty} \frac{\epsilon_j}{\rho_j} = 0 \label{eqn:rho_j-4}\, .
		\end{align}
		The hypothesis of the diagonalization lemma involves a double limit in which $\epsilon_j$ tends to $0$ first, then
		$\rho$ tends to $0$. Therefore in the present step we lay the groundwork for \eqref{eqn:rho_j} and \eqref{eqn:rho_j-2}
		by showing that
		\begin{enumerate}
			\item[(a)] if $Q_{2\rho}(x_0) \subset \Omega$ then
			\begin{equation} \label{eqn:prep-for-rho_j}
				\lim_{j \rightarrow \infty} \int_{Q_1} |w_{j}^{\rho}(x) - w_0(x)|^2 \, dx =
				\frac{1}{\rho^{N+2}} \int_{Q_\rho(x_0)} |u(x)-u(x_0)- \nabla u(x_0) \cdot (x-x_0)|^2 \, dx \, ;
			\end{equation}
			
			\item[(b)] if in addition $\mu \big( \partial Q_\rho(x_0) \big) = 0 $ then
			\begin{equation} \label{eqn:prep-for-rho_j-2}
				\lim_{j \rightarrow \infty} \mu_{j} \big( Q_{\rho}(x_0) - \epsilon_j \xi_j \big) = \mu \big( Q_\rho (x_0) \big) \, .
			\end{equation}
		\end{enumerate}
		
		For \eqref{eqn:prep-for-rho_j}, we start by changing variables in \eqref{eqn:w_j} to get
		$$
		\int_{Q_1} |w_{j}^{\rho}(x) - w_0(x)|^2 \, dx =
		\frac{1}{\rho^{N+2}} \int_{Q_\rho(x_0)}
		|u^{\epsilon_j}(x - \epsilon_j \xi_j)-u(x_0)- \nabla u(x_0) \cdot (x-x_0)|^2 \, dx \, .
		$$
		Our task is thus to show that $u^{\epsilon_j}(x - \epsilon_j \xi_j) - u(x)$ converges to $0$ in
		$L^2(Q_\rho(x_0))$. By the triangle inequality
		\begin{equation} \label{eqn:prep-for-rho_j-triangle-ineq}
			|u^{\epsilon_j}(x - \epsilon_j \xi_j) - u(x)| \leq
			|u^{\epsilon_j}(x - \epsilon_j \xi_j) - u (x-\epsilon_j \xi_j)| + |u(x - \epsilon_j \xi_j) - u(x)| \, .
		\end{equation}
		The first term on the right tends to $0$ in $L^2(Q_\rho(x_0))$ since
		$Q_\rho(x_0) - \epsilon_j \xi_j \subset Q_{2\rho}(x_0) \subset \Omega$ when $\epsilon_j$ is sufficiently small, and
		$u^{\epsilon_j}$ tends weakly to $u$ in $H^1(\Omega)$ (which implies strong convergence in $L^2(\Omega)$).
		To deal with the second term on the right side of \eqref{eqn:prep-for-rho_j-triangle-ineq}
		we use the fact that
		$$
		\int_{Q_\rho(x_0)} |u(x-a) - u(x)|^2 \, dx \leq C a^2 \int_{Q_{2\rho}(x_0)} |\nabla u|^2 \, dx
		$$
		when $a$ is sufficiently small. Applying this with $a = \epsilon_j \xi_j$, we
		see that the second term also tends to $0$ in $L^2(Q_\rho(x_0))$.
		This completes the proof of \eqref{eqn:prep-for-rho_j}.
		
		For \eqref{eqn:prep-for-rho_j-2} we observe that
		\begin{equation} \label{eqn:prep-for-rho_j-2-triangle-ineq}
			\big| \mu_j \big( Q_\rho (x_0) - \epsilon_j \xi_j \big) \big| \leq
			\big| \mu_j \big( Q_\rho (x_0) - \epsilon_j \xi_j \big) - \mu_j \big( Q_\rho(x_0) \big) \big| +
			\big| \mu_j \big( Q_\rho (x_0)) - \mu(Q_\rho(x_0) \big) \big| \, .
		\end{equation}
		The second term on the right tends to $0$ because the measures $\mu_j$ converge weakly to $\mu$ and we have assumed that $\mu(\partial Q_\rho(x_0)) = 0$. Indeed, weak convergence implies that
		$\liminf_{j \rightarrow \infty} \mu_j(O) \geq \mu(O)$ when $O$ is open and
		$\limsup_{j \rightarrow \infty} \mu_j(C) \leq \mu(C)$ when $C$ is closed, so
		\begin{equation} \label{eqn:limiting-measure}
			\mu \big( Q_\rho(x_0) \big) \leq \liminf_{j \rightarrow \infty} \mu_j \big( Q_\rho(x_0) \big) \leq
			\limsup_{j \rightarrow \infty} \mu_j \big( Q_\rho(x_0) \big) \leq
			\limsup_{j \rightarrow \infty} \mu_j \big( \overline{Q}_\rho(x_0) \big) \leq
			\mu \big( \overline{Q}_\rho(x_0) \big) \, .
		\end{equation}
		When $\mu(\partial Q_\rho(x_0)) = 0$ the far left and far right expressions are equal, so each inequality is actually
		an equality.
		To deal with the first term on the right side of \eqref{eqn:prep-for-rho_j-2-triangle-ineq} we observe that for
		any pair of sets $A$ and $B$,
		$$
		\big| \mu_j(A) - \mu_j(B) \big| \leq \mu_j(A \triangle B)
		$$
		where $A \triangle B = (A \setminus B) \cup (B \setminus A)$ is the symmetric difference of $A$ and $B$. Applying this
		with $A = Q_{\rho}(x_0) - \epsilon_j \xi_j$ and $B = Q_{\rho}(x_0)$, we conclude that for any $\lambda > 0 $
		$$
		\big| \mu_j (Q_{\rho}(x_0) - \epsilon_j \xi_j) - \mu_j(Q_\rho(x_0)) \big| \leq
		\mu_j \big( Q_{\rho + \lambda}(x_0) \setminus Q_{\rho - \lambda} (x_0) \big)
		$$
		when $\epsilon_j$ is sufficiently small. Since $\mu_j$ converges weakly to $\mu$ we conclude that
		$$
		\limsup_{j \rightarrow \infty} \big| \mu_j (Q_{\rho}(x_0) - \epsilon_j \xi_j) - \mu_j(Q_\rho(x_0)) \big| \leq
		\mu \big( \overline{Q}_{\rho + \lambda}(x_0) \setminus Q_{\rho - \lambda} (x_0) \big) \, .
		$$
		Now taking the limit $\lambda \rightarrow 0$ and using that $\mu (\partial Q_\rho(x_0)) = 0$ we see that the
		first term on the right in \eqref{eqn:prep-for-rho_j-2-triangle-ineq} tends to $0$. This completes the proof of
		\eqref{eqn:prep-for-rho_j-2}.
		\medskip
		
		\noindent {\sc Step 4: Applying the diagonalization lemma.} We need to avoid the (at most countably many)
		values of $\rho$ where either $\partial Q_\rho (x_0)$ or $\partial Q_{2\rho}(x_0)$ has nonzero measure under $\mu$. It
		is therefore convenient to use the discrete version of our diagonalization lemma
		(see \cref{rmk:diagonalization-lemma-for-sequences}), using a sequence $\rho_k$ converging monotonically to $0$ such that
		\begin{equation} \label{eqn:initial-sequence-rho-k}
			\mu \big( \partial Q_{\rho_k} (x_0) \big) = 0 \quad \mbox{and} \quad
			\mu \big( \partial Q_{2\rho_k} (x_0) \big) = 0 \quad \mbox{for all $k \, .$}
		\end{equation}
		We start by observing that
		\begin{align*}
			& \lim_{k \rightarrow 0} \lim_{j \rightarrow \infty} \int_{Q_1} |w_{j}^{\rho_k}(x) - w_0(x)|^2 \, dx = 0 \, , \\
			& \lim_{k \rightarrow 0} \lim_{j \rightarrow \infty} \rho_k^{-N} \mu_{j} \big( Q_{\rho_k}(x_0) - \epsilon_j \xi_j \big) =
			\frac{d \mu}{dx}(x_0) \, , \\
			& \lim_{k \rightarrow 0} \lim_{j \rightarrow \infty} (2\rho_k)^{-N} \mu_{j} \big( Q_{2\rho_k}(x_0) \big)=
			\frac{d \mu}{dx}(x_0) \, , \quad \text{and} \\
			& \lim_{k \rightarrow 0} \lim_{j \rightarrow \infty} \frac{\epsilon_j}{\rho_k} = 0 \, .
		\end{align*}
		Indeed, the first line is immediate from \eqref{eqn:good-pt-for-u} and \eqref{eqn:prep-for-rho_j}; the second is
		immediate from \eqref{eqn:lebesgue-pt-of-mu} and \eqref{eqn:prep-for-rho_j-2}; the justification of the third is
		similar to (but easier than) that of the second; and the last line is obvious. The diagonalization lemma is thus
		applicable with
		\begin{multline*}
			f(\rho_k,\epsilon_j) = \int_{Q_1} |w_{j}^{\rho_k}(x) - w_0(x)|^2 \, dx  +
			\big| \rho_k^{-N} \mu_{j} \big( Q_{\rho_k}(x_0) - \epsilon_j \xi_j \big) - \frac{d \mu}{dx}(x_0) \big| + \\
			\big| (2\rho_k)^{-N} \mu_{j} \big( Q_{2\rho_k}(x_0) \big) - \frac{d \mu}{dx}(x_0) \big| +
			\frac{\epsilon_j}{\rho_k} \, .
		\end{multline*}
		It supplies a correspondence $j \mapsto k(j)$ such that \eqref{eqn:rho_j}--\eqref{eqn:rho_j-4} hold when $\rho_j$ is replaced
		by $\rho_{k(j)}$. To simplify the notation, \emph{we shall henceforth denote $\rho_{k(j)}$ by $\rho_j$}.
		(This will lead to no confusion, since we shall make no further use of the original sequence $\{\rho_k\}$ introduced
		in \eqref{eqn:initial-sequence-rho-k}.)
		
		We claim that $w^{\rho_j}_j$ converges weakly in $H^1(Q_1)$ to $w_0$. Since we already know $L^2$ convergence from
		\eqref{eqn:rho_j}, it suffices to show that $\int_{Q_1} |\nabla w^{\rho_j}_j|^2 \, dx$ remains uniformly bounded as
		$j \rightarrow \infty$. To this end we observe that
		$$
		\int_{Q_1} |\nabla w^{\rho_j}_j|^2 \, dx =
		\rho_j^{-N} \int_{Q_{\rho_j}(x_0) - \epsilon_j \xi_j} |\nabla u^{\epsilon_j}|^2 \, dx \, .
		$$
		It is by now familiar that this can be bounded using the key property of our energy that
		$$
		C_2 \int_{\epsilon_j U + \alpha} |\nabla u^{\epsilon_j}|^2 \, dx \leq
		E^{\epsilon_j}( u^{\epsilon_j}, \epsilon_j U + \alpha) + D_2 |\epsilon_j U + \alpha| \, .
		$$
		Indeed, adding this estimate over all cells $\epsilon_j U + \alpha$ of the $\epsilon_j$-scaled lattice that
		meet $Q_{\rho_j}(x_0) - \epsilon_j \xi_j$ and using that $\epsilon_j / \rho_j \rightarrow 0$, we obtain an estimate
		of the form
		$$
		\int_{Q_{\rho_j}(x_0) - \epsilon_j \xi_j} |\nabla u^{\epsilon_j}|^2 \, dx \leq
		C [ E^{\epsilon_j}(u^\epsilon_j,Q_{2\rho_j}(x_0)) + \rho_j^N  ]\
		$$
		with a constant $C$ that's independent of $j$. Finally, we note that
		$$
		\rho_j^{-N} E^{\epsilon_j}(u^\epsilon_j,Q_{2\rho_j}(x_0) ) \leq \rho_j^{-N} \mu_j \big( Q_{2 \rho_j}(x_0) \big) \, ,
		$$
		which remains bounded as $j \rightarrow 0$ by \eqref{eqn:rho_j-3}. These estimates combine to give the
		desired uniform upper bound on $\int_{Q_1} |\nabla w^{\rho_j}_j|^2 \, dx \, .$
		\medskip
		
		\noindent {\sc Step 5: Putting it all together.} In Step 1 we reduced our task to showing that
		$d\mu/dx \geq \overline{W}(\nabla u)$ almost everywhere. By combining the preceding results, we now show
		that it holds at $x_0$. Since $w^{\rho_j}_j$ is defined on the $\epsilon_j/\rho_j$ lattice,
		$\epsilon_j/\rho_j \rightarrow 0$, and $w^{\rho_j}_j $ converges weakly to $w_0(x) = \nabla u (x_0) \cdot x$
		in $H^1(Q_1)$, we know from \cref{lemma:affine} that
		$$
		\liminf_{j \rightarrow 0} E^{\epsilon_j/\rho_j}(w^{\rho_j}_j,Q_1) \geq \overline{W}(\nabla u(x_0)) \, .
		$$
		By \eqref{eqn:correspondence-of-energies} this can be rewritten as
		$$
		\liminf_{j \rightarrow 0} \rho_j^{-N} E^{\epsilon_j}(u^{\epsilon_j},Q_{\rho_j}(x_0) - \epsilon_j \xi_j) \geq
		\overline{W}(\nabla u(x_0)) \, .
		$$
		Now we evaluate the $\liminf$ using \eqref{eqn:rho_j-2} to obtain the desired conclusion
		$$
		\frac{d\mu}{dx} (x_0) \geq \overline{W}(\nabla u(x_0)) \, .
		$$
	\end{proof}
	
	\subsection{The proof of Theorem \ref{thm:theorem-dirichlet}}\label{subsec:dirichlet-proof}
	Theorem \ref{thm:theorem-dirichlet} asserts that when a Dirichlet boundary condition is imposed, the
	$\Gamma$-limit is again given by the same effective energy $\int_\Omega \overline{W}(\nabla u) \, dx$. This
	section provides the proof.
	
	\begin{proof}[Proof of \cref{thm:theorem-dirichlet}]
		The statement of the theorem requires that the boundary condition $\psi \! : \! \partial \Omega \rightarrow \mathbb{R}^N$
		be Lipschitz continuous. But by Kirzbraun's theorem, such $\psi$ can be extended to a Lipschitz function defined on
		on all $\mathbb{R}^N$. Therefore we may (and we will) consider that $\psi$ is defined everywhere rather than
		just on $\partial \Omega$. (Actually, our argument only uses it on a neighborhood of $\Omega$.)
		
		Let us start with the lower bound. It asserts that if $u^\epsilon - \psi \in \mathcal{A}^0_\epsilon (\Omega)$ and
		$u^\epsilon \rightharpoonup u$ in $H^1(\Omega)$ then
		$$
		\liminf_{\epsilon \rightarrow 0} E^\epsilon (u^\epsilon, \Omega) \geq \int_\Omega \overline{W}(\nabla u) \, dx \quad
		\text{and} \quad \text{$u = \psi$ at $\partial \Omega$} \, .
		$$
		The first assertion follows from \cref{thm:main-theorem}, so we only need to prove the second one.
		Let $\psi^\epsilon$ be the piecewise linearization of $\psi$. (More precisely,
		$\psi^\epsilon$ is the piecewise linearization of the deformation which
		takes the value $\psi(x^\epsilon)$ at each node $x^\epsilon$ of the $\epsilon$-scale lattice.) We know from
		\cref{lemma:interp-of-lip-fn} that $|\nabla \psi^\epsilon|$ is uniformly bounded (independent of $\epsilon$) and
		$|\psi^\epsilon - \psi| \leq C \epsilon$, so it is immediately clear that $\psi^\epsilon \rightharpoonup \psi$ in $H^1(\Omega)$.
		Since $u^\epsilon - \psi^\epsilon \in \mathcal{A}_\epsilon^0 (\Omega)$, this piecewise linear function
		vanishes at $\partial \Omega$, i.e. it is in $H^1_0(\Omega)$. Since
		$u^\epsilon - \psi^\epsilon$ converges weakly to $u - \psi$ in $H^1(\Omega)$ and $H^1_0(\Omega)$ is closed
		under weak $H^1$ convergence, we conclude that $u = \psi$ at $\partial \Omega$, as desired.
		
		We turn now to finding a recovery sequence.
		Given any $u \in H^1(\Omega)$ with $u=\psi$ at $\partial \Omega$, we must
		show the existence of a sequence $u^\epsilon$ such that
		$$
		u^\epsilon - \psi \in \mathcal{A}^0_\epsilon \quad \mbox{and} \quad
		E^\epsilon(u^\epsilon,\Omega) \rightarrow \int_\Omega \overline{W}(\nabla u) \, dx \, .
		$$
		The sequence provided by the proof of \cref{thm:main-theorem} is not sufficient, since it doesn't
		satisfy the first condition. We shall proceed in two steps. In the first, which assumes that $u = \psi$ near
		$\partial \Omega$, we shall modify the recovery sequence from \cref{thm:main-theorem} using the method of
		de Giorgi. In the second step we handle the general case using a density argument. (These arguments are parallel to
		ones used in \cite{alicandro2004general} for a similar purpose.)
		\medskip
		
		\noindent {\sc Step 1:} Suppose $u=\psi$ in a neighborhood of $\partial \Omega$, and let $u^\epsilon \rightharpoonup u$
		satisfy $E^\epsilon(u^\epsilon,\Omega) \rightarrow\int_\Omega \overline{W}(\nabla u) \, dx$. (We showed the existence of
		such $u^\epsilon$ when we proved \cref{thm:main-theorem}.) We now sketch how the method of appendix \ref{appendix:degiorgi} lets us
		modify $u^\epsilon$ to obtain a new sequence $\tilde{u}^\epsilon$ with the desired properties.
		
		A key point is that since $u = \psi$ near $\partial \Omega$, we can (and do) choose the set
		$\Omega_0'$ in \eqref{eqn:shell-near-bdry} so that $u = \psi $ in $\Omega \setminus \Omega'_0$. Since the desired boundary
		condition is now $\psi$ rather than $\lambda x$, we consider the deformation defined at each node of the
		$\epsilon$-scale lattice by
		\begin{equation} \label{eqn:w_i_dir_bc}
			w_i^\epsilon(x) = \psi(x) + \varphi_i(x)(u^\epsilon(x)-\psi(x)) = \varphi_i(x) u^\epsilon(x) + \big(1-\varphi_i(x)\big) \psi(x)
		\end{equation}
		rather than the one defined by \eqref{eqn:w_i}. As usual, the piecewise linearization of this deformation will also be
		called $w_i^\epsilon$.
		
		The arguments that led us to \eqref{eqn:appendix-b-almost-done} extend easily to this setting. Minor adjustments are needed
		since in appendix \ref{appendix:degiorgi} the function $\lambda x$ was its own piecewise linearization,
		while in the present context
		$\psi^\epsilon \neq \psi$. However, \cref{lemma:interp-of-lip-fn} shows that $|\nabla \psi^\epsilon|$ is uniformly bounded,
		and this is what the argument needs. Consolidating constants, the analogue of \eqref{eqn:appendix-b-almost-done} is
		$$
		E^\epsilon (w^\epsilon_{i(\epsilon)}, \Omega) \leq E^\epsilon (u^\epsilon, \Omega) + C \delta +
		\frac{C'}{\nu} \left( |\nabla u^\epsilon - \nabla \psi^\epsilon |_{L^2(\Omega \setminus \Omega_0')} +
		\frac{4\nu^2}{R^2} |u^\epsilon - \psi^\epsilon |_{L^2(\Omega \setminus \Omega_0')}  \right)
		$$
		where $C$ and $C'$ do not depend on $\epsilon$, $\delta$, or $\nu$. Since $u^\epsilon$ remains
		bounded in $H^1(\Omega)$, we may (and do) choose $\nu = \nu(\delta)$ so that
		$$
		\frac{C'}{\nu} |\nabla u^\epsilon - \nabla \psi^\epsilon |_{L^2(\Omega \setminus \Omega_0')} \leq \delta
		$$
		for all $\epsilon$. The previous estimate then simplifies to
		\begin{equation} \label{eqn:analogue-of-est-from-appendix}
			E^\epsilon (w^\epsilon_{i(\epsilon)}, \Omega) \leq E^\epsilon (u^\epsilon, \Omega) + (C+1) \delta +
			\frac{4C'\nu^2}{R^2} |u^\epsilon - \psi^\epsilon |_{L^2(\Omega \setminus \Omega_0')} \, .
		\end{equation}
		
		We claim that $w^\epsilon_{i(\epsilon)}$ converges weakly to $u$ in $H^1(\Omega)$. To show this, we first observe
		that $E^\epsilon(w^\epsilon_{i(\epsilon)}, \Omega)$ stays uniformly bounded, by \eqref{eqn:analogue-of-est-from-appendix}.
		So the lower bound on our discrete energy can be used to control the $L^2$ norm of
		$\nabla  w^\epsilon_{i(\epsilon)}$ by arguing as
		we did for \eqref{eqn:after-adding-lower-bounds}. (Since
		$w^\epsilon_{i(\epsilon)} = \psi^\epsilon$ near $\partial \Omega$, we can consider its extension by $\psi^\epsilon$ and
		work in a domain slightly larger than $\Omega$; thus the issue that troubled us in
		\eqref{eqn:L2-bound-misses-bdry-layer} is not present here.) Therefore to show weak convergence to $u$, it suffices to show that
		\begin{equation} \label{eqn:w-i-eps-L2-convergence}
			\lim_{\epsilon \rightarrow 0} | w^\epsilon_{i(\epsilon)} - u |_{L^2(\Omega)} = 0 \, .
		\end{equation}
		This is relatively easy. Notice that in $\Omega_0'$ we have $w^\epsilon_{i(\epsilon)} = u^\epsilon$, while in
		$\Omega \setminus \Omega_0'$, we have $u=\psi$ and $w^\epsilon_{i(\epsilon)}$ is the piecewise linearization of
		$\psi + \phi_{i(\epsilon)}(u^\epsilon - \psi)$. Let us write $h^\epsilon$ for the piecewise linearization of
		$\phi_{i(\epsilon)}(u^\epsilon - \psi)$, or (equivalently) the piecewise linearization of
		$\phi_{i(\epsilon)}(u^\epsilon - \psi^\epsilon)$. Then
		$$
		\int_\Omega |w^\epsilon_{i(\epsilon)} - u|^2 \, dx = \int_{\Omega \setminus \Omega_0'} |u^\epsilon - u|^2 \, dx +
		\int_{\Omega_0'} |\psi^\epsilon - \psi + h^\epsilon |^2 \, dx \, .
		$$
		The first term on the right tends to zero since $u^\epsilon \rightarrow u$ in $L^2(\Omega)$, and the second term
		tends to zero by combining Lemmas \ref{lemma:interp-of-lip-fn} and \ref{lemma:interp-of-two-defs} with the $L^2$
		convergence of $u^\epsilon$ to $u$ and the triangle inequality.
		
		Next, we claim that
		\begin{equation} \label{eqn:w-i-eps-energy-double-limit}
			\limsup_{\delta \rightarrow 0} \limsup_{\epsilon \rightarrow 0} E^\epsilon (w^\epsilon_{i(\epsilon)}, \Omega) =
			\int_\Omega \overline{W}(\nabla u) \, dx \, .
		\end{equation}
		This follows easily from \eqref{eqn:analogue-of-est-from-appendix}, since $|\psi^\epsilon - \psi| \leq C \epsilon$ by
		\cref{lemma:interp-of-lip-fn}, while $u^\epsilon \rightarrow u$ in $L^2$ and $u=\psi$ in $\Omega \setminus \Omega_0'$.
		
		We now apply the diagonalization \cref{lemma:diagonal-argument} with
		$$
		f(\epsilon,\delta) = E^\epsilon (w^\epsilon_{i(\epsilon)}, \Omega) - \int_\Omega \overline{W}(\nabla u) \, dx \, .
		$$
		To be clear about the respective roles of $\epsilon$ and $\delta$, we recall that in the definition \eqref{eqn:w_i_dir_bc} of
		$w^\epsilon_i$ only $\varphi_i$ depends on $\delta$. So the dependence of $w^\epsilon_{i(\epsilon)}$ on $\delta$ is
		that
		$$
		w^\epsilon_{i(\epsilon)} = \phi^\delta_{i(\epsilon,\delta)} u^\epsilon +
		(1 - \phi^\delta_{i(\epsilon,\delta)}) \psi \quad \mbox{at lattice nodes.}
		$$
		The sequence $v^\epsilon$ provided by the diagonalization lemma is obtained by simply taking $\delta$ to be a suitable
		function of $\epsilon$. It is clear that
		$$
		\lim_{\epsilon \rightarrow 0} | v^\epsilon - u |_{L^2(\Omega)} = 0
		$$
		since our proof of \eqref{eqn:w-i-eps-L2-convergence} works uniformly in $\delta$. The diagonalization lemma assures
		us that
		\begin{equation} \label{eqn:limsup-of-energies-v-epsilon}
			\limsup_{\epsilon \rightarrow 0} E^\epsilon (v^\epsilon, \Omega) \leq  \int_\Omega \overline{W}(\nabla u) \, dx \, .
		\end{equation}
		It follows that $\int_\Omega |\nabla v^\epsilon|^2 \, dx$ remains bounded, by arguing as we did for
		$w^\epsilon_{i(\epsilon)}$ a little earlier. Thus $v^\epsilon$ converges weakly to $u$ in $H^1(\Omega)$. Now combining
		\eqref{eqn:limsup-of-energies-v-epsilon} with lower bound part of \cref{thm:main-theorem} gives
		$$
		\limsup_{\epsilon \rightarrow 0} E^\epsilon(v^\epsilon,\Omega) \leq  \int_\Omega \overline{W}(\nabla u) \, dx \leq
		\liminf_{\epsilon \rightarrow 0} E^\epsilon(v^\epsilon,\Omega) \, .
		$$
		Since $\liminf \leq \limsup$, we conclude that
		$\lim_{\epsilon \rightarrow 0} E^\epsilon(v^\epsilon,\Omega) = \int_\Omega \overline{W}(\nabla u) \, dx$. Thus we have
		achieved the goals of Step 1.
		\medskip
		
		\noindent {\sc Step 2:} Now consider any $u \in H^1(\Omega)$ satisfying $u = \psi$ at $\partial \Omega$. Since
		$u - \psi \in H^1_0(\Omega)$ and compactly supported functions are dense in $H^1_0(\Omega)$, for $k=1,2,\ldots$ we
		can choose $u_k \in H^1(\Omega)$ such that
		$$
		|u_k - u |_{H^1(\Omega)} \leq 2^{-k} \quad \text{and} \quad \text{$u_k = \psi$ in a neighborhood of $\partial \Omega \, .$}
		$$
		By Step 1 there is a sequence $u^\epsilon_k$ converging weakly to $u_k$ in $H^1(\Omega)$ such that
		$u^\epsilon_k - \psi \in \mathcal{A}^0_\epsilon(\Omega)$ and
		$$
		\lim_{\epsilon \rightarrow 0} E^\epsilon (u^\epsilon_k, \Omega) = \int_\Omega \overline{W}(\nabla u_k) \, dx \, .
		$$
		Using the quadratic growth and Lipschitz properties of $\overline{W}$ we have
		$$
		\lim_{k \rightarrow \infty} \int_\Omega \overline{W}(\nabla u_k) \, dx = \int_\Omega \overline{W}(\nabla u) \, dx \, .
		$$
		Therefore we can apply the discrete version of the diagonalization \cref{lemma:diagonal-argument} with
		$$
		f(\epsilon, 1/k) = \big| E^\epsilon (u^\epsilon_k, \Omega) - \int_\Omega \overline{W}(\nabla u) \, dx \big|+
		\int_\Omega |u^\epsilon_k - u|^2 \, dx
		$$
		to get a sequence $u^\epsilon_{k(\epsilon)}$ that satisfies our Dirichlet boundary condition and converges to $u$ in
		$L^2(\Omega)$, with
		$$
		\lim_{\epsilon \rightarrow 0} E^\epsilon (u^\epsilon_{k(\epsilon)}, \Omega) = \int_\Omega \overline{W}(\nabla u) \, dx \, .
		$$
		Moreover, since the discrete energy of $u^\epsilon_{k(\epsilon)}$ stays bounded as $\epsilon \rightarrow 0$, we get a uniform
		$H^1$ bound on this sequence by arguing as in Step 1. Thus $u^\epsilon_{k(\epsilon)}$ converges weakly to $u$ and its energy
		converges to the associated effective energy, fullfilling the obligations of a recovery sequence.
	\end{proof}
	
	
	\section{Applications to 2D lattice systems of springs}\label{sec:examples}
	Our framework is applicable to a broad variety of periodic lattice systems. To provide guidance about its use, this
	section discusses its application to four specific two-dimensional examples. The key point is always to choose a unit cell $U$
	and an appropriate energy $E(u,U)$ whose scaled version $E^\epsilon$ satisfies our basic conditions
	\eqref{eqn:unit-cell-eps-periodicity}-\eqref{eqn:unit-cell-lower}. We must also identify, for each example, the mesh to be
	used for our piecewise linearization scheme.
	
	As we explained in \cref{subsec:intro-effective}, to avoid unintended degeneracy the energy should include a term
	penalizing change of orientation. We implement this additively: throughout this section
	\begin{equation} \label{eqn:total-energy}
		E(u,U) = E_\text{spr}(u, U) + E_\text{pen}(u,U) \, ,
	\end{equation}
	where $E_\text{spr}$ is a sum of spring energies and $E_\text{pen}$ is a change-of-orientation penalty. As discussed in
	\cref{subsec:intro-effective}, it is natural for $E_\text{pen}(u,U)$ to have the form
	\begin{equation} \label{eqn:penalty-unit-cell}
		E_\text{pen}(u, U) = \sum_{T \in \mathscr{T}} f^\eta \big( \text{det}(\nabla u|_T) \big) |T| \,
	\end{equation}
	where $\mathscr{T}$ is an appropriately chosen collection of triangles from the mesh used for piecewise linearization and
	$f^\eta$ is the piecewise constant function defined by \eqref{eqn:discontinuous-penalization-term}. However other forms
	are permitted. For our framework to apply, it is sufficient (though not necessary) that $E_\text{pen}$ be
	translation-invariant, nonnegative, and bounded above:
	\begin{equation}  \label{eqn:penalty-bound}
		\begin{split}
			&\mbox{$E_\text{pen}(u+c,U) = E_\text{pen}(u,U)$ when $c$ is constant, and}\\
			&\mbox{$0 \leq E_\text{pen}(u,U) \leq M$ for some finite $M$ (independent of $u$).}
		\end{split}
	\end{equation}
	
	Concerning the spring energy $E_{\text{spr}}$: all our examples involve Hookean springs joining selected pairs of lattice
	nodes. We gave two examples in \cref{subsec:lattice-nodes-etc}, for the Kagome lattice \eqref{eqn:kagome-intro-energy}
	and for a square lattice with long-range interactions \eqref{eqn:square-long-intro-energy}; our other examples
	will be similar. Since the energy of a
	Hookean spring is automatically translation-invariant and nonnegative, the only nontrivial requirements on $E_\text{spr}$
	are that it satisfy our upper and lower bounds:
	\begin{align}
		& E_\text{spr}(u,U) \leq C_1 \Big( |\nabla u|^2_{L^2(U_n)} + |U_n| \Big) \ \mbox{and} \label{eqn:unit-cell-upper-spring}\\
		& E_\text{spr}(u,U) \geq C_2 \Big( |\nabla u|^2_{L^2(U)} - D_2 |U| \Big) \, , \label{eqn:unit-cell-lower-spring}
	\end{align}
	where $U_n$ is defined by \eqref{eqn:U_n} and the constants $C_1,C_2,$ and $D_2$ must of course be independent of $u$.
	Notice that since $E_{\text{pen}}$ is assumed to be nonnegative and bounded above, if $E_\text{spr} $ satisfies
	these conditions then so does total energy $E = E_\text{str} + E_\text{pen}$.
	
	\begin{remark} \label{rmk:review-affine-interpolant}
		A review about the term $|\nabla u|^2_{L^2(U_n)}$ introduced in \cref{sec:setup}: while a deformation $u$ takes values
		only at the nodes of the lattice, we want to treat it as an everywhere-defined piecewise linear function. This is done by
		triangulating the unit cell $U$, specifying how $u$ is defined at any ``ghost vertices'' (see \cref{eqn:piecewise-linearization-rule-a}), then using the triangulation
		to define a piecewise linear function. (In our examples, there will actually be no ghost vertices.) The terms
		$|\nabla u|^2_{L^2(U_n)}$ and $|\nabla u|^2_{L^2(U)}$ on the right hand sides of \eqref{eqn:unit-cell-upper-spring}
		and \eqref{eqn:unit-cell-lower-spring} refer to this piecewise linear function.
	\end{remark}
	
	For a given lattice system, it is in general a nontrivial task to identify a suitable spring energy $E_\text{spr}(u,U)$.
	If there are springs connecting nodes in the unit cell to nodes in other cells, then $E_\text{spr}$ should include the energies
	of some springs of this type, and the value of $n$ in \eqref{eqn:unit-cell-upper-spring} will be bigger than $1$.
	(The square lattice with long-range interactions in \cref{fig:square-long} has this character.) To satisfy the lower bound \eqref{eqn:unit-cell-lower-spring}, $E_\text{spr}$ must include the energies of sufficiently many springs. But it cannot
	include \emph{too} many, since the total of it and its translates must be the energy of the full lattice system. In some cases,
	this dichotomy is best handled by letting $E_\text{spr}$ include just \emph{part} of the energies of some springs. This must,
	of course, be done with care so that the sum of $E_\text{spr}$ and its translates counts the energy of each spring exactly
	once. (We shall proceed this way for the square lattice in \cref{subsec:square}.)
	
	\begin{figure}[!htb]
		\centering
		\includegraphics[scale=.35]{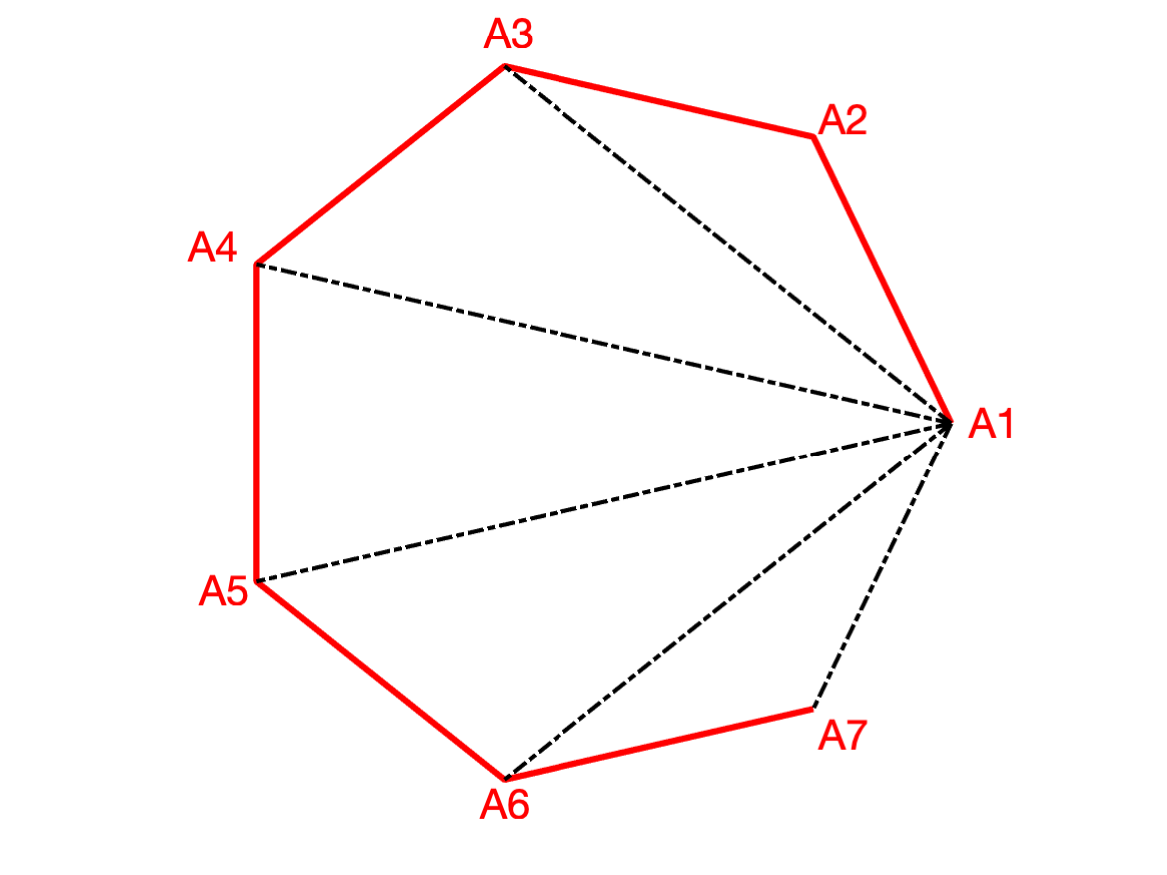}
		\caption{A polygon with $n=7$: the energy on the red solid edges are counted in
			$E_{\text{poly}}(u, P_n)$, while the dotted edges are artificial edges to indicate the triangular mesh.
			Here the vertices are numbered counter-clockwise, however our upper and lower bounds are also
			valid (with the same proofs) when the vertices are numbered clockwise.}
		\label{fig:polygon}
	\end{figure}
	
	In all our examples, the proofs of the essential inequalities \eqref{eqn:unit-cell-upper-spring}--\eqref{eqn:unit-cell-lower-spring} rely on a result concerning the
	spring energy of a convex polygon. We discuss it now in fairly general terms, since this result is also useful for
	other examples. Consider an $n$-sided convex polygon $P_n$ as shown in \cref{fig:polygon}, with vertices $A_1,  A_2, \dots, A_n$
	where $A_1 \sim A_2$, $A_2 \sim A_3$, $\dots$, $A_{n-1} \sim A_n$, $A_n \sim A_1$ . For a given deformation $u$ that has values at $A_1, A_2, \dots, A_n$, we consider the following energy
	\begin{align} \label{eqn:poly-energy}
		E_{\text{poly}}(u, P_n) &:= \sum_{i=1}^{n-1} \Big(|u(A_{i+1}  - u(A_i))|  - |A_{i+1} - A_i|\Big)^2 \, ,
	\end{align}
	which is the sum of the energies of $n-1$ springs (all except the one connecting $A_1$ and $A_n$).
	We show in appendix \ref{appendix-polygon} that this energy has the following upper and
	lower bounds: for any deformation $u$ that has values at $A_1, \dots, A_n$,
	\begin{itemize}
		\item there is an upper bound
		\begin{align}
			E_{\text{poly}}(u, P_n) &\leq c_1 \Big(|\nabla u|_{L^2(P_n)}^2 + |P_n|\Big) \quad \text{and} \label{eqn:poly-upper}
		\end{align}
		
		\item there is a lower bound
		\begin{align}
			E_{\text{poly}}(u, P_n) & \geq c_2 |\nabla u|_{L^2(P_n)}^2 - c_3 |P_n| \, , \label{eqn:poly-lower}
		\end{align}
	\end{itemize}
	with the understanding that $|\nabla u|_{L^2(P_n)}$ is determined by the nodal values of $u$ using
	the mesh that we are about to discuss. The constants $c_1, c_2, c_3$ are positive and depend only on the geometry of
	the polygon. On the right hand side, $|\nabla u|^2_{L^2(P_n)}$ refers to the piecewise triangularization of $u$
	using the mesh shown in \cref{fig:polygon}; it consists of $n-2$ triangles:
	$\Delta A_1 A_2 A_3$, $\Delta A_1 A_3 A_4$, $\dots$, and $\Delta A_1 A_{n-1} A_n$.
	
	A key feature of estimates \eqref{eqn:poly-upper} and \eqref{eqn:poly-lower} is that
	\emph{we only need the spring energy on $n-1$ edges} to upper
	and lower bound the $L^2$ norm of $|\nabla u|$ on an $n$-sided polygon. We shall apply these estimates
	to the four examples considered in this section, and similar arguments work for many other lattice
	systems of springs.
	
	\begin{remark}
		When using the bounds \eqref{eqn:poly-upper} and \eqref{eqn:poly-lower}, it is important to keep in mind that they
		are not asserted for an arbitrary piecewise linearization scheme; rather, they are asserted only when the right hand
		side is evaluated using the piecewise linearization scheme specified above.
	\end{remark}
	
	\subsection{The Kagome and \edit{Rotating Squares} metamaterials, viewed as lattices of springs}\label{subsec:rotating-squa}
	We start with the Kagome metamaterial and the \edit{Rotating Squares} metamaterial as our first illustrative examples,
	since they have mechanisms but are not entirely degenerate. As we explained in \cref{subsec:mechanism-based}, we believe
	that the soft modes of such systems are best understood as the macroscopic deformations whose effective energy vanishes.
	It is therefore important to know that there is indeed a well-defined effective energy.
	
	Another interesting feature of these two systems is that besides our spring model, there is also a cut-out model (as we
	discussed in \cref{subsec:mechanism-based}. As a result, it is natural to only penalize change of orientation
	on \emph{some} of the triangles in our mesh -- specifically, those that lie within the material that has been kept
	in the cut-out model.
	
	\subsubsection{The Kagome metamaterial}\label{subsubsec:kagome}
\begin{figure}[!htb]
		\begin{minipage}{.45\linewidth}
			\centering
			\subfloat[]{\label{fig:kagome-unit-3}\includegraphics[scale=.35]{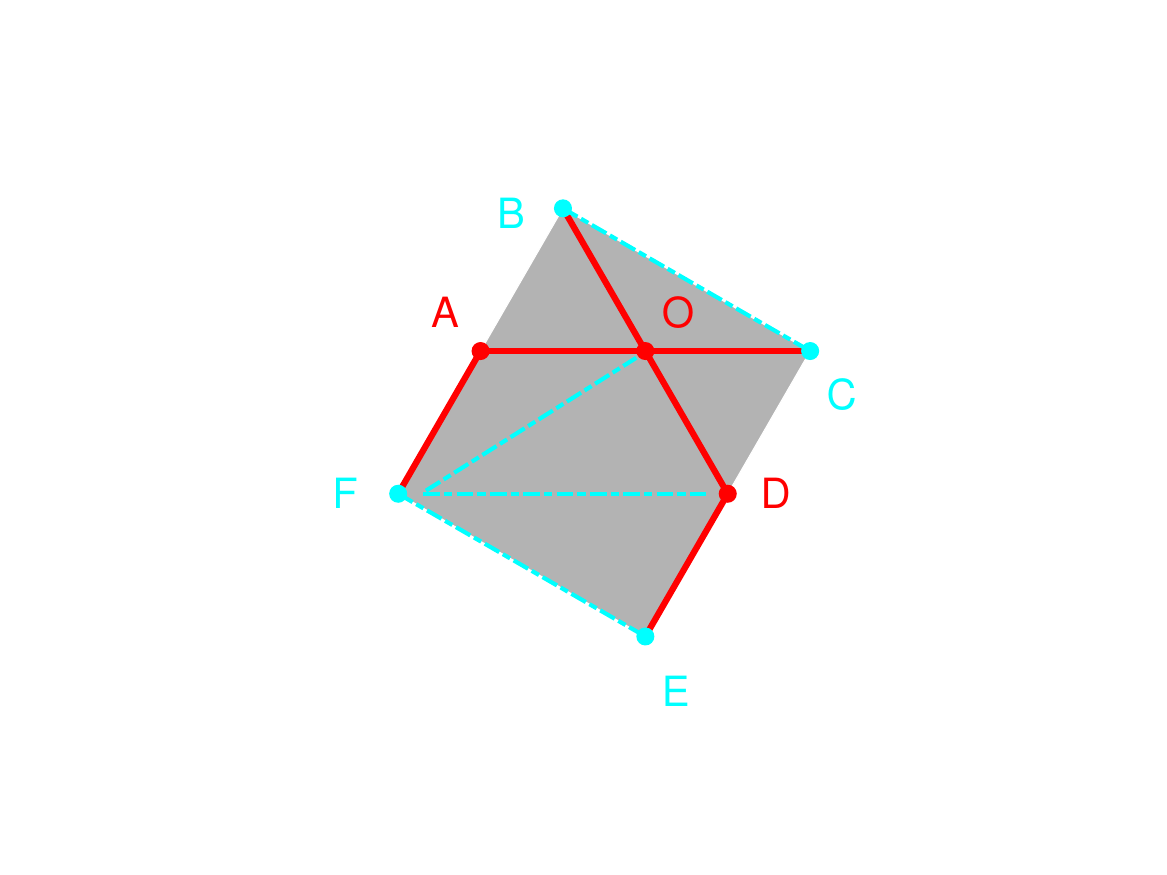}}
		\end{minipage}
		\begin{minipage}{.45\linewidth}
			\centering
			\subfloat[]{\label{fig:rotating-square-unit}\includegraphics[scale=.35]{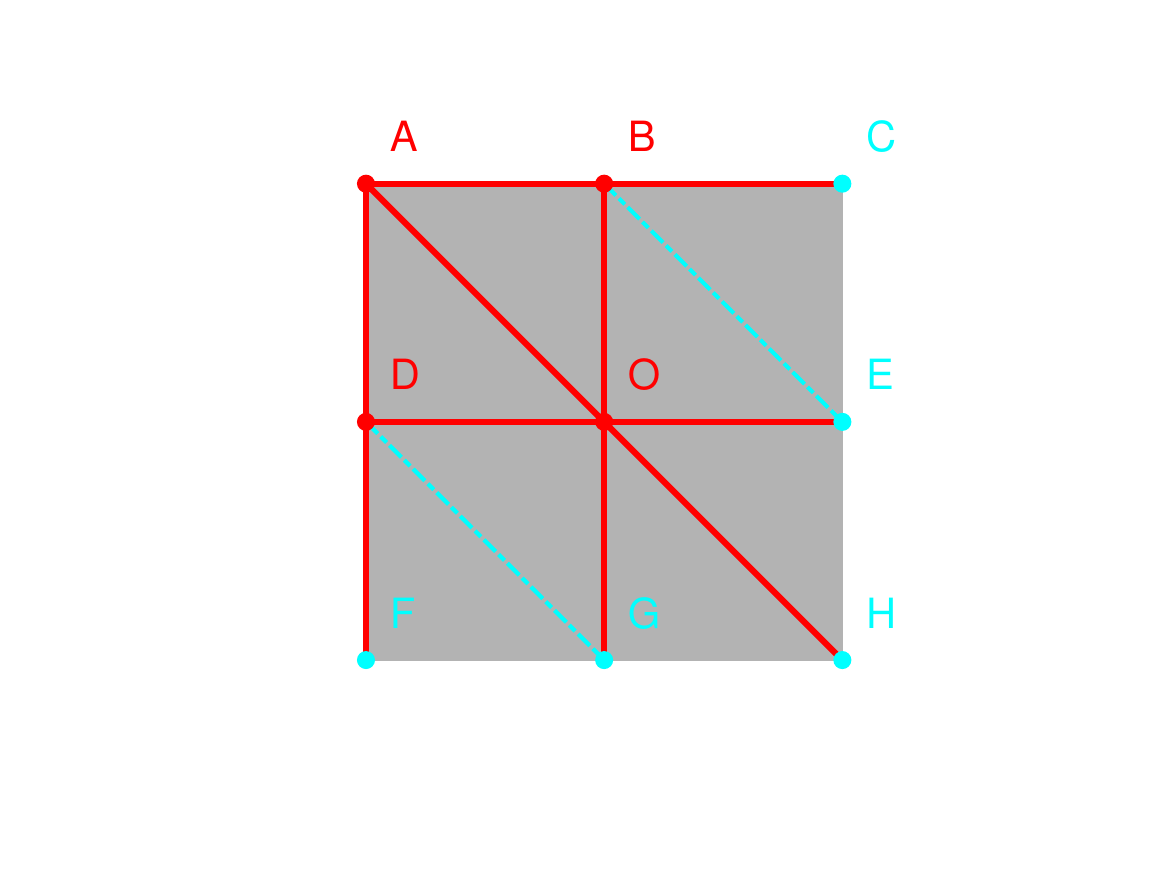}}
		\end{minipage}
		\caption{The unit cells of the Kagome lattice and the \edit{Rotating Squares} lattice: nodes in $\mathcal{V}$ are marked in red;
			nodes not in $\mathcal{V}$ but used in the energy $E(u,U)$ are marked in cyan; edges whose springs are counted in
			$E(u,U)$ are marked in red solid lines; artificial edges used to define the triangular mesh are marked in
			cyan dotted lines; the shaded area is $U$.}
	\end{figure}

	The Kagome metamaterial was already introduced in \cref{subsec:lattice-nodes-etc}. Our unit
	cell and triangular mesh for the Kagome lattice were identified in \cref{fig:kagome-3}; for the reader's
	convenience, they are shown again in \cref{fig:kagome-unit-3}. In the cut-out model of this metamaterial the hexagonal regions
	in \cref{fig:kagome-3} are holes, leaving material only in the equilateral triangles. Therefore it is physically
	natural to penalize change of orientation only in the triangles. Since the unit cell contains two such triangles,
	$\Delta AOB$ and $\Delta DOC$, the set $\mathscr{T}$ in \eqref{eqn:penalty-unit-cell} should
	contain just these triangles; thus we propose
	\begin{align} \label{eqn:penalty-kagome}
		E_{\text{pen}}(u,U) = f^\eta (\det(\nabla u|_{\Delta AOB})) +  f^\eta (\det(\nabla u|_{\Delta DOC}))
	\end{align}
	where $f^\eta$ is given by \eqref{eqn:discontinuous-penalization-term} with $\eta$ sufficiently small.
	
	For $E_\text{spr}(u,U)$ we use the energies of the springs associated with the unit cell, which are
	marked in red in \cref{fig:kagome-unit-3}; they lie along $AO$, $BO$, $CO$, $DO$, $DE$, and $AF$. The formula
	for $E_\text{spr}$ is thus given by \eqref{eqn:kagome-intro-energy}. To show that these choices fit our framework we
	must show that $E_\text{spr}$ satisfies the upper
	and lower bounds \eqref{eqn:unit-cell-upper-spring}--\eqref{eqn:unit-cell-lower-spring}. This will be done by bounding
	$E_\text{spr}$ above or below by sums of energies of polygons then using \eqref{eqn:poly-upper} or \eqref{eqn:poly-lower}.
	
	For the upper bound, we observe that all the springs associated with the unit cell $U$ have both ends
	in $\overline{U}$. Therefore we can take $n=1$. Since our mesh does not use ghost vertices, we
	also have $m=1$.
	For the upper bound, we observe that the spring energy can be written as the sum of two polygon energies (associated with
	the triangle $\Delta BOC$ and the pentagon $P_{FAODE}$ with vertices $F,A,O,D,E$), then we apply the upper
	bound \eqref{eqn:poly-upper}:
	\begin{align*}
		E_\text{spr}(u,U) &=  E_{\text{poly}}(u,\Delta BOC) + E_\text{poly}(u,P_{FAODE})  \\
		& \leq c_1(\Delta BOC) \Big(|\nabla u|^2_{L^2(\Delta BOC)} + |\Delta BOC |\Big) +
		c_1(P_{FAODE}) \Big(|\nabla u|^2_{L^2(P_{FAODE})} + |P_{FAODE}|\Big)  \\
		&\leq \widetilde{c}_1 \Big(|\nabla u|^2_{L^2(U)} + |U|\Big) \, .
	\end{align*}
	
	For the lower bound we must use a different decomposition, since the one used above does not include
	triangles $\Delta AOB$ and $\Delta COD$. To cover the missing triangles, we use the decomposition
	\begin{align*}
		E_\text{spr}(u,U) &= \frac{1}{2} E_{\text{poly}}(u,\Delta AOB) + \frac{1}{2} E_{\text{poly}}(u,\Delta BOC) +
		\frac{1}{2} E_{\text{poly}}(u,\Delta COD) \\
		& \quad + \frac{1}{2} E_{\text{poly}}(u,P_{FAODE}) + \frac{1}{2} E_{AF}(u) + \frac{1}{2} E_{DE}(u) \, ,
	\end{align*}
	where $E_{AF}(u) = (|u(A)-u(F)| - |A-F|)^2$ is the energy of the spring $AF$ and $E_{DE}(u)$ \edit{is similarly} the
	energy of the spring $DE$. Using this, we obtain the desired lower bound
	\begin{align*}
		E_\text{spr}(u,U) &\geq \frac{1}{2} \Big( c_2(\Delta AOB) |\nabla u|_{L^2(\Delta AOB)}^2 - c_3(\Delta AOB)  |\Delta AOB|\Big) \\
		& \quad + \frac{1}{2} \Big( c_2(\Delta BOC) |\nabla u|_{L^2(\Delta BOC)}^2 - c_3(\Delta BOC) |\Delta BOC|\Big)\\
		& \quad + \frac{1}{2} \Big( c_2(\Delta COD) |\nabla u|_{L^2(\Delta COD)}^2 - c_3(\Delta COD)  |\Delta COD|\Big) \\
		& \quad + \frac{1}{2} \Big( c_2(P_{FAODE}) |\nabla u|_{L^2(P_{FAODE})}^2 - c_3 (P_{FAODE}) |P_{FAODE}|\Big)\\
		& \geq \frac{1}{2} \Big( c_{2, \, \rm min} |\nabla u|_{L^2(U)}^2 - c_{3, \, \rm max} |U|\Big)
	\end{align*}
	where $c_{2, \, \rm min}$ and $c_{3, \, \rm max}$ are respectively the min and max of the corresponding constants for
	the polygons under consideration. Since the last expression can easily be rewritten in the desired form
	$C_2 \Big( |\nabla u|_{L^2(U)}^2 - D_2 |U| \Big)$, we have established the lower bound.
	Thus the spring model of the Kagome metamaterial fits our framework.
	
	\subsubsection{The \edit{Rotating Squares} metamaterial}\label{subsubsec:rotating}
	This is perhaps the best-understood mechanism-based mechanical metamaterial, see e.g.
	\cite{czajkowski2022conformal,deng2020characterization,nassar2020microtwist,zheng2022continuum,zheng2022modeling} for
	recent developments and many references. Like the Kagome metamaterial, the \edit{Rotating Squares} example has both a
	cut-out model and a spring model. The cut-out model is obtained by patterning a 2D elastic sheet like a checkerboard
	then removing the white squares. This leaves the black squares meeting one another at corners, which we
	idealize as hinges where rotation is free. (In a more realistic model the black squares would meet at thin
	necks, which would permit rotation with very little elastic energy.) This metamaterial has a single
	mechanism -- that is, a one-parameter family of deformations that moves each black square by a rigid motion.
	Under the mechanism, the originally-square holes become parallelograms (see e.g. Figure 1 in
	\cite{czajkowski2022conformal}).
	
\begin{figure}[!htb]
		\begin{minipage}{.48\linewidth}
			\centering
			\subfloat[]{\label{fig:square}\includegraphics[scale=.3]{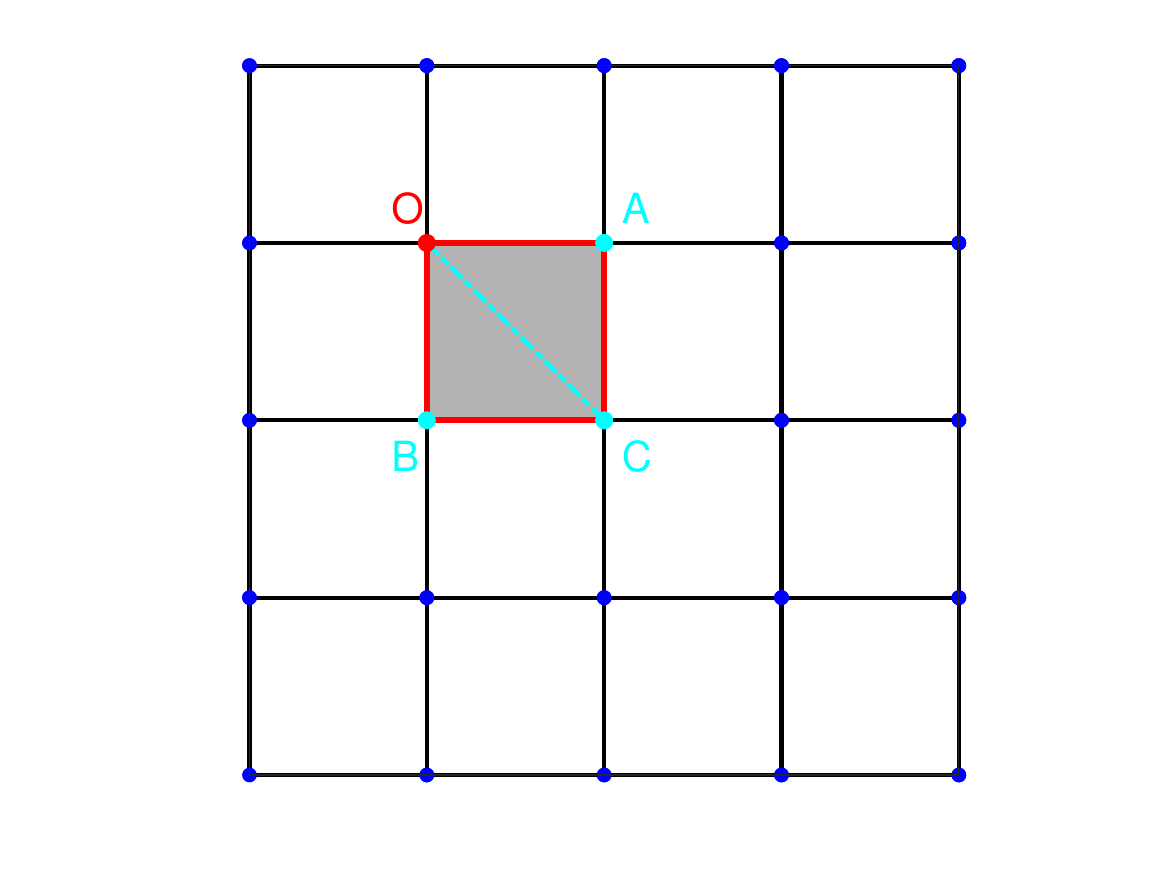}}
		\end{minipage}
		\begin{minipage}{.48\linewidth}
			\centering
			\subfloat[]{\label{fig:rotating-square}\includegraphics[scale=.3]{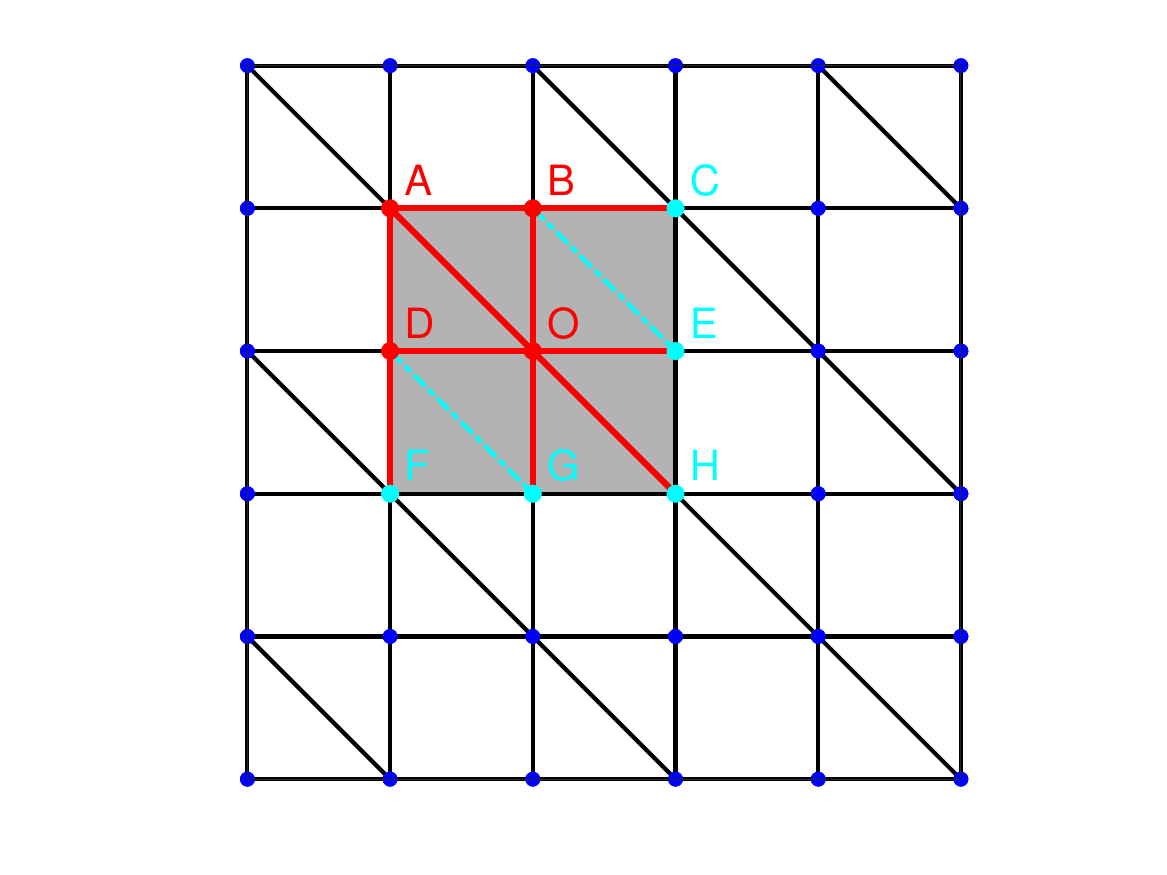}}
		\end{minipage}
		\caption{Two examples whose nodes are those of a square lattice: (a) the square lattice of springs; (b) the \edit{Rotating Squares}
			lattice. The nodes associated with the unit cell (our set $\mathcal{V}$) are marked in red; nodes not in
			$\mathcal{V}$ but used in the energy $E(u,U)$ are marked in cyan; springs counted in $E(u,U)$ are marked by
			red solid lines; artificial edges used only for the triangularization of $U$ are marked by cyan dotted lines;
			the shaded area is $U$.}
	\end{figure}

	The spring lattice version of this structure is obtained by starting with a square lattice then adding diagonals
	in the ``black squares'' (to give them rigidity) but not in the ``white squares'' (which play the role of the holes).
	The result is shown in Figure \ref{fig:rotating-square}, in which the solid edges are all Hookean springs.
	We like to call this the \emph{\edit{Rotating Squares} lattice}. As the unit cell $U$, it is convenient to use
	the square with vertices $A,C,H,F$. As the mesh for piecewise linearizing deformations, we choose the one shown
	as shown in the figure, which decomposes $U$ into $8$ triangles. (Note that the dotted segments $BE$ and $DG$ are
	not springs; rather, they are merely edges of triangles used for piecewise linearization.)
	
	The natural choice of $E_\text{pen}(u,U)$ penalizes change of orientation only in the ``black squares'' that belong
	to $U$. Those squares are $P_{BCEO}$ and $P_{DOGF}$, so change of orientation should be penalized only on the four
	triangles $\Delta ABO, \Delta ADO, \Delta OEH$ and $\Delta OGH$.
	
	The spring energy in this example is the aggregate energy of the $10$ springs $AB$, $AO$, $AD$, $BC$, $BO$, $DO$,
	$EO$, $DF$, $OG$, and $OH$:
	\begin{align*}
		E_\text{spr}(u,U) &= E_{AB}(u) + E_{AO}(u) + E_{BO}(u) + E_{DO}(u) + E_{AD}(u)\\
		& \quad + E_{BC}(u) + E_{DF}(u) + E_{OG}(u) + E_{OH}(u) + E_{OE}(u)
	\end{align*}
	where $E_{AB}(u) = (|u(A)-u(B)|-|A-B|)^2$ is the energy of the spring connecting $A$ and $B$, etc.
	It has the feature that when added to the energies of all periodic translates of $U$ we get, as desired, the
	total energy of the lattice. We note that $n=m=1$ for this example, since $E_\text{spr}(u,U)$ depends only on
	the values of $u$ in $\overline{U}$ and our piecwise linearization scheme has no ghost vertices.
	
	To show that our framework applies to this example, we must show that $E_\text{spr}$ satisfies our basic upper and
	lower bounds \eqref{eqn:unit-cell-upper-spring} and \eqref{eqn:unit-cell-lower-spring}. The arguments are similar to
	those in \cref{subsubsec:kagome}. We start by rewriting $E_\text{spr}$ as
	\begin{align*}
		E_\text{spr}(u,U) &= \Big[E_{AD}(u) + \frac{1}{2} E_{AO}(u)\Big] + \Big[E_{AB}(u) +
		\frac{1}{2} E_{AO}(u)\Big] + \Big[E_{BC}(u) + E_{BO}(u) + \frac{1}{2} E_{OE}(u)\Big]\\
		& \quad + \Big[E_{OD}(u) + E_{DF}(u) + \frac{1}{2} E_{OG}(u)\Big] +
		\frac{1}{2}\Big[E_{OG}(u) +  E_{OH}(u)\Big] + \frac{1}{2}\Big[E_{OH}(u) +  E_{OE}(u)\Big]\, .
	\end{align*}
	Each term in brackets is bounded above by the energy $E_\text{poly}(u,P_n)$ (defined by \eqref{eqn:poly-energy}) with $P_n$ being
	either a triangle or a quadrilateral; similarly, each term in brackets is bounded below by $1/2$ times the energy of a
	polygon. Using our upper and lower bounds \eqref{eqn:poly-upper}--\eqref{eqn:poly-lower}
	and adding, we easily deduce that
	$$
	E_\text{spr}(u,U) \leq c_{1,\, {\rm max}} \big( |\nabla u|^2_{L^2(U)} + |U| \big) \quad \text{and} \quad
	E_\text{spr}(u,U) \geq c_{2,\, {\rm min}} |\nabla u|^2_{L^2(U)} - c_{3, \, {\rm max }} |U|
	$$
	where $c_{1,\, {\rm max}}$, $c_{2,\, {\rm min}}$, and $c_{3,\, {\rm max}}$ are the max or min of the corresponding
	constants for the polygons under consideration. The first inequality has exactly the desired form \eqref{eqn:unit-cell-upper-spring} and the second can be rewritten in the desired form \eqref{eqn:unit-cell-lower-spring}.
	Thus our framework applies to the \edit{Rotating Squares} lattice.
	
	\subsection{The square lattice}\label{subsec:square}
	The square lattice with only nearest-neighbor connections is another interesting example.
	Unlike the \edit{Rotating Squares} lattice depicted in \cref{fig:rotating-square}, in this example \emph{none}
	of the squares have diagonal springs (see \cref{fig:square}). This system has a huge variety of mechanisms.
	The simplest is a uniform shear, which deforms each square to a parallelogram (all the parallelograms being
	identical in shape). But there are also many periodic mechanisms, which deform the squares to different-shaped parallelograms.
	Using suitable periodic mechanisms, one can achieve different compression ratios in the vertical and horizontal directions;
	moreover, this can be done without any local change of orientation (see for example Figure 1 of \cite{milton2013complete}). The reader might wonder why it is worthwhile to consider such a degenerate example. The answer lies in the connection between
	homogenization and soft modes that we discussed in \cref{subsec:mechanism-based}. There are systems (including the square lattice
	and the Kagome lattice) whose mechanisms are not easily enumerated. In such systems, it is not obvious how to define a soft mode.
	We think a macroscopic deformation $u$ should be considered a soft mode when its effective energy vanishes, i.e.
	when \edit{$\overline{W}(\nabla u)$} vanishes everywhere in $\Omega$. For this proposal to be meaningful, the effective energy must be
	well-defined even for systems with many mechanisms. The square lattice is a natural example of such a system.
	
\begin{figure}[!htb]
		\begin{minipage}{.45\linewidth}
			\centering
			\subfloat[]{\label{fig:square-unit}\includegraphics[scale=.35]{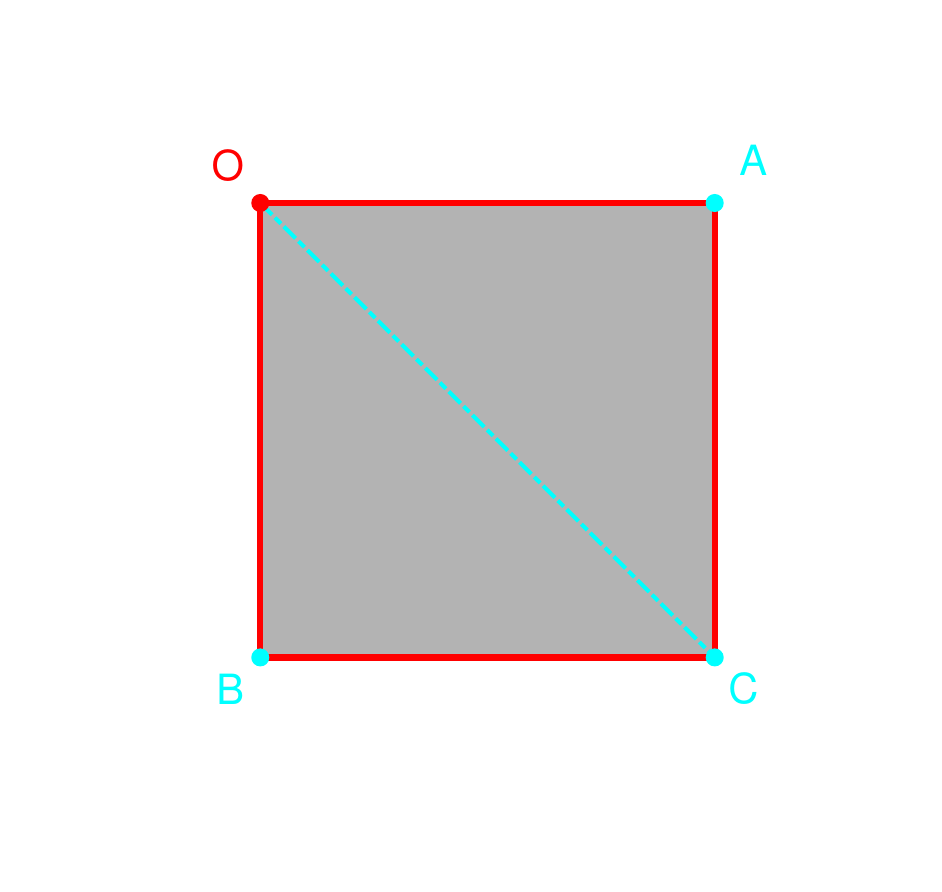}}
		\end{minipage}
		\begin{minipage}{.45\linewidth}
			\centering
			\subfloat[]{\label{fig:square-long-unit}\includegraphics[scale=.35]{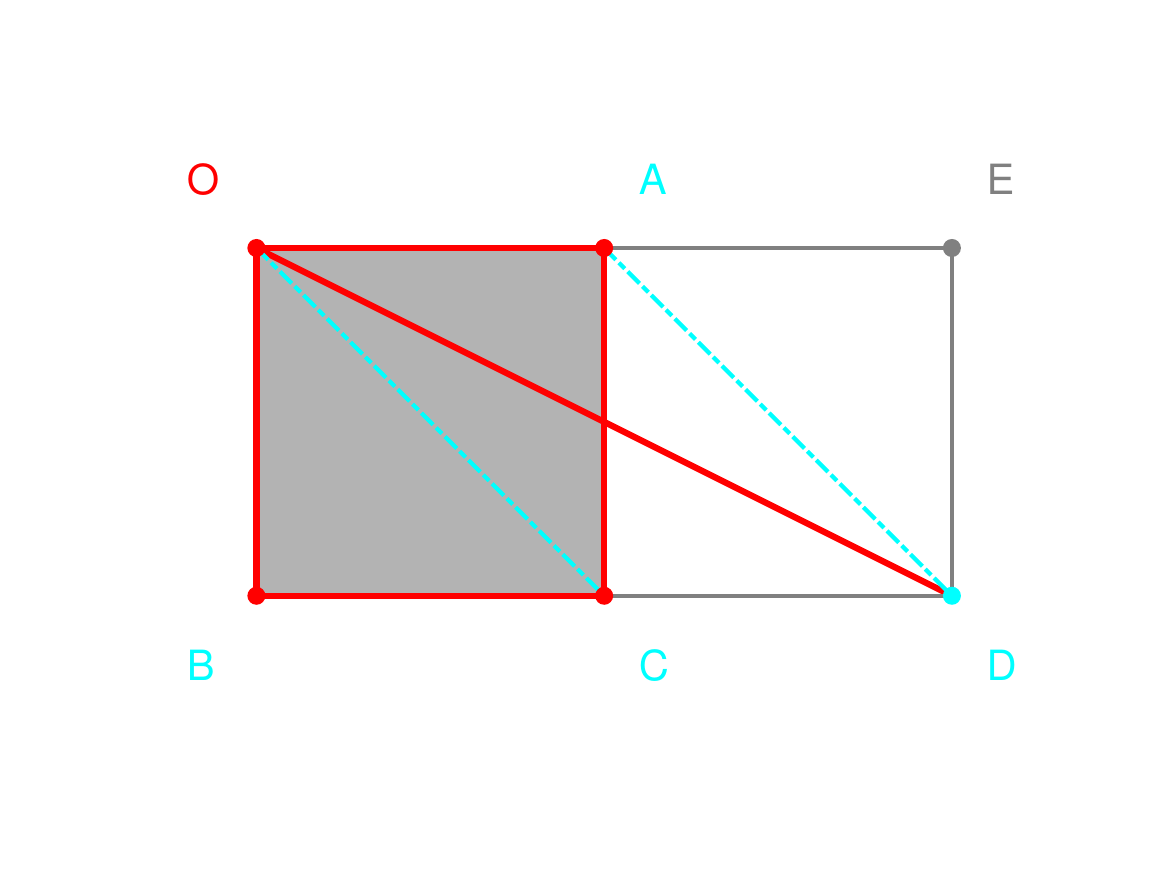}}
		\end{minipage}
		\caption{The unit cells of the square lattice and a square lattice with long-range and periodic edges:
			the nodes associated with the unit cell (that is, the ones in $\mathcal{V}$) are marked in red; nodes
			not in $\mathcal{V}$ but used in the energy $E(u,U)$ are marked in cyan; edges counted in $E(u,U)$ are
			marked by red solid lines; artificial edges used only for the triangular mesh are marked by cyan dotted
			lines; the shaded area is $U$. The gray edges in (b) are edges of the neighboring unit cell.}
	\end{figure}

	Our choice of the unit cell $U$ for this example is shown in \cref{fig:square} and again in \cref{fig:square-unit}. The mesh
	used for our piecewise linearization scheme has only two triangles: $\Delta OAC$ and $\Delta OBC$. To avoid
	folding deformations it is natural for $E_\text{pen}$ to penalize change of orientation on both of these triangles:
	\begin{equation} \label{eqn:E-pen-square-lattice}
		E_{\text{pen}}(u,U) = f^\eta(\det(\nabla u|_{\Delta OAC})) + f^\eta(\det(\nabla u|_{\Delta OBC}))
	\end{equation}
	where $f^\eta$ is defined by \eqref{eqn:discontinuous-penalization-term}.
	
	The spring energy for this example is
	\begin{equation*}
		E_\text{spr}(u,U) = \frac{1}{2} \Big(E_{AO}(u) + E_{BO}(u) + E_{AC}(u) + E_{BC}(u) \Big) \, .
	\end{equation*}
	The weight $1/2$ assures that when we add the energy of the unit cell and all its translates we count each spring
	exactly once. (One might ask why we don't take $E_\text{spr}$ to be the energy of just two springs, for example
	$E_{OA} + E_{OB}$ or $E_{OB} + E_{BC}$. The answer is that while those choices also get the aggregate spring
	energy right, neither one satisfies the crucial lower bound \eqref{eqn:unit-cell-lower-spring}.) Since
	$E(u,U)$ depends only on nodal values of $u$ in $\overline{U}$ and our piecewise linearization
	scheme involves no ghost vertices, this example has
	$n=m=1$.
	
	We claim that $E_\text{spr}$ satisfies the required
	bounds \eqref{eqn:unit-cell-lower-spring}-\eqref{eqn:unit-cell-upper-spring}.
	To see why, we observe that the spring energy can be rewritten as
	\begin{equation} \label{eqn:square-lattice-sum-of-polygon-energies}
		E_\text{spr}(u,U) =  \frac{1}{2} E_{\text{poly}} (u, \Delta OAC) + \frac{1}{2} E_{\text{poly}} (u, \Delta OBC) \, ,
	\end{equation}
	in which each term is our polygon energy \eqref{eqn:poly-energy} specialized to the indicated triangle:
	\begin{align*}
		E_{\text{poly}} (u, \Delta OAC) &= E_{AO}(u) + E_{AC}(u) \, ,\\
		E_{\text{poly}} (u, \Delta OBC) &= E_{BO}(u) + E_{BC}(u) \,  .
	\end{align*}
	Now the upper and lower bounds \eqref{eqn:poly-upper}--\eqref{eqn:poly-lower} on the polygon energies combine
	with \eqref{eqn:square-lattice-sum-of-polygon-energies} to give the desired inequalities for
	$E_\text{spr}$, exactly as they did in our discussion of the Kagome and \edit{Rotating Squares} lattices.
	Thus our framework applies to the square lattice.
	
	\subsection{The square lattice with long-range springs}\label{subsec:square-long}
	This example is illustrated in \cref{fig:square-long}. It has all the springs of the square lattice, plus additional springs
	whose endpoints are \emph{not} in the same translation of $\overline{U}$. Our goal in discussing this example is to show how
	our framework deals with the presence of long-range springs.
	
	As shown in \cref{fig:square-long-unit}, we take the unit cell $U$ and the mesh used for piecewise linearization to be
	exactly as they were for the square lattice. To avoid folding, it is natural to again penalize change of orientation on
	each triangle, i.e. to again take use \eqref{eqn:E-pen-square-lattice} for $E_\text{pen}$.
	
	The spring energy was already identified in \cref{subsec:lattice-nodes-etc}; we repeat it here for the reader's convenience:
	\begin{equation*}
		E_\text{spr}(u,U) = \frac{1}{2} \Big(E_{AO}(u) + E_{BO}(u) + E_{AC}(u) + E_{BC}(u) \Big) + E_{OD}(u)
	\end{equation*}
	Since $E_\text{spr}(u,U)$ depends on $u(D)$ as well as on $u(A)$, $u(B)$, $u(C)$, and $u(D)$, the value of $n$ for this example
	is clearly greater than $1$. Based on the definition \eqref{eqn:U_n}, we see that $n=2$. Since our piecewise
	linearization scheme has no ghost vertices, the value of $m$ is also $2$.
	
	As usual, we must show that $E_\text{spr}$ satisfies the upper and lower bounds
	\eqref{eqn:unit-cell-upper-spring} and \eqref{eqn:unit-cell-lower-spring}. No new work is needed for the lower bound,
	since the spring energy under discussion here is strictly larger than that of the square lattice, which we have already shown
	to satisfy the lower bound. (Note that the right side of the lower bound involves only $|\nabla u|_{L^2(U)}$ regardless of
	the value of $n$.)
	
	Turning now to the upper bound, we recall that its right hand side involves $|\nabla u|_{L^2(U_2)}$, where $U_2$ is the
	union of $U$ and its eight neighbors. We shall actually prove
	\begin{equation} \label{eqn:square-long-upper-desired}
		E_\text{spr}(u,U) \leq C_1 \Big(|\nabla u|_{L^2(U \cup P_{ACD})}^2 + |U \cup P_{ACD}|\Big) \, ,
	\end{equation}
	where $P_{ACD}$ is the triangle with corners $A,C,D$ in \cref{fig:square-long-unit}. This is stronger than
	\eqref{eqn:unit-cell-upper-spring}, since $U \cup P_{ACD}$ is a subset of $U_2$.
	We begin with the observation that the energy on $OD$ is bounded by
	\begin{align*}
		(|u(O) - u(D)| - |O-D|)^2 &\leq |u(O) - u(D)|^2 + |O-D|^2\\
		&\leq 2|u(O)-u(A)|^2 + 2|u(A) - u(D)|^2 + |O-D|^2 \, .
	\end{align*}
	Since $u$ is affine on each triangle of our mesh, it is elementary to see that $|u(O)-u(A)|^2$$\leq c |\nabla u|^2_{L^2(\Delta OAC)}$
	and $|u(A)-u(D)|^2 \leq c' |\nabla u|^2_{L^2(\Delta ADC)}$ where $c, c'$ are suitable constants (independent of $u$). Since
	$|O-D|^2$ is also a constant, we easily obtain an inequality of the form
	$$
	E_{OD}(u) \leq C \Big(|\nabla u|_{L^2(U \cup P_{ACD})}^2 + |U \cup P_{ACD}| \Big)
	$$
	Combining this with the upper bound proved earlier for the
	square lattice leads easily to \eqref{eqn:square-long-upper-desired}, completing the proof that this example fits
	into our framework.
    
\edit{
\section{Conclusions and a look forward} \label{sec:conclusions}

Motivated by the emerging literature on mechanism-based mechanical metamaterials, we have studied the sense in which 
a discrete, spatially periodic, geometrically nonlinear mechanical system (for example a lattice of springs) has a
well-defined macroscopic behavior. We have shown that when a macroscopic domain $\Omega$ is filled by a scaled version
of the mechanical system (with length scale $\epsilon$), the resulting discrete energy minimization problem gamma-converges (as $\epsilon \rightarrow 0$) to minimization of an effective nonlinear elastic energy 
$\int_\Omega \overline{W}(\nabla u) \, dx$. Moreover, we have given a variational characterization of the effective energy
density $\overline{W}$. 

While results of this type have been proven before for specific classes of systems, our approach is different 
from those in the literature. It has the advantage of being very general; in particular, it makes no hypothesis about
the system's geometry or topology aside from periodicity. Our approach uses piecewise linear interpolation 
to make the discrete problem resemble a continuous one. This lets us study the limit $\epsilon \rightarrow 0$
using methods from the homogenization literature concerning the effective behavior of spatially periodic nonlinearly
elastic solids. 

When using this framework to study a mechanism-based mechanical metamaterial, it is important to include -- as part of the
discrete energy -- a term that penalizes change of orientation. The details of this penalization term should reflect the 
physics of the problem. In particular, for the Kagome metamaterial (shown in 
\cref{fig:kagome-one-periodic}(a) and \cref{fig:modulation}(a)), 
change of orientation should be penalized only on the triangles (not the hexagons), since in the cut-out model 
of the Kagome metamaterial the hexagons are actually holes. 

As we discussed in the Introduction, mechanism-based mechanical metamaterials typically have 
\emph{soft modes} -- that is, discrete deformations whose elastic energy is very small (though it 
is nonzero, so they are not mechanisms). Such deformations
can be obtained by modulating a mechanism, as shown in fig. \ref{fig:modulation}(b) and 
\ref{fig:modulation}(c). We are, however,
not aware of a system-independent \emph{definition} of the term soft mode -- one that applies even to the Kagome 
metamaterial (which has many mechanisms -- raising the possibility that a soft mode might use multiple mechanisms,
rather than modulating just one). We propose that soft modes should be viewed as macroscopic objects, not microscopic
ones: a macroscopic deformation $u$ is a soft mode defined on a region $\Omega$ if its (appropriately-defined) effective
energy vanishes. Put differently: $u$ is a soft mode if there is a sequence $u^\epsilon$ converging (weakly) to $u$ 
such that $E^\epsilon(u^\epsilon,\Omega) \rightarrow 0$ as $\epsilon \rightarrow 0$. 
In plain English: a soft mode is the limiting behavior of a sequence of deformations $u^\epsilon$ whose discrete
energy tends to $0$ in the continuum limit $\epsilon \rightarrow 0$.

This paper shows that our proposed definition of a soft mode makes sense. It is, however, natural to ask: 

\begin{enumerate}
\item[(a)] Can we characterize the soft modes of some systems explicitly? 

\item[(b)] Does knowing the soft modes permit one to predict the results of mechanical experiments? 
\end{enumerate}

Our forthcoming paper \cite{liforthcoming} addresses (a) for the Kagome and the \edit{Rotating Squares} 
metamaterials (and also for some related examples). We show there -- using the discrete models discussed in 
this paper's \cref{subsec:rotating-squa}, with a sufficiently large choice of $\eta$ in the term that penalizes change of
orientation -- that for both Kagome and \edit{Rotating Squares}, the effective energy satisfies
\begin{equation} \label{eqn:eff-energy-lower-bound}
 \begin{aligned}
            \overline{W}(\lambda) & \geq 
			C \Big[(\lambda_1 - \lambda_2)^2 + (\lambda_1 - 1)_+^2 + (\lambda_2 - 1)_+^2\Big] & \mbox{if} \ \det(\lambda) \geq 0,\\
            \overline{W}(\lambda) & \geq  C \Big[(\lambda_1 + \lambda_2)^2 + (\lambda_1 - 1)_+^2 + (\lambda_2 - 1)_+^2\Big] & \mbox{if} \  \det(\lambda) < 0 .
        \end{aligned}
\end{equation}
Here $\lambda_1$ and $\lambda_2$ are the principal stretches of the $2 \times 2$ matrix $\lambda$ (the eigenvalues of
$(\lambda^T \lambda)^{1/2}$), and we use the notation $(x-1)_+^2 = (\max\{x-1,0\})^2$. 
The right side vanishes only when $\lambda = c R$ with $0 \leq c \leq 1$ and $R \in SO(2)$. Thus if a 2D domain 
$\Omega$ is filled with scaled copies of the Kagome or \edit{Rotating Squares} metamaterials, then (in the continuum limit)
its soft modes are the deformations $u : \Omega \rightarrow \mathbb{R}^2$ such that $\nabla u(x) = c(x) R(x)$ with 
$0 \leq c(x) \leq 1$ and $R(x) \in SO(2)$. Indeed, one deduces easily from \eqref{eqn:eff-energy-lower-bound} that soft 
modes must have this character; moreover, any such $u$ is indeed a soft mode since it can be approximated by 
modulating a periodic mechanism as shown in \cref{fig:modulation}(b) and (c). (The lower bound on $c(x)$ is $0$ not $1/2$ because 
our model permits the triangles to overlap.) 

Such functions $u$ have an alternative characterization: $u$ is a soft mode exactly if $f=u_1 + iu_2$ is a complex analytic
function of $z=x_1+ix_2$ with $|f'(z)| \leq 1$. (This, too, is explained in \cite{liforthcoming}; see also \cite{czajkowski2022conformal}.) 

The preceding discussion doesn't fully capture the relationship between the present paper and \cite{liforthcoming}. As we have already noted, 
the bound \eqref{eqn:eff-energy-lower-bound} makes sense because the left side is well-defined -- a key result of the 
present paper. But equally important: the variational characterization of $\overline{W}$ provided by \cref{lemma:periodic-bc} is crucial to our proof of 
\eqref{eqn:eff-energy-lower-bound}. It assures us that to prove a lower bound on $\overline{W}$ it is sufficient to consider
deformations of the form $u(x) = \lambda x + \psi(x)$ where $\psi$ is $k$-periodic for some $k$. This is true despite the 
fact that not all low-energy deformations have this form. (In fact, for the Kagome system there are even \emph{mechanisms} that 
don't have this form \cite{li2023some,liforthcoming}.)

Turning now to question (b): a complex analytic function defined on a domain $\Omega \subset \mathbb{C}$ is 
fully determined by its values at $\partial \Omega$. This suggests that when $\Omega$ is filled with the Kagome metamaterial
(with microstructural length scale $\epsilon$ sufficiently small) and $u$ is a soft mode, imposing a Dirichlet boundary condition that's close to 
$u|_{\partial \Omega}$ should lead to a deformation $u^\epsilon$ that's close to $u$ throughout $\Omega$. There is, however, some 
room for doubt: the discrete energy $E^\epsilon$ is nonconvex, and every \emph{local} minimum is a stable state. Our theory 
is based on gamma-convergence, so it describes the limiting behavior of $u^\epsilon$ only if this energy is asymptotically 
optimal -- in other words (since $u$ is a soft mode) if $\lim_{\epsilon \rightarrow 0} E^\epsilon(u^\epsilon,\Omega) = E_\text{eff}(u,\Omega)=0$. 
Depending upon the protocol by which the boundary condition is imposed, one can imagine that the $\epsilon$-scale structures 
could get stuck in local minima that don't have this property.

What about traction boundary conditions? Alas, it appears that while it is useful to know the soft modes, this alone 
might not be enough information to predict the macroscopic consequences of loading. Indeed, the papers \cite{czajkowski2022conformal,zheng2022continuum,zheng2022modeling} 
consider problems of this type, for the \edit{Rotating Squares} metamaterial and some generalizations thereof. Their results 
suggest that to match the results of physical experiments or finite-element simulations, one should minimize 
(over the soft modes) the \emph{leading-order elastic energy of the $\epsilon$-scale system plus the work done by the loads}. 
Moreover, these papers include in their elastic energies a term associated with the bending of the thin necks that we have been treating as nodes (where rotation is free). 

The preceding paragraphs have focused on the \edit{Rotating Squares} and Kagome examples, since they are the focus of \cite{liforthcoming}. 
We note, however, that \cite{zheng2022continuum} has characterized the soft modes of some other 2D systems, obtaining a first-order system for
$\nabla u$ (see also \cite{czajkowski2024duality}). Depending upon the details of the microstructure, this system can be either elliptic or hyperbolic. The qualitative 
features of the soft modes are of course very different in the elliptic vs hyperbolic cases. 

We have focused on the role of soft modes, however there are plenty of Dirichlet boundary conditions that are not consistent with 
any soft mode. (For the Kagome and \edit{Rotating Squares} systems, if $u_1 |_{\partial \Omega}$ and $u_2 |_{\partial \Omega}$ are not the 
boundary values of conjugate harmonic functions then they are inconsistent with $u$ being a soft mode.) For such data it would
be natural to minimize the effective energy directly. Alas, we know of no example where a formula for the effective energy has 
been derived in a systematic way. The right hand side of our inequality \eqref{eqn:eff-energy-lower-bound} could perhaps be 
useful as a toy model. We note in this regard that the expression $(\lambda_1 - \lambda_2)^2$ can be written much more 
explicitly: when $\lambda = [(\nabla u)^T \nabla u]^{1/2}$ it equals 
$\mbox{tr}(\lambda^2) - 2 \det(\lambda) = |\nabla u|^2 - 2 \det(\nabla u)$.
}
	
	\appendix
	\section*{Appendices}
	\addcontentsline{toc}{section}{Appendices}
	\renewcommand{\thesubsection}{\Alph{subsection}}
	
	\renewcommand{\theequation}{A.\arabic{equation}}
	\setcounter{equation}{0}
	\subsection{Proofs of Lemma \ref{lemma:interp-of-lip-fn} and Lemma \ref{lemma:interp-of-two-defs}} \label{appendix:piecewise-linearization-lemmas}
	
	This appendix provides the proofs of lemmas \ref{lemma:interp-of-lip-fn} and \ref{lemma:interp-of-two-defs}.
	
	\begin{proof}[Proof of \cref{lemma:interp-of-lip-fn}]
		The lemma is stated for any cell $\epsilon U + \alpha$ of the scaled lattice, however to simplify the
		notation we shall (without loss of generality) take $\alpha = 0$. Our goal is to prove  
		\eqref{eqn:interp-of-lip-fn-sup-norm} -- \eqref{eqn:interp-of-lip-fn-convergence},
		which we repeat (with $\alpha = 0 $) for the reader's convenience:
		$$
		| u^\epsilon |_{L^\infty(\epsilon U_n)} \leq |\phi |_{L^\infty(\epsilon U_m)} \, , \ \
		| \nabla u^\epsilon |_{L^\infty(\epsilon U_n)} \leq C |\nabla \phi |_{L^\infty(\epsilon U_m)} \, ,  \ \ \mbox{and} \ \
		| u^\epsilon - \phi |_{L^\infty(\epsilon U_n )} \leq C' \epsilon |\nabla \phi |_{L^\infty(\epsilon U_m)} \, ,
		$$
		where $u^\epsilon$ refers to the piecewise linearization of the deformation that equals $\phi$ at each
		node of the scaled lattice. We recall that our piecewise
		linearization scheme uses the scaled version of a triangulation of $U$ that was fixed as part of our
		framework. The nodes of the lattice must be vertices of the triangulation, but the triangulation can
		also have other vertices (which we call ``ghost vertices'' in \cref{sec:setup}). If there are ghost vertices, the
		rules for determining $u^\epsilon$ there must also be fixed as part of the framework; moreover they must
		have the form \eqref{eqn:piecewise-linearization-rule-a}--\eqref{eqn:piecewise-linearization-rule-b}. The
		definition \eqref{eqn:max-dependence-of-grad-u} of $U_m$ assures that the rules for determining the value of
		$u^\epsilon$ at ghost vertices in $\epsilon U_n$ use only its values at nodes of the scaled lattice that lie in
		$\epsilon \overline{U}_m$.
		
		We start by observing that the estimate on $|u^\epsilon - \phi|$ follows easily from the one on $|\nabla u^\epsilon|$,
		since $u^\epsilon - \phi$ vanishes at each node of the scaled lattice. Thus, it suffices to prove the other two estimates.
		
		For simplicity, we shall discuss the 2D case; it will be clear, however, that the same ideas can be
		used in any space dimension. The estimates are scale-invariant, so it would be sufficient to present
		the proof only for $\epsilon = 1$; however we shall keep the scale factor $\epsilon$, since setting $\epsilon = 1$
		doesn't really simplify matters. Since our triangulation of $U_n$ uses finitely many triangles, it is
		sufficient to prove
		\begin{equation} \label{eqn:interp-of-lip-fn-sufft}
			| u^\epsilon |_{L^\infty(\epsilon T)} \leq
			|\phi |_{L^\infty(\epsilon U_m)} \quad \mbox{and} \quad
			| \nabla u^\epsilon |_{L^\infty(\epsilon T)} \leq C_T |\nabla \phi |_{L^\infty(\epsilon U_m)}
		\end{equation}
		for each of the triangles $T$ in the triangulation of $U_n$. (The constant $C_T$ must of course
		be independent of $\epsilon$.)
		
		The first estimate is easier, so we start with it. As a warm-up, consider the case when all the vertices
		$x^\epsilon_i$ of $\epsilon T$ are nodes of the scaled lattice. Since $\epsilon T$ is convex, any point
		$x^\epsilon \in \epsilon T$ is a convex combination of the vertices:
		$x^\epsilon = \sum_i \rho_i x^\epsilon_i$ with $0 \leq \rho_i \leq 1$ and $\sum_i \rho_i = 1$.
		Since $u^\epsilon$ is affine on $\epsilon T$, we have
		\begin{align*}
			u^\epsilon (x^\epsilon ) & = \sum_i \rho_i u^\epsilon(x^\epsilon_i)\\
			& = \sum_i \rho_i \phi(x^\epsilon_i) \, .
		\end{align*}
		Since each vertex belongs to $\epsilon U_m$, we conclude the desired estimate
		$|u^\epsilon(x^\epsilon)| \leq |\phi |_{L^\infty (\epsilon U_m ) }$. For the general case, we must allow some
		or all of the vertices $x^\epsilon_i$ to be ghost nodes. If, say, $x^\epsilon_1$ is a ghost node, then (by definition)
		\begin{equation} \label{eqn:rule-for-x1-a}
			x^\epsilon_1 = \sum_j \theta_j y_j^\epsilon
		\end{equation}
		where each $y_j^\epsilon$ is a node of the scaled lattice in $\epsilon U_m$,
		and the associated evaluation rule is
		\begin{equation} \label{eqn:rule-for-x1-b}
			u^\epsilon (x_1^\epsilon) = \sum_j \theta_j u^\epsilon (y^\epsilon_j) \, .
		\end{equation}
		Since $u^\epsilon = \phi$ at nodes of the scaled lattice, it follows that
		$$
		|u^\epsilon (x_1)| \leq \sum_j \theta_j |\phi (y^\epsilon_j)| \leq |\phi |_{L^\infty(\epsilon U_m)} \, .
		$$
		Thus: we have $|u^\epsilon(x_i^\epsilon)| \leq |\phi |_{L^\infty(\epsilon U_m)} $
		at \emph{each} vertex of $\epsilon T$, whether or not it is a ghost vertex.
		The desired inequality now follows, by the argument we used to warm up.
		
		We turn now to the estimate on $|\nabla u^\epsilon |_{L^\infty(\epsilon T)}$. Let us start once again
		with the case when all the vertices $\{x_i^\epsilon\}$ of $\epsilon T$ are nodes of the scaled lattice.
		Since $u^\epsilon$ is affine on $\epsilon T$, its gradient on $\epsilon T$ is characterized by
		\begin{equation} \label{eqn:simplest-case-a}
			\nabla u^\epsilon =
			\begin{pmatrix} u^\epsilon (x^\epsilon_1)-u^\epsilon(x^\epsilon_2) & u^\epsilon (x^\epsilon_1)-u^\epsilon(x^\epsilon_3) \end{pmatrix}
			\begin{pmatrix}	x^\epsilon_1-x^\epsilon_2 & x^\epsilon_1-x^\epsilon_3 \end{pmatrix}^{-1} \, .
		\end{equation}
		Since the angles of $T$ are bounded away from $0$ there is a bounded, nonsingular matrix $M_T$ such that
		\begin{equation} \label{eqn:simplest-case-b}
			M_T \begin{pmatrix}	x_1-x_2 & x_1-x_3 \end{pmatrix} =
			\begin{pmatrix} |x_1-x_2| & 0 \\ 0 & |x_1-x_3| \end{pmatrix} \, ,
		\end{equation}
		and combining these equations gives
		\begin{equation} \label{eqn:simplest-case-c}
			\nabla u^\epsilon =
			\begin{pmatrix}
				\frac{u^\epsilon(x^\epsilon_1) - u^\epsilon(x^\epsilon_2)}{|x^\epsilon_1-x^\epsilon_2|} &
				\frac{u^\epsilon(x^\epsilon_1) - u^\epsilon(x^\epsilon_3)}{|x^\epsilon_1-x^\epsilon_3|}
			\end{pmatrix}
			M_T \, .
		\end{equation}
		Since
		\begin{equation} \label{eqn:simplest-case-d}
			\frac{|u^\epsilon(x^\epsilon_i) - u^\epsilon(x^\epsilon_j)|}{|x^\epsilon_i-x^\epsilon_j|} =
			\frac{|\phi(x^\epsilon_i) - \phi(x^\epsilon_j)|}{|x^\epsilon_i-x^\epsilon_j|} \leq
			|\nabla \phi|_{L^\infty(\epsilon T)}
		\end{equation}
		we conclude that the (constant) value of $\nabla u^\epsilon$ on $\epsilon T$ satisfies
		$$
		|\nabla u^\epsilon| \leq C_T |\nabla \phi|_{L^\infty(\epsilon T)}
		$$
		with a constant $C_T$ that depends only on the matrix $M_T$.
		This is, of course, stronger than \eqref{eqn:interp-of-lip-fn-sufft} since the $L^\infty$ norm on the right
		is restricted to $\epsilon T$.
		
		Now suppose two vertices (say, $x_1^\epsilon$ and $x_2^\epsilon$) are nodes of the scaled lattice but the
		third ($x_3^\epsilon$) is a ghost vertex. We recall that $u^\epsilon (x^\epsilon_3)$ is then
		determined by a specific representation of $x_3^\epsilon$ as a convex combination of scaled lattice nodes
		in $\overline{U}_m$
		\begin{equation} \label{eqn:rule-for-x3-a}
			x^\epsilon_3 = \sum_j \theta_j y_j^\epsilon \, ,
		\end{equation}
		and the associated evaluation rule is
		\begin{equation} \label{eqn:rule-for-x3-b}
			u^\epsilon (x^\epsilon_3) = \sum_j \theta_j u^\epsilon (y^\epsilon_j) \, .
		\end{equation}
		Equations \eqref{eqn:simplest-case-a}--\eqref{eqn:simplest-case-c} are still applicable, and the estimate
		\eqref{eqn:simplest-case-d} is still available for the difference quotient involving $x^\epsilon_1$ and
		$x^\epsilon_2$, however we must proceed differently for the one that involves $x^\epsilon_3$. We have
		\begin{align} \label{eqn:difference-of-ueps}
			|u^\epsilon (x^\epsilon_1) - u^\epsilon (x^\epsilon_3)| &=
			\big| \sum_j \theta_j (\phi(x^\epsilon_1) - \phi(y^\epsilon_j)) \big| \nonumber\\
			& \leq  |\nabla \phi|_{L^\infty(\epsilon U_m)} \sum_j \theta_j |x^\epsilon_1 - y^\epsilon_j| \nonumber \\
			& \leq \epsilon d_m |\nabla \phi|_{L^\infty(\epsilon U_m)}
		\end{align}
		since $x^\epsilon_1$ and all the $y^\epsilon_j$'s are in $\epsilon \overline{U}_m$. (We have used that since $\phi$ is
		assumed to be Lipschitz continuous, its restriction to $\epsilon \overline{U}_m$ is Lipschitz with constant
		$|\nabla \phi|_{L^\infty(\epsilon U_m)}$.) Now let
		\begin{equation} \label{eqn:ell-min-subT}
			\ell^{\rm min}_T = \mbox{minimum length of the sides of $T$} \, ,
		\end{equation}
		so that
		$$
		|x^\epsilon_1 - x^\epsilon_3| \geq \epsilon \ell^{\rm min}_T \, .
		$$
		Then combining the preceding inequalities gives
		$$
		\frac{|u^\epsilon (x^\epsilon_1) - u^\epsilon (x^\epsilon_3)|}{|x^\epsilon_1 - x^\epsilon_3|}
		\leq \frac{d_m}{\ell_T^{\rm min}} |\nabla \phi|_{L^\infty(\epsilon U_m)} \, .
		$$
		Proceeding as before, we obtain an inequality of the form \eqref{eqn:interp-of-lip-fn-sufft} with a constant $C_T$
		that depends on $T$ only through $M_T$ and $\ell_T^{\rm min}$.
		
		When the triangle has more than one ghost vertex most of the previous calculation remains intact, however
		we need an estimate for the difference of $u^\epsilon$ between two ghost vertices (as a substitute for \eqref{eqn:difference-of-ueps}). Suppose, for example, that $x^\epsilon_1$ and $x^\epsilon_3$ are both
		ghost vertices. Keeping our prior notation \eqref{eqn:rule-for-x3-a}--\eqref{eqn:rule-for-x3-b} for the rule determining
		$u^\epsilon(x_3^\epsilon)$, we suppose the corresponding rule determining $u^\epsilon (x_1^\epsilon)$ comes from
		the expression for $x_1^\epsilon$ as a convex combination of scaled lattice nodes in $\epsilon \overline{U}_m$:
		$$
		x^\epsilon_1 = \sum_k  \theta'_k  z_k^\epsilon \, .
		$$
		To take advantage of our previous calculation, we choose a nearby scaled lattice node
		$x_4^\epsilon \in \epsilon \overline{U}_m$ and use it ``as a bridge'' between $x_1^\epsilon$
		and $x_3^\epsilon$:
		\begin{align*}
			u^\epsilon(x^\epsilon_1) - u^\epsilon (x^\epsilon_3) & =
			\big( u^\epsilon(x^\epsilon_1) - u^\epsilon(x^\epsilon_4) \big) +
			\big( u^\epsilon(x^\epsilon_4) - u^\epsilon (x^\epsilon_3) \big) \\
			& = \sum_k \theta'_k \big( \phi(z^\epsilon_k) - \phi(x^\epsilon_4) \big)
			+ \sum_j \theta_j \big( \phi(x^\epsilon_4) - \phi(y^\epsilon_j) \big) \, .
		\end{align*}
		Each term on the right has the form we considered in \eqref{eqn:difference-of-ueps}. Therefore proceeding as
		before leads to
		$$
		\frac{|u^\epsilon (x^\epsilon_1) - u^\epsilon (x^\epsilon_3)|}{|x^\epsilon_1 - x^\epsilon_3|}
		\leq \frac{2d_m}{\ell_T^{\rm min}} |\nabla \phi|_{L^\infty(\epsilon U_m)} \, ,
		$$
		from which we once again deduce an inequality of the form \eqref{eqn:interp-of-lip-fn-sufft}.
		
		The case when $x^\epsilon_1$, $x^\epsilon_2$, and $x^\epsilon_3$ are all ghost vertices is essentially the same.
		(It is natural to use the same ``bridge'' $x^\epsilon_4$ in the estimation of both
		$u^\epsilon(x^\epsilon_1) - u^\epsilon (x^\epsilon_3)$ and $u^\epsilon(x^\epsilon_1) - u^\epsilon (x^\epsilon_2)$,
		however our argument does not require this.)
	\end{proof}
	\medskip
	
	\begin{proof}[Proof of \cref{lemma:interp-of-two-defs}]
		We once again take $\alpha = 0$ without loss of generality, and we focus again on the 2D
		setting though it will be clear that the same ideas can be used in any space dimension. Focusing first on
		\eqref{eqn:interpolate-scale-eps}, we shall show that for any triangle $T$ in our triangulation of $U_n$,
		\begin{equation} \label{eqn:interp-of-two-defs-sufft}
			|\nabla h^\epsilon|^2_{L^2(\epsilon T)} \leq C_T \Big(
			|u^\epsilon|^2_{L^2(\epsilon U_m)} |\nabla \varphi|^2_{L^\infty(\epsilon U_m)} +
			|\nabla u^\epsilon|^2_{L^2(\epsilon T)} |\varphi|^2_{L^\infty(\epsilon T)}
			\Big) \, .
		\end{equation}
		Here $u^\epsilon$ is the piecewise linearization of a deformation defined at all lattice nodes in
		$\epsilon \overline{U}_m$, $\phi$ is a Lipschitz continuous function, and $h^\epsilon$ is the piecewise linearization of
		a deformation that equals $u^\epsilon \phi $ at each lattice node in $\epsilon \overline{U}_m$.
		The constant $C_T$ in \eqref{eqn:interp-of-two-defs-sufft} will depend on the shape of $T$ and
		the details of our piecewise linearization scheme, but it will be independent of $\epsilon$,
		$u^\epsilon$, and $\phi$. The estimate \eqref{eqn:interpolate-scale-eps} follows immediately
		by summing \eqref{eqn:interp-of-two-defs-sufft} over the (finitely many) triangles $T$ in the triangulation of $U_n$.
		
		We start with some preliminary observations. The first is that if the triangle $\epsilon T$ has vertices
		$\{x^\epsilon_i\}_{i=1}^3$ and $u^\epsilon$ is an affine function on $\epsilon T$ then $\nabla u^\epsilon$
		(which is constant) satisfies
		\begin{equation} \label{eqn:interp-of-two-defs-first-observation}
			|\nabla u^\epsilon|^2 \sim \sum_{i \neq j}
			\frac{|u^\epsilon (x^\epsilon_i) - u^\epsilon (x^\epsilon_j)|^2}{|x^\epsilon_i - x^\epsilon_j|^2}
		\end{equation}
		in the sense that each side is less than or equal to an $\epsilon$-independent constant times the other.
		This is an immediate consequence of \eqref{eqn:simplest-case-c}.
		
		Our second observation is that
		\begin{equation} \label{eqn:interp-of-two-defs-second-observation}
			\frac{1}{|\epsilon T|} \int_{\epsilon T} |u^\epsilon|^2 \, dx \sim \sum_i |u^\epsilon(x_i^\epsilon)|^2 \, .
		\end{equation}
		Since this estimate is scale invariant, it suffices to prove it when $\epsilon = 1$. Writing
		$x_i$ for the vertices and $u$ for the affine deformation, we may represent $u$ using introduce
		barycentric coordinates. This means writing
		$$
		u(x) = \sum_i u(x_i) \lambda_i(x)
		$$
		where $\lambda_i$ is the affine function on $T$ with value
		$1$ at $x_i$ and $0$ at the other vertices. Evidently
		\begin{equation} \label{eqn:barycentric-quadratic-form-a}
			\int_T |u(x)|^2 \, dx = \sum_{i,j} u(x_i) \cdot u(x_j) \int_T \lambda_i (x) \lambda_j (x) \, dx \, .
		\end{equation}
		The right side is a quadratic form in $\{ u(x_i) \}$. We see from
		\eqref{eqn:barycentric-quadratic-form-a} that it is positive definite, so
		\begin{equation} \label{eqn:barycentric-quadratic-form-b}
			\sum_{i,j} u(x_i) \cdot u(x_j) \int_T \lambda_i (x) \lambda_j (x) \, dx \sim \sum_i |u(x_i)|^2 \, .
		\end{equation}
		Combining \eqref{eqn:barycentric-quadratic-form-a} and \eqref{eqn:barycentric-quadratic-form-b}, we
		get the $\epsilon = 1$ version of \eqref{eqn:interp-of-two-defs-second-observation}.
		
		Turning now to our main task, we begin by considering the case when all three vertices
		$x_i^\epsilon$ of $\epsilon T$ are nodes of the scaled lattice. Then by
		\eqref{eqn:interp-of-two-defs-first-observation}
		$$
		|\nabla h^\epsilon|^2 \sim \sum_{i \neq j}
		\frac{|u^\epsilon (x^\epsilon_i) \phi (x^\epsilon_i) - u^\epsilon (x^\epsilon_j) \phi(x^\epsilon_j)|^2}
		{|x^\epsilon_i - x^\epsilon_j|^2} \, .
		$$
		For each pair $i \neq j$ we have
		\begin{align*}
			|u^\epsilon (x^\epsilon_i) \phi (x^\epsilon_i) - u^\epsilon (x^\epsilon_j) \phi(x^\epsilon_j)| &\leq
			|u^\epsilon(x^\epsilon_i)| \, |\phi (x^\epsilon_i) - \phi (x^\epsilon_j)| +
			|u^\epsilon(x^\epsilon_i) - u^\epsilon(x^\epsilon_j)| \, |\phi(x^\epsilon_j)| \\
			&\leq |u^\epsilon (x^\epsilon_i)| \, |\nabla \phi|_{L^\infty(\epsilon T)} | \, |x^\epsilon_i - x^\epsilon_j| +
			|u^\epsilon(x^\epsilon_i) - u^\epsilon(x^\epsilon_j)| \, |\phi|_{L^\infty(\epsilon T)} \, .
		\end{align*}
		Since $u^\epsilon$ is affine on $\epsilon T$ our first observation applies to it, so
		$$
		\frac{|u^\epsilon (x^\epsilon_i) \phi (x^\epsilon_i) - u^\epsilon (x^\epsilon_j) \phi(x^\epsilon_j)|}
		{|x^\epsilon_i - x^\epsilon_j|} \leq
		C_T \Big( |u^\epsilon (x^\epsilon_i)| \, |\nabla \phi|_{L^\infty(\epsilon T)} +
		|\nabla u^\epsilon| \, |\phi|_{L^\infty(\epsilon T)} \Big) \, ,
		$$
		where on the right hand side $|\nabla u^\epsilon|$ is the norm of the constant matrix $\nabla u^\epsilon \big|_T$ .
		Squaring both sides and applying our second observation to $u^\epsilon$, we conclude that
		\begin{equation} \label{eqn:grad-h-with-no-ghost-vertex}
			\frac{|u^\epsilon (x^\epsilon_i) \phi (x^\epsilon_i) - u^\epsilon (x^\epsilon_j) \phi(x^\epsilon_j)|^2}
			{|x^\epsilon_i - x^\epsilon_j|^2} \leq
			C_T \Big( \frac{1}{|\epsilon T|} |u|_{L^2(\epsilon T)}^2 |\nabla \phi|_{L^\infty(\epsilon T)}^2 +
			|\nabla u^\epsilon|^2 |\phi|_{L^\infty(\epsilon T)}^2 \Big) \, .
		\end{equation}
		(Here and below, we permit the constant $C_T$ to change from line to line, however it always represents an $\epsilon$-independent constant depending only on the triangle $T$ and our piecewise linearization scheme.)
		We integrate over $\epsilon T$ and sum over $i \neq j$ to get
		$$
		|\nabla h^\epsilon|^2_{L^2(\epsilon T)} \leq C_T \Big(
		|u^\epsilon|^2_{L^2(\epsilon T)} |\nabla \varphi|^2_{L^\infty(\epsilon T)} +
		|\nabla u^\epsilon|^2_{L^2(\epsilon T)} |\varphi|^2_{L^\infty(\epsilon T)}
		\Big) \, .
		$$
		This is stronger than \eqref{eqn:interp-of-two-defs-sufft}, since on the right the $L^\infty$ norms
		are only over $\epsilon T$.
		
		It remains to consider the case when some or all the vertices of $\epsilon T$ are ghost vertices. We still have
		$$
		|\nabla h^\epsilon|^2 \sim \sum_{i \neq j}
		\frac{|h^\epsilon(x^\epsilon_i) - h^\epsilon (x^\epsilon_j)|^2}
		{|x^\epsilon_i - x^\epsilon_j|^2}
		$$
		but the evaluation of $h^\epsilon(x^\epsilon_i) - h^\epsilon (x^\epsilon_j)$ must proceed differently when one or both
		of $x^\epsilon_i , x^\epsilon_j$ are ghost nodes. To simplify the notation let us take $i=1$ and $j=3$, and to see the
		idea in its simplest form let us suppose for now that $x^\epsilon_3$ is a ghost node but $x^\epsilon_1$ is not.
		Recall that the rule determining $h^\epsilon(x_3)$ is then of the form \eqref{eqn:rule-for-x3-a}--\eqref{eqn:rule-for-x3-b},
		so
		\begin{equation} \label{eqn:one-ghost-vertex-a}
			h^\epsilon (x^\epsilon_1) - h^\epsilon (x^\epsilon_3) = u^\epsilon (x^\epsilon_1) \phi (x^\epsilon_1) -
			\sum_j \theta_j u^\epsilon (y^\epsilon_j) \phi(y^\epsilon_j)
		\end{equation}
		where $\{ y^\epsilon_j \}$ are certain nodes of the scaled lattice that lie in $\overline{U}_m$. Since $u^\epsilon$ is
		itself the piecewise linearization of a deformation defined at scaled lattice nodes, we also have
		$$
		u^\epsilon (x^\epsilon_3) = \sum_j \theta_j u^\epsilon (y^\epsilon_j) \, .
		$$
		Combining these relations gives
		\begin{equation} \label{eqn:one-ghost-vertex-b}
			h^\epsilon (x^\epsilon_1) - h^\epsilon (x^\epsilon_3) =
			\big[ u^\epsilon (x^\epsilon_1) \phi (x^\epsilon_1) - u^\epsilon (x^\epsilon_3) \phi (x^\epsilon_3) \big] +
			\sum_j \theta_j u^\epsilon (y^\epsilon_j) \big( \phi(x^\epsilon_3) - \phi(y^\epsilon_j) \big) \, .
		\end{equation}
		The term in square brackets is of the form we considered when there were no ghost nodes, so we have already estimated it.
		As for the other term: we have
		$$
		\Big| \sum_j \theta_j u^\epsilon (y^\epsilon_j) \big( \phi(x^\epsilon_3) - \phi(y^\epsilon_j) \big) \Big|
		\leq \epsilon d_m |\nabla \phi|_{L^\infty(\epsilon U_m)} \max_j |u^\epsilon (y_j^\epsilon)| \, ,
		$$
		whence (using our second observation and also \eqref{eqn:ell-min-subT})
		$$
		\frac{1}{|x^\epsilon_1 - x^\epsilon_3|^2}
		\left( \sum_j \theta_j u^\epsilon (x^\epsilon_3) \big( \phi(x^\epsilon_3) - \phi(y^\epsilon_j) \big) \right)^2 \leq
		C_T |\nabla \phi|^2_{L^\infty(\epsilon U_m)}
		\sum_{\substack{T' \in \, {\rm triangulation}\\{\rm of} \, U_m}}
		\frac{1}{|\epsilon T'|} \int_{\epsilon T'} |u^\epsilon(x)|^2 \, dx \, .
		$$
		Combining this with \eqref{eqn:grad-h-with-no-ghost-vertex} and \eqref{eqn:one-ghost-vertex-b} gives
		\begin{equation} \label{eqn:one-ghost-vertex-c}
			\frac{|h^\epsilon(x^\epsilon_1) - h^\epsilon (x^\epsilon_3)|^2}
			{|x^\epsilon_1 - x^\epsilon_3|^2} \leq
			C_T \Big( \max_{\substack{T' \in \, {\rm triangulation}\\{\rm of \, U_m}}}\frac{1}{|\epsilon T'|}
			|u^\epsilon|_{L^2(\epsilon T')}^2 |\nabla \phi|_{L^\infty(\epsilon U_m)}^2 +
			|\nabla u^\epsilon|^2 |\phi|_{L^\infty(\epsilon T)}^2 \Big)
		\end{equation}
		where on the right hand side $|\nabla u^\epsilon|$ is the norm of the constant matrix $\nabla u^\epsilon |_T$.
		
		The same estimate \eqref{eqn:one-ghost-vertex-c} is also valid when both $x_1$ and $x_3$ are ghost nodes. Indeed,
		if the rule defining $h^\epsilon(x^\epsilon_1)$ is associated with \eqref{eqn:rule-for-x1-a} then \eqref{eqn:one-ghost-vertex-a}
		is replaced by
		$$
		h^\epsilon (x^\epsilon_1) - h^\epsilon (x^\epsilon_3) = \sum_k \theta_k' u^\epsilon (z^\epsilon_k) \phi (z^\epsilon_k) -
		\sum_j \theta_j u^\epsilon (y^\epsilon_j) \phi(y^\epsilon_j) \, .
		$$
		Since we also have $u^\epsilon (x^\epsilon_1) = \sum_k \theta'_k u^\epsilon (z^\epsilon_k)$, this can be rewritten as
		\begin{align*}
			h^\epsilon (x^\epsilon_1) - h^\epsilon (x^\epsilon_3) = &
			\big[ u^\epsilon (x^\epsilon_1) \phi (x^\epsilon_1) - u^\epsilon (x^\epsilon_3) \phi (x^\epsilon_3) \big] + \\
			& \sum_j \theta_j u^\epsilon (y^\epsilon_j) \big( \phi(x^\epsilon_3) - \phi(y^\epsilon_j) \big) +
			\sum_k \theta'_k u^\epsilon (z_k^\epsilon) \big( \phi(z^\epsilon_k) - \phi(x^\epsilon_1) \big) \, .
		\end{align*}
		Each term is of a type we have already discussed, so arguing as before we obtain once again an estimate of the form
		\eqref{eqn:one-ghost-vertex-c}.
		
		Finally, our assertion \eqref{eqn:interp-of-two-defs-sufft} follows easily from these results. Indeed, we have shown that
		the right hand side of \eqref{eqn:one-ghost-vertex-c} estimates
		$|h^\epsilon(x^\epsilon_i) - h^\epsilon (x^\epsilon_j)|^2/|x^\epsilon_i - x^\epsilon_j|^2$ for every pair of vertices of $\epsilon T$.
		Integrating this estimate over $\epsilon T$ and using our first observation leads immediately to \eqref{eqn:interp-of-two-defs-sufft},
		since
		$$
		\max_{\substack{T' \in \, {\rm triangulation}\\{\rm of} \, U_m}}\frac{|\epsilon T|}{|\epsilon T'|}
		$$
		is an $\epsilon$-independent constant that depends only on the triangle $T$ and our piecewise linearization scheme.
		The proof of \eqref{eqn:interpolate-scale-eps} is now complete.
		
		We turn now to the Lemma's second assertion, \eqref{eqn:interpolate-scale-eps-L2}. It clearly suffices to show
		that for each triangle $T$ in our triangulation of $U_n$ we have
		\begin{equation} \label{eqn:interpolate-scale-eps-L2-sufft}
			|h^\epsilon|^2_{L^2(\epsilon T)} \leq
			C |u^\epsilon|^2_{L^2(\epsilon U_m)} |\varphi|^2_{L^\infty(\epsilon U_m)} \, .
		\end{equation}
		By \eqref{eqn:interp-of-two-defs-second-observation} we have
		\begin{equation} \label{eqn:interp-L2-getting-started}
			|h^\epsilon|^2_{L^2(\epsilon T)} \leq C |\epsilon T| \sum_i |h^\epsilon (x_i^\epsilon)|^2
		\end{equation}
		where $x_i^\epsilon$ are the vertices of $\epsilon T$. If all the $x_i^\epsilon$ are nodes of the lattice then things
		are very simple: $|h^\epsilon(x_i^\epsilon)| \leq |u^\epsilon(x_i^\epsilon)| |\phi|_{L^\infty(\epsilon T)} $, so another
		application of \eqref{eqn:interp-of-two-defs-second-observation} gives
		$$
		|h^\epsilon|^2_{L^2(\epsilon T)} \leq C |u^\epsilon|^2_{L^2(\epsilon T)} |\varphi|^2_{L^\infty(\epsilon T)} \, ,
		$$
		which is better than \eqref{eqn:interpolate-scale-eps-L2-sufft}. If, however, some or all the $x_i^\epsilon$
		are ghost vertices, then we must work a little harder. Suppose, for example, that $x_1^\epsilon$ is a ghost vertex, and that
		the piecewise linearization rule uses \eqref{eqn:rule-for-x1-a}--\eqref{eqn:rule-for-x1-b}. Then
		\begin{align*}
			|h^\epsilon(x_1^\epsilon)| &= \big| \sum_j \theta_j \varphi(y_j^\epsilon) u^\epsilon(y_j^\epsilon) \big| \\
			& \leq \big( \sum_j \theta_j^2 \big)^{1/2} \, |\varphi|_{L^\infty (\epsilon U_m) } \,
			\big( \sum_j |u^\epsilon (y_j^\epsilon)|^2 \big)^{1/2}  \\
			& \leq |\varphi|_{L^\infty (\epsilon U_m) } \, \big( \sum_j |u^\epsilon (y_j^\epsilon)|^2 \big)^{1/2}
		\end{align*}
		using that $\sum_j \theta_j^2 \leq \sum_j \theta_j = 1$ since $0 \leq \theta_j \leq 1$. Squaring this,
		using that the vertices of our triangulation include all lattice nodes, and using
		\eqref{eqn:interp-of-two-defs-second-observation} once again, we have
		$$
		|h^\epsilon(x_1^\epsilon)|^2 \leq C |\varphi|^2_{L^\infty (\epsilon U_m) }
		\sum_{\substack{T' \in \, {\rm triangulation}\\{\rm of} \, U_m}}
		\frac{1}{|\epsilon T'|} \int_{\epsilon T'} |u^\epsilon(x)|^2 \, dx \, .
		$$
		Treating each ghost vertex of $\epsilon T$ this way and using \eqref{eqn:interp-L2-getting-started} we conclude that
		$$
		|h^\epsilon|^2_{L^2(\epsilon T)} \leq C |\varphi|^2_{L^\infty (\epsilon U_m) }
		\int_{\epsilon U_m} |u^\epsilon|^2 \, dx
		\left( \max_{\substack{T' \in \, {\rm triangulation}\\{\rm of} \, U_m}} \frac{|\epsilon T|}{|\epsilon T'|} \right) \, .
		$$
		Since the max that appears on the far right is a constant independent of $\epsilon$, this establishes
		\eqref{eqn:interpolate-scale-eps-L2-sufft}.
	\end{proof}
	
	\renewcommand{\theequation}{B.\arabic{equation}}
	\setcounter{equation}{0}
	\subsection{Details of Step 3 in the proof of Lemma \ref{lemma:affine}}\label{appendix:degiorgi}
	We recall that Step 2 in the proof of \cref{lemma:affine} part (b) established the lower bound \eqref{eqn:liminf} when
	$u^\epsilon - \lambda x \in \mathcal{A}^\epsilon_0 (\Omega)$. Our task here is to prove that the same lower bound holds
	without imposing such an ``affine boundary condition,'' provided that $u^\epsilon$ converges weakly to $u(x) = \lambda x$
	in $H^1(\Omega)$. As already noted in \cref{subsec:affine-limits}, the proof uses a well-known argument due to
	De Giorgi, whose main novelty in our setting is its need for \cref{lemma:interp-of-two-defs}.
	
\begin{figure}[!htb]
		\centering
		\begin{tikzpicture}[scale=0.75]
			\def\amp{1} 
			\def\freq{2} 
			\draw[domain=0:360,thick,smooth,variable=\t] plot ({(6+\amp*sin(\freq*\t))*cos(\t)}, {(3+\amp*sin(\freq*\t))*sin(\t)});
			\node[black] at (2.3*6/4,2*6/4) {$\Omega$};
			
			\def\amp{1/6*5} 
			\def\freq{2} 
			\draw[domain=0:360,red,thick,smooth,variable=\t] plot ({(5+\amp*sin(\freq*\t))*cos(\t)}, {(2.5+\amp*sin(\freq*\t))*sin(\t)});
			\node[red] at (2.3*5/4,2*5/4) {$\Omega_{i}$};
			
			\def\amp{1/6*4} 
			\def\freq{2} 
			\draw[domain=0:360,blue,thick,smooth,variable=\t] plot ({(4+\amp*sin(\freq*\t))*cos(\t)}, {(2+\amp*sin(\freq*\t))*sin(\t)});
			\node[blue] at (2.3,2) {$\Omega_{i-1}$};
			
			\def\amp{1/6*3} 
			\def\freq{2} 
			\draw[domain=0:360,black,thick,smooth,variable=\t] plot ({(3+\amp*sin(\freq*\t))*cos(\t)}, {(1.5+\amp*sin(\freq*\t))*sin(\t)});
			\node[black] at (-2.8,-0.3) {$\Omega_0$};
			
			\def\amp{1/6*2.5} 
			\def\freq{2} 
			\draw[domain=0:360,black,dotted,thick,smooth,variable=\t] plot ({(2.5+\amp*sin(\freq*\t))*cos(\t)}, {(1.25+\amp*sin(\freq*\t))*sin(\t)});
			\node[black] at (2.3*3/4,-2*3/4+2) {$\Omega'_0$};
			
			\def\sidelength{0.85}
			
			\pgfmathsetmacro{\height}{sqrt((\sidelength)^2-(\sidelength/2)^2)}
			\draw[thick] (-4.5,-1) -- ++(0:\sidelength) -- ++(120:\height) -- ++(180:\sidelength) -- cycle;
			
			\coordinate (A) at (-4.5,-1);
			\coordinate (B) at ($(A)+(0:\sidelength)$);
			\coordinate (C) at ($(B)+(120:\height)$);
			\coordinate (D) at ($(A)+(120:\height)$);
			\coordinate (E) at ($(B)!0.5!(D)$);
			\coordinate (F) at ($(E)+(0,-0.75)$);
			
			\draw[dotted, thick] (B) -- (D);
			\node[below] at (F) {$\epsilon d_m$};
			
			\draw[-stealth, thick] (E) -- (F);
		\end{tikzpicture}
		\caption{An illustratration of the nested sets $\{\Omega_i\}, i=1,2,\dots,\nu-3$. The parallelogram (lower left)
			is a scaled cell $\epsilon U_{m} + \alpha$ with $\alpha \in R_\epsilon(S_i^\epsilon)$; the dotted diagonal line
			indicates the largest distance between two points in the $\epsilon U_{m} + \alpha$, which is $\epsilon d_m$.}
		\label{fig:covering}
	\end{figure}
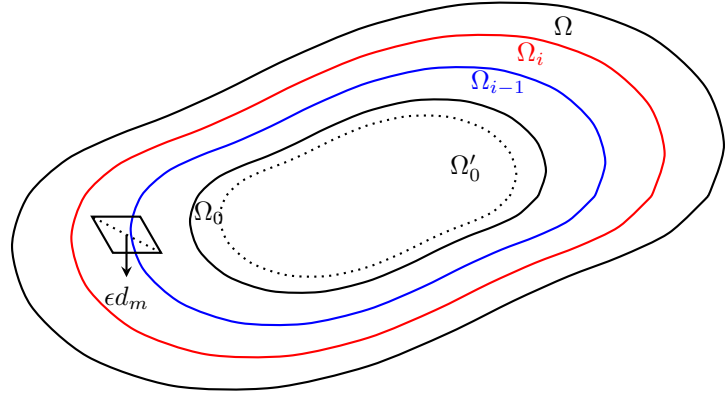

	We begin with some preliminaries. First, for arbitrarily small $\delta$, we choose two open subsets of
	$\Omega$ that are both compactly supported in $\Omega$ with the following two properties:
	\begin{equation} \label{eqn:shell-near-bdry}
		\Omega_0' \ssubset \Omega_0 \ssubset \Omega \qquad \text{and} \qquad |\Omega \setminus \Omega_0'| \leq \delta \, .
	\end{equation}
	Second, we introduce a nested family of sets $\Omega_i$ that contain $\Omega_0$, by taking (for any positive
	integer $\nu$, which will eventually tend to infinity)
	\begin{align*}
		\Omega_i &= \{x \in \Omega \: \big| \:\text{dist}(x,\Omega_0) < \frac{i}{\nu} R\} \qquad i = 1,2,\dots, \nu \, ,
		\quad \text{where} \quad R = \frac{1}{2}\text{dist}(\partial \Omega, \Omega_0) \, .
	\end{align*}
	We thus obtain a sequence of sets with
	$\Omega_0 \subset \Omega_1 \subset \Omega_2 \subset \dots \subset \Omega_\nu \subset \Omega$, as shown in
	Figure \ref{fig:covering}. Next, we choose corresponding cut-off functions $\varphi_i(x)$ such that $0 \leq \varphi \leq 1$ with
	\begin{align*}
		& \varphi_i(x) =
		\begin{cases}
			1 \qquad & x \in \Omega_{i-1}\\
			0 \qquad & x \in \Omega \setminus \Omega_{i}
		\end{cases}
		\qquad \text{and} \qquad \big|\nabla \varphi_i\big|_{L^\infty(\Omega)}  \leq \frac{2\nu}{R} \, .
	\end{align*}
	The upper bound on the $L^\infty$ norm of $\nabla \varphi$ is feasible, since for every pair $x \in \Omega_{i-1}$ and
	$ y \in \partial \Omega_i$, their distance is lower bounded by $R/\nu$.
	
	\noindent Finally, we introduce the deformations $w_i^\epsilon(x)$, defined at nodes of our $\epsilon$-scale lattice by
	\begin{align}
		w_i^\epsilon(x) &= \lambda x + \varphi_i(x)(u^\epsilon-\lambda x) = \varphi_i(x) u^\epsilon + \big(1-\varphi_i(x)\big) \lambda x \, . \label{eqn:w_i}
	\end{align}
	Notice that each $w_i^\epsilon(x)$ is ``affine at $\partial \Omega$'' (in the sense that
	$w^\epsilon_i - \lambda x \in \mathcal{A}_\epsilon^0 (\Omega)$), since
	\begin{align*}
		\quad w_i^\epsilon(x) &=
		\begin{cases}
			u^\epsilon \qquad & x \in \Omega_{i-1}\\
			\lambda x \qquad & x \in \Omega \setminus \Omega_{i} \, .
		\end{cases}
	\end{align*}
	By Step 2 of the proof of \cref{lemma:affine} part (b), we know that
	\begin{align}\label{eqn:step4-lower-1}
		\liminf_{\epsilon \rightarrow 0} E^\epsilon(w_i^\epsilon, \Omega) \geq \big|\Omega\big| \, \overline{W}(\lambda) \, .
	\end{align}
	
	Our plan is to show that if $i$ is chosen properly, $E^\epsilon (w^\epsilon_i, \Omega)$ is close to $E^\epsilon(u^\epsilon, \Omega)$.
	To prepare for this argument, we observe that the energy $E^\epsilon(w_i^\epsilon, \Omega)$ has (roughly speaking) three parts:
	the energy of $u^\epsilon$ inside $\Omega_{i-1}$, the energy of $\lambda x$ outside $\Omega_i$, and the energy
	associated with $w_i^\epsilon$ in the layer $\Omega_i \setminus \Omega_{i-1}$. To make this more precise, we introduce the sets
	\begin{align*}
		S_i^\epsilon = \{x \in \Omega \: \big| \: \text{dist}(x, \overline{\Omega_i}\setminus \Omega_{i-1}) < 2\epsilon d_m \},
		\qquad i = 1,2,\dots, \nu-3 \, ,
	\end{align*}
	where $d_m $ is defined in \eqref{eqn:d_m}. They are useful because
	\begin{align}
		E^\epsilon(w_i^\epsilon, \Omega) & \leq
		E^\epsilon(w_i^\epsilon, \Omega_{i-1}) + E^\epsilon(w_i^\epsilon, \Omega \setminus \overline{\Omega}_i) +
		E^\epsilon(w_i^\epsilon, S_i^\epsilon) \nonumber\\
		&= E^\epsilon(u^\epsilon, \Omega_{i-1}) + E^\epsilon(\lambda x, \Omega \setminus \overline{\Omega}_i) +
		E^\epsilon(w_i^\epsilon, S_i^\epsilon).\label{eqn:step-4-energy}
	\end{align}
	This upper bound holds since if a scaled cell $\epsilon U + \alpha$ has the property that $\epsilon U_m + \alpha$ is
	neither compactly included in $\Omega_{i-1}$ nor in $\Omega \setminus \overline{\Omega_i}$, then $\epsilon \overline{U}_m + \alpha$
	must intersect $\overline{\Omega_i} \setminus \Omega_{i-1}$. When this happens, $\alpha \in R_\epsilon(S_i^\epsilon)$
	(see Figure \ref{fig:covering} for a visualization of this). These sets have the following properties
	for $\epsilon$ sufficiently small:
	\begin{enumerate}[(i)]
		\item $\cup_{i=1}^{\nu -3} S_i^\epsilon \subset \Omega \setminus \Omega_0'$ \, ;
		\item $R_\epsilon(S^\epsilon_i) \cap R_\epsilon(S^\epsilon_j) \neq \emptyset$ if and only if $|i-j| = 1$ \, ;
		\item $\cup_{i=1}^{\nu -3} R_\epsilon(S_i^\epsilon) \subset R_\epsilon(\Omega \setminus \Omega_0')$ \, ;
		\item $S_i^\epsilon \cap S_j^\epsilon \neq \emptyset$ if and only if $|i-j| = 1$ \, ;
		\item $\sum_{i=1}^{\nu-3} |S_i^\epsilon| \leq 2 |\Omega \setminus \Omega_0'|$ \, .
	\end{enumerate}
	
	We are now ready to show the desired lower bound. The key idea is to show that the right hand side of
	\eqref{eqn:step-4-energy} is upper bounded by $E^\epsilon(u^\epsilon, \Omega)$ and some small terms.
	Combining this with \eqref{eqn:step4-lower-1} will then give the desired lower bound for the energy of $u^\epsilon$.
	So our task is to estimate the right hand side of \eqref{eqn:step-4-energy}. The first term is easy: we have
	\begin{equation} \label{eqn:first-term-of-w-estimate}
		E^\epsilon(u^\epsilon, \Omega_{i-1}) \leq E^\epsilon(u^\epsilon, \Omega) \, .
	\end{equation}
	The second term is also easy: by \cref{lemma:constant-gradient-energy-bound} we have
	\begin{equation} \label{eqn:second-term-of-w-estimate}
		E^\epsilon(\lambda x, \Omega \setminus \overline{\Omega_i}) \leq
		E^\epsilon(\lambda x, \Omega \setminus \overline{\Omega_0'}) \leq
		C_1  (2n-1)^N (|\lambda|^2 + 1) |\Omega \setminus \overline{\Omega_0'}| \leq  C_1 (2n-1)^N \delta (|\lambda|^2 + 1) \, .
	\end{equation}
	The third term in \eqref{eqn:step-4-energy} is, however, more difficult; in fact, if $i$ is held fixed then it cannot be
	adequately controlled. However, we will show the existence of a choice $i(\epsilon)$ for every $\epsilon$ such that
	$E^\epsilon(w^\epsilon_{i(\epsilon)}, S^\epsilon_{i(\epsilon)})$ is adequately controlled. The idea is relatively simple:
	we first estimate the average $\frac{1}{\nu-3} \sum_{i=1}^{\nu - 3} E^\epsilon(w_i^\epsilon, S_i^\epsilon)$, then
	take $i(\epsilon)$ corresponding to the smallest of the terms that were averaged.
	
	We start by estimating $E^\epsilon(w_i^\epsilon, S_i^\epsilon)$ using our basic upper bound \eqref{eqn:unit-cell-upper} on the
	energy of the unit cell:
	\begin{align*}
		E^\epsilon(w_i^\epsilon, S_i^\epsilon) & =
		\sum_{\alpha \in R_\epsilon(S_i^\epsilon)}  E^\epsilon(w_i^\epsilon, \epsilon U+\alpha) \nonumber\\
		& \leq C_1 \sum_{\alpha \in R_\epsilon(S_i^\epsilon)} \Big(\big|\epsilon U_n \big| +
		\big|\nabla w_i^\epsilon\big|^2_{L^2(\epsilon U_n+\alpha)}\Big)\, . 
	\end{align*}
	The right hand side of this bound refers, as usual, to the piecewise linearization of $w_i^\epsilon$. Remembering from
	\eqref{eqn:w_i} that $w_i^\epsilon (x) = \lambda x + \phi_i(x) (u^\epsilon - \lambda x)$ at the nodes of the scaled lattice,
	and recalling that the linear function $\lambda x$ is its own piecewise linearization, we apply \cref{lemma:interp-of-two-defs}
	to get
	$$
	|\nabla w_i^\epsilon|^2_{L^2(\epsilon U_n+\alpha)} \leq
	C \Big(|\lambda|^2 |\epsilon U_m| + |\nabla u^\epsilon - \lambda|^2_{L^2(\epsilon U_m + \alpha)} +
	\frac{4\nu^2}{R^2} |u^\epsilon - \lambda x|^2_{L^2(\epsilon U_m + \alpha)}\Big) \, ,
	$$
	where the constant $C \geq 1$ depends only on our piecewise linearization scheme. Combining the preceding estimates gives
	\begin{align*}
		E^\epsilon(w_i^\epsilon, S_i^\epsilon) & \leq C C_1 \sum_{\alpha \in R_\epsilon(S_i^\epsilon)} \Big((1+|\lambda|^2)\big|\epsilon U_m\big| + |\nabla u^\epsilon - \lambda|^2_{L^2(\epsilon U_m + \alpha)} +
		\frac{4\nu^2}{R^2} |u^\epsilon - \lambda x|^2_{L^2(\epsilon U_m + \alpha)} \Big)\\
		& \leq C C_1 (2m-1)^N  \Big((1+|\lambda|^2)\big|S_i^\epsilon\big| + |\nabla u^\epsilon - \lambda|^2_{L^2(S_i^\epsilon)} + \frac{4\nu^2}{R^2} |u^\epsilon - \lambda x|^2_{L^2(S_i^\epsilon)} \Big) \, .
	\end{align*}
	The second line holds since each $|\nabla u^\epsilon|^2_{L^2(\epsilon U+\alpha)}$ with $\alpha \in R_\epsilon(S_i^\epsilon)$ is
	appears in $|\nabla u^\epsilon|^2_{L^2(\epsilon U_{m}+\beta)}$ for some $\beta$ at most $(2m-1)^N$ times. We now use that for sufficiently small $\epsilon$ the sets $S_i^\epsilon$ and $S_j^\epsilon$ are disjoint unless $|i-j|=1$ (see bullet (iv)),
	and similarly $R_\epsilon(S_i^\epsilon)$ and $R_\epsilon(S_j^\epsilon)$ are disjoint unless $|i-j|=1$ (see bullet (ii)). This
	leads to the following estimate on the average of $E^\epsilon(w_i^\epsilon, S_i^\epsilon)$ from $i=1$ to $\nu-3$:
	\begin{multline}\label{eqn:est-average-w_i}
		\textstyle \frac{1}{\nu-3} \sum_{i=1}^{\nu-3} E^\epsilon(w_i^\epsilon, S_i^\epsilon) \leq \frac{C C_1 (2m-1)^N}{\nu-3} \sum\limits_{i=1}^{\nu-3} \Big((1+|\lambda|^2)|S_i^\epsilon| + |\nabla u^\epsilon - \lambda|^2_{L^2(S_i^\epsilon)} +
		\frac{4\nu^2}{R^2} |u^\epsilon - \lambda x|^2_{L^2(S_i^\epsilon)} \Big) \\
		\textstyle \leq \: \frac{2C C_1(2m-1)^N}{\nu-3} \Big((1+|\lambda|^2) \big|\Omega\setminus \Omega_0'\big| +
		|\nabla u^\epsilon - \lambda|^2_{L^2(\Omega\setminus \Omega_0')} +
		\frac{4\nu^2}{R^2} |u^\epsilon - \lambda x|^2_{L^2(\Omega\setminus \Omega_0')}\Big) \, .
	\end{multline}
	There is a factor of $2$ in \eqref{eqn:est-average-w_i} because we at most cover the whole area
	$\Omega\setminus\Omega_0'$ twice when summing over all $i=1,2,\dots, \nu-3$ (see bullet (v)). Finally, we choose
	\begin{align*}
		i(\epsilon) = \argmin  E^\epsilon(w_i^\epsilon, S_i^\epsilon)
	\end{align*}
	so that
	\begin{equation} \label{eqn:special-w_i}
		E^\epsilon(w_{i(\epsilon)}^\epsilon, S_i^\epsilon) \leq
		\frac{1}{\nu-3} \sum_{i=1}^{\nu-3} E^\epsilon(w_i^\epsilon, S_i^\epsilon) \, .
	\end{equation}
	Combining \eqref{eqn:step-4-energy}--\eqref{eqn:special-w_i}, we obtain
	\begin{multline} \label{eqn:appendix-b-almost-done}
		E^\epsilon(w^\epsilon_{i(\epsilon)}, \Omega) \leq \:  E^\epsilon(u^\epsilon,\Omega) + C_1 \delta (|\lambda|^2 + 1) \\
		+\frac{2C C_1(2m-1)^N}{\nu-3} \Big((1+|\lambda|^2) \big|\Omega\setminus \Omega_0'\big| +
		|\nabla u^\epsilon - \lambda|^2_{L^2(\Omega\setminus \Omega_0')} +
		\frac{4\nu^2}{R^2} |u^\epsilon - \lambda x|^2_{L^2(\Omega\setminus \Omega_0')}\Big) \, .
	\end{multline}
	We know from Step 2 of the proof of \cref{lemma:affine}  part (b) that
	$$
	|\Omega| \overline{W}(\lambda) \leq \liminf_{\epsilon \rightarrow 0} E^\epsilon ( w^\epsilon_{i(\epsilon)}, \Omega)
	$$
	since $w^\epsilon_{i(\epsilon)} - \lambda x \in \mathcal{A}_\epsilon^0$. Moreover, our hypothesis that
	$u^\epsilon \rightharpoonup \lambda x$ in $H^1(\Omega)$ assures that
	$|\nabla u^\epsilon - \lambda|^2_{L^2(\Omega\setminus \Omega_0')}$ remains uniformly bounded, and Rellich's lemma
	assures that $\lim_{\epsilon \rightarrow 0} |u^\epsilon - \lambda x|^2_{L^2(\Omega\setminus \Omega_0')} =0$. Therefore
	we can conclude from \eqref{eqn:appendix-b-almost-done} that
	$$
	|\Omega| \, \overline{W}(\lambda) \leq \liminf_{\epsilon \rightarrow 0} E^\epsilon ( u^\epsilon, \Omega)
	$$
	by taking the $\liminf$ in $\epsilon$, then the limit $\nu \rightarrow \infty$, then finally the limit $\delta \rightarrow 0$. The
	proof of \cref{lemma:affine} is now complete.
	\renewcommand{\theequation}{C.\arabic{equation}}
	\setcounter{equation}{0}
	\renewcommand{\theequation}{\thesubsection.\arabic{equation}}
	\subsection{Approximating \texorpdfstring{$u\in H^1(\Omega)$}{Lg} by piecewise affine functions}\label{appen:approximation}
	Near the beginning of \cref{subsec:main-proof} we asserted that for any Lipschitz domain $\Omega$ and any
	$u \in H^1(\Omega)$, there is a piecewise linear approximation $u_\delta$ of $u$ satisfying
	\eqref{eqn:udelta-approaches-utilde} and \eqref{eqn:uniform-bounds-for-udelta}, which we repeat for the reader's
	convenience:
	\begin{gather*}
		|\tilde{u} - u_\delta|_{H^1(\R^N)} \rightarrow 0
		\mbox{ as $\delta \rightarrow 0$ \, and} \\
		|u_\delta|_{L^\infty(\R^N)} +
		|\nabla u_\delta|_{L^\infty(\R^N)} \leq c_u \delta^{-a} \, .
	\end{gather*}
	Here $\tilde{u}$ is a compactly supported extension of $u$ satisfying
	$|\tilde{u}|_{H^1(\mathbb{R}^N)} \leq C |u|_{H^1(\Omega)}$, $c_u$ is a constant (depending on $|u|_{H^1(\Omega)}$),
	and $a$ is a positive constant depending only on the spatial dimension $N$. This appendix provides a detailed
	justification of that assertion.
	
	Let $u^\eta = \varphi^\eta * \tilde{u}$ be the smooth approximation to $\tilde{u}$ obtained by mollification with
	$\varphi^\eta = \frac{1}{\eta^N} \varphi\left(\frac{x}{\eta}\right)$, where $\varphi$ is smooth and compactly
	supported with integral $1$. It is standard that $u^\eta \rightarrow u$ in $H^1(\mathbb{R}^N)$, with
	\begin{equation} \label{eqn:u-eta-bound}
		|u^\eta - \tilde{u}|_{L^2(\mathbb{R}^N)} \leq  C \eta |\nabla \tilde{u}|_{L^2(\mathbb{R}^N)}
		\leq C \eta |u|_{H^1(\Omega)}
		\qquad \text{and} \qquad
		\lim_{\eta \rightarrow 0} |\nabla u^\eta - \nabla \tilde{u}|_{L^2(\mathbb{R}^N)} = 0 \, .
	\end{equation}
	Moreover, we can bound the $L^\infty$ norms of $u^\eta$, $\nabla u^\eta$ and $\nabla \nabla u^\eta$ by
	\begin{align}
		|u^\eta|_{L^\infty(\mathbb{R}^N)} & \leq \frac{M_1}{\eta^{\frac{N}{2}}} |\tilde{u}|_{L^2(\mathbb{R}^N)}
		\leq C \frac{M_1}{\eta^{\frac{N}{2}}} |u|_{H^1(\Omega)}
		\quad \text{with } M_1 = | \varphi|_{L^2(\mathbb{R}^N)} \, ; \label{eqn:u-eta} \\
		|\nabla u^\eta|_{L^{\infty}(\mathbb{R}^N)} & \leq \frac{M_1}{\eta^{\frac{N}{2}}}
		|\nabla \tilde{u}|_{L^2(\mathbb{R}^N)}
		\leq C \frac{M_1}{\eta^{\frac{N}{2}}} |u|_{H^1(\Omega)} \, ; \quad \text{and}  \label{eqn:nabla-u-eta}\\
		|\nabla \nabla u^\eta|_{L^{\infty}(\mathbb{R}^N)} & \leq
		\frac{M_2}{\eta^{\frac{N+2}{2}}} |\nabla  \tilde{u}|_{L^2(\mathbb{R}^N)}
		\leq C \frac{M_2}{\eta^{\frac{N+2}{2}}} |u|_{H^1(\Omega)}
		\quad \text{with } M_2 = |\nabla \varphi|_{L^2(\mathbb{R}^N)}\, . \label{eqn:nabla-nabla-u-eta}
	\end{align}
	The proofs of \eqref{eqn:u-eta}-\eqref{eqn:nabla-nabla-u-eta} are straightforward; to indicate the method, we
	provide the details for \eqref{eqn:nabla-nabla-u-eta}.
	Since $\nabla \nabla u^\eta = (\nabla \varphi^\eta) * \nabla \tilde{u}$, Young's inequality gives
	$$
	|\nabla \nabla u^\eta|_{L^{\infty}(\mathbb{R}^N)} \leq
	|\nabla \varphi^\eta|_{L^2(\mathbb{R}^N)} |\nabla \tilde{u}|_{L^2(\mathbb{R}^N)} \leq
	C |\nabla \varphi^\eta|_{L^2(\mathbb{R}^N)} |u|_{H^1(\Omega)} \,
	$$
	and $|\nabla \varphi^\eta|_{L^2(\mathbb{R}^N)}$ can be computed directly:
	\begin{align*}
		\int_{\mathbb{R}^N} |\nabla \varphi^\eta|^2 \: dx & =
		\int_{\mathbb{R}^N} \frac{1}{\eta^{2N+2}} \Big|\nabla \varphi \left(\frac{x}{\eta}\right)\Big|^2\: dx \qquad
		\text{since }\nabla \varphi^\eta = \frac{1}{\eta^{N+1}} \nabla \varphi\left(\frac{x}{\eta}\right)\\
		&= \int_{\mathbb{R}^N} \frac{1}{\eta^{N+2}} |\nabla \varphi(y)|^2 \: dy \qquad \quad (x = \eta y) \, .
	\end{align*}
	
	As already discussed in \cref{subsec:main-proof}, we choose $u_\delta$ to be the piecewise affine interpolant
	of $u^\eta$ using a mesh of order $\delta$. We claim that it has the desired properties when $\eta$ is chosen to
	depend appropriately on $\delta$; in particular, it is sufficient that
	\begin{equation} \label{eqn:choice-of-t}
		\eta = \delta^t \quad \text{with} \quad t = \frac{1}{N+2} \, .
	\end{equation}
	Indeed, if $u^{\eta}_{\delta}$ is the piecewise affine interpolant of $u^\eta$ then we can estimate
	$u^{\eta}_\delta - u^\eta$ using \eqref{eqn:nabla-u-eta} and \eqref{eqn:nabla-nabla-u-eta}:
	\begin{align}
		|\nabla u^{\eta}_{\delta} - \nabla u^\eta|_{L^{\infty}(\mathbb{R}^N)} & \leq
		C' \delta |\nabla \nabla u^\eta|_{L^\infty(\mathbb{R}^N)} \leq
		C' C M_2 \frac{\delta}{\eta^{\frac{N+2}{2}}} |u|_{H^1(\Omega)} \quad \text{and} \label{eqn:u-eta-difference-1}\\
		|u^{\eta}_{\delta}  -  u^{\eta}|_{L^{\infty}(\mathbb{R}^N)} & \leq
		C' \delta |\nabla u^\eta|_{L^\infty(\mathbb{R}^N)} \leq
		C' C M_1 \frac{\delta}{\eta^{\frac{N}{2}}} |u|_{H^1(\Omega)} \, . \label{eqn:u-eta-difference-2}
	\end{align}
	where $C'$ depends on the geometry of the mesh used to define $u^\eta_\delta$ (but not
	on $\eta$ or $\delta$). Now,
	\begin{equation} \label{eqn:udelta-vs-utilde-in-H1}
		|\tilde{u} - u^\eta_\delta|_{H^1(\mathbb{R}^N)} \leq
		|\tilde{u} - u^\eta|_{H^1(\mathbb{R}^N)} + |u^\eta - u^\eta_\delta|_{H^1(\mathbb{R}^N)} \, .
	\end{equation}
	The first term on the right tends to $0$ as $\eta \rightarrow 0$ by \eqref{eqn:u-eta-bound}. Since
	$\tilde{u}$ is compactly supported in $\mathbb{R}^N$, the functions $u^\eta$ and $u^\eta_\delta$ also have
	compact support (on sets whose volumes are uniformly bounded for $\delta \leq 1$ and $\eta \leq 1$). Combining this
	with \eqref{eqn:u-eta-difference-1} and \eqref{eqn:u-eta-difference-2}, we see that the second term on the right
	side of \eqref{eqn:udelta-vs-utilde-in-H1} tends to zero provided that $\delta/\eta^{(N+2)/2} \rightarrow 0$.
	Thus the choice $\eta = \delta^t$ assures that $|\tilde{u} - u^\eta_\delta|_{H^1(\mathbb{R}^N)} \rightarrow 0$ as
	$\delta \rightarrow 0$ provided that
	$$
	0 < t < 1 \quad \text{and} \quad t < \frac{2}{N+2} \, .
	$$
	For any such choice of $t$, we get estimates on the $L^\infty$ norms of $u_\delta$ and $\nabla u_\delta$ from
	\eqref{eqn:u-eta} and \eqref{eqn:nabla-u-eta}; in particular, the choice $t=1/(N+2)$ gives
	$$
	|u_\delta|_{L^\infty(\R^N)} +
	|\nabla u_\delta|_{L^\infty(\R^N)} \leq c_u \delta^{-a} \quad \text{with} \quad a = \frac{N}{2(N+2)} \quad
	\mbox{and} \quad c_u = C |u|_{H^1(\Omega)} \, .
	$$
	
	\renewcommand{\theequation}{D.\arabic{equation}}
	\setcounter{equation}{0}
	\renewcommand{\theequation}{\thesubsection.\arabic{equation}}
	\subsection{The upper and lower bounds for convex 2D polygons}\label{appendix-polygon}
	This appendix proves our upper and lower bounds \eqref{eqn:poly-upper}-\eqref{eqn:poly-lower} for convex
	2D polygons. The argument is by induction. The initial step (showing the bounds for triangles) is presented
	in \cref{appendix-triangle-1}. The inductive step is then treated in \cref{appendix:quadrilateral}.
	
	\begin{figure}[!htb]
		\centering
		\begin{minipage}{.45\linewidth}
			\centering
			\subfloat[]{\label{fig:poly-triangle}\includegraphics[scale=.23]{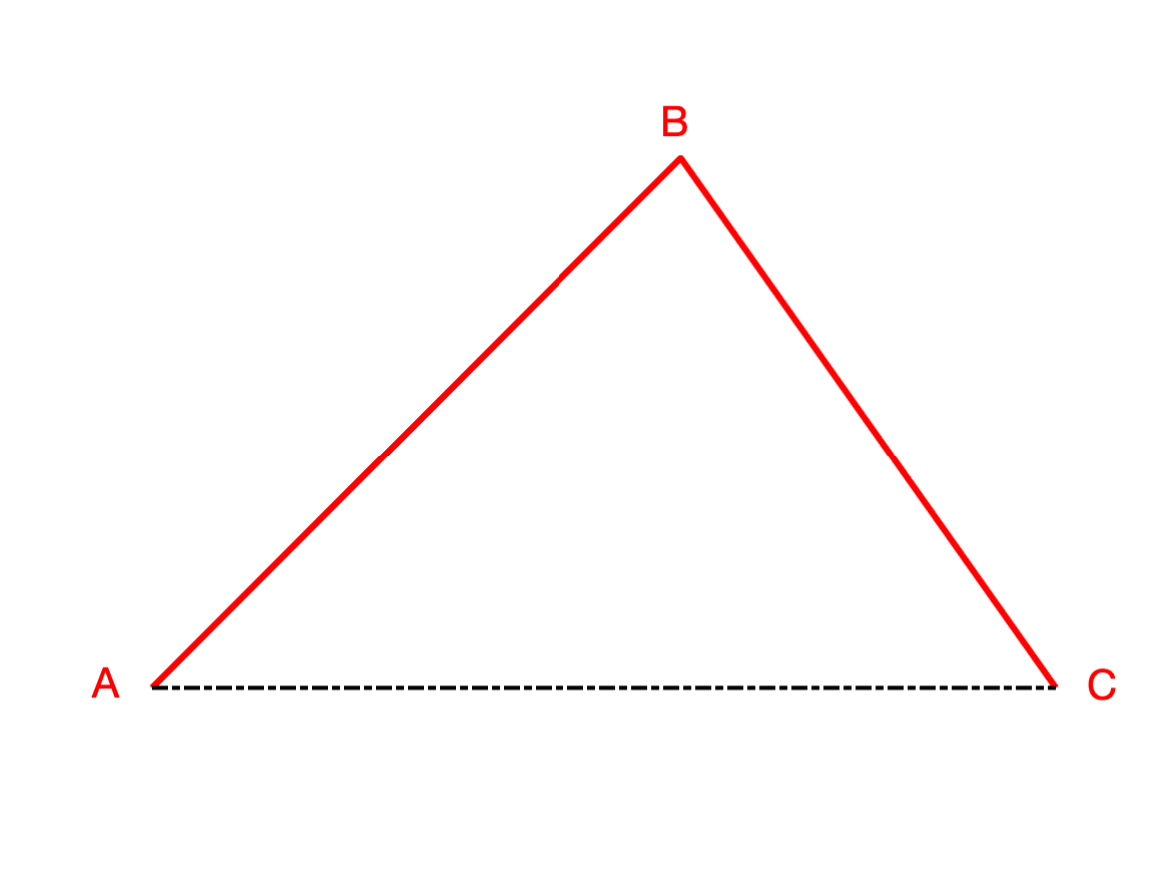}}
		\end{minipage}
		\begin{minipage}{.45\linewidth}
			\centering
			\subfloat[]{\label{fig:poly-quad}\includegraphics[scale=.23]{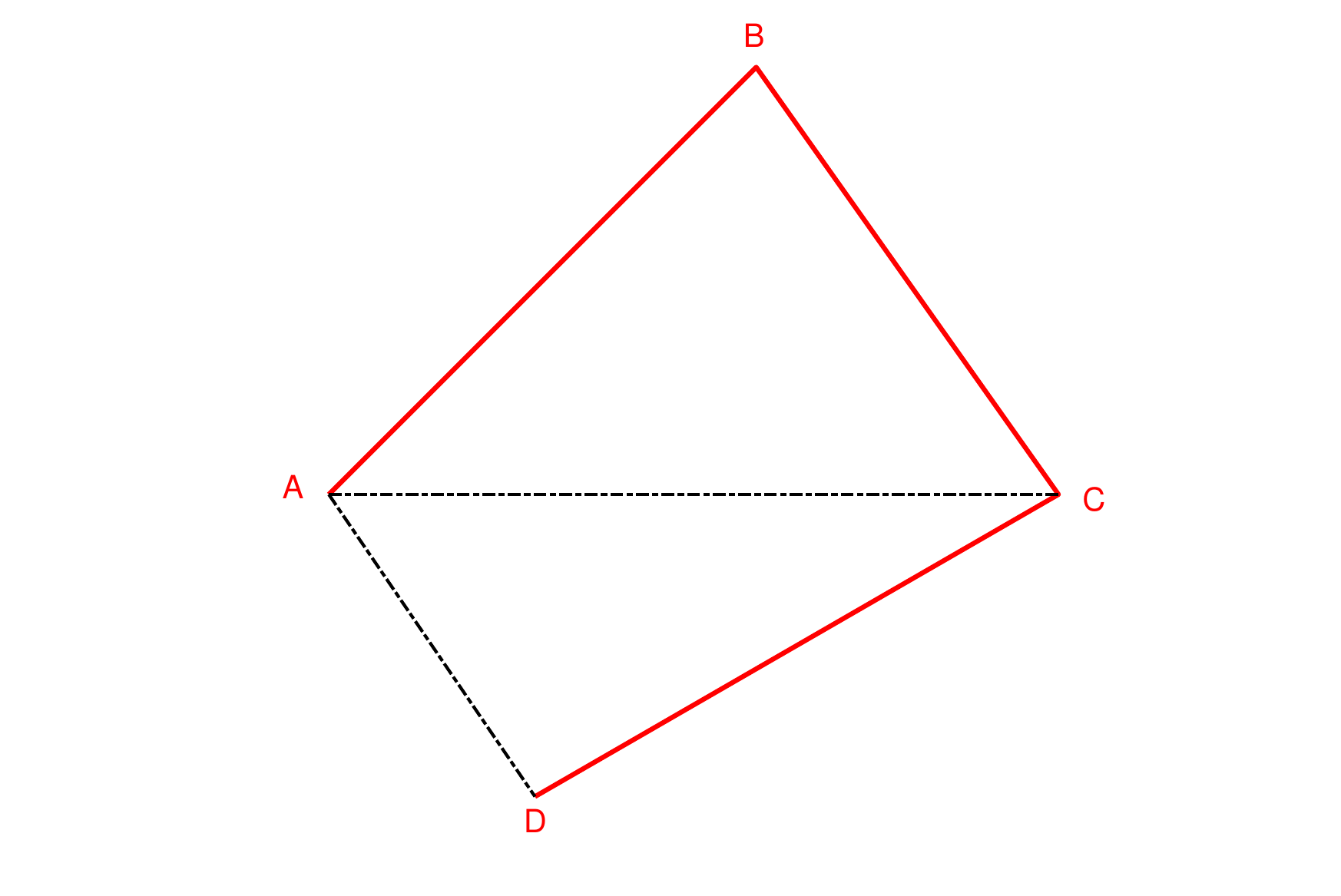}}
		\end{minipage}
		\caption{(a) a general triangle; (b) a convex quadrilateral: the red solid edges are
			counted in the energy $E_{\text{poly}}(u,P_n)$ and the dotted edges indicate the triangular mesh.}
	\end{figure}
	
	\subsubsection{Triangles}\label{appendix-triangle-1}
	When specialized to the triangle $\Delta ABC$ shown in \cref{fig:poly-triangle}, our polygon energy reduces to
	\begin{equation} \label{eqn:triangle-energy}
		E_{\text{poly}}(u,\Delta ABC) = (|u(A) - u(B)| - |A-B|)^2 + (|u(B) - u(C)| - |B-C|)^2
	\end{equation}
	and $\nabla u$ becomes the constant matrix characterized by
	\begin{align}
		\nabla u &=
		\begin{pmatrix} u(A) - u(B) & u(B) - u(C) \end{pmatrix}
		\begin{pmatrix}	A - B & B - C \end{pmatrix}^{-1} =
		M_1 M_2^{-1}, \label{eqn:gradient}
	\end{align}
	where we have set $M_1 = \begin{pmatrix} u(A) - u(B) & u(B) - u(C)	\end{pmatrix}$ and
	$M_2 = \begin{pmatrix} A - B & B - C \end{pmatrix}$.
	
	We start with the following elementary upper and lower bounds:
	\begin{align}
		E_{\text{poly}}(u,\Delta ABC) &\leq |u(A) - u(B)|^2 + |u(B) - u(C)|^2 + |A-B|^2 + |B-C|^2
		\quad \text{and} \label{eqn:elementary-upper-triangle} \\
		E_{\text{poly}}(u,\Delta ABC) &\geq \frac{1}{2} \Big(|u(A) - u(B)|^2 + |u(B) - u(C)|^2\Big)  - 2 \Big(|A-B|^2 + |B-C|^2\Big)
		\label{eqn:elementary-lower-triangle}
	\end{align}
	(using for the latter the fact that $(x-y)^2 \geq \frac{1}{2}x^2 - 2 y^2$ for any $x,y \in \mathbb{R}$). Our main task
	is evidently to control the term that's quadratic in $u$,
	\begin{equation} \label{eqn:u-vs-M1}
		|u(A) -u(B)|^2  + |u(B) - u(C)|^2 = |M_1|^2
	\end{equation}
	in terms of $|\nabla u|^2$. This is easy:
	we are of course assuming that the triangle is nondegenerate (that is, none of
	its sides has length zero), so the matrix $M_2$ is nonsingular. Therefore $|M|$ and $|M M_2^{-1}|$ are both norms on a
	$2 \times 2$ matrix $M$. Since all norms are equivalent in finite dimensions, there are constants $\alpha$ and $\beta$
	(depending on the geometry of the triangle) such that $|M M_2^{-1}|^2 \geq \alpha |M|^2$ and
	$|M M_2^{-1}|^2 \leq \beta |M|^2$. Taking $M=M_1$ and integrating, this gives
	\begin{equation}
		|\nabla u|^2_{L^2(\Delta ABC)} \geq \alpha |M_1|^2 |\Delta ABC| \quad \text{and} \quad
		|\nabla u|^2_{L^2(\Delta ABC)} \leq \beta |M_1|^2 |\Delta ABC| \, .
	\end{equation}
	Combining this with \eqref{eqn:elementary-upper-triangle} -- \eqref{eqn:u-vs-M1} gives
	\begin{align*}
		E_{\text{poly}}(u,\Delta ABC) &\leq \frac{1}{\alpha |\Delta ABC|} |\nabla u|^2_{L^2{\Delta ABC}} +
		\big(|A-B|^2 + |B-C|^2\big) \quad \text{and}\\
		E_{\text{poly}}(u,\Delta ABC) &\geq \frac{1}{2\beta |\Delta ABC|} |\nabla u|^2_{L^2{\Delta ABC}} -
		2\big(|A-B|^2 + |B-C|^2\big) \, ,
	\end{align*}
	which clearly imply inequalities of the desired form \eqref{eqn:poly-upper}--\eqref{eqn:poly-lower}.
	
	\subsubsection{The inductive step}\label{appendix:quadrilateral}
	The passage from triangles to quadrilaterals provides the main ideas of the inductive step, so we shall start
	by discussing that. Afterward we'll briefly explain how a similar argument handles the passage from
	$n-1$ sided polygons to $n$ sided ones.
	
	Consider the quadrilateral $P_4$ in \cref{fig:poly-quad}, whose energy is
	$$
	E_\text{poly}(u,P_4) = E_{AB}(u) + E_{BC}(u) + E_{CD}(u)
	$$
	where $E_{AB}(u) = (|u(A)  -  u(B)|- |A-B|)^2$, etc. The associated upper bound is easy: we have
	\begin{equation} \label{eqn:inductive-step-quad-upper}
		E_\text{poly}(u,P_4) \leq \Big( E_{AB}(u) + E_{BC}(u) \Big) + \Big( E_{AC}(u) + E_{CD}(u) \Big)
	\end{equation}
	since $E_{AC}(u) \geq 0$. Our result for triangles estimates each of the terms on the right:
	\begin{align*}
		E_{AB}(u) + E_{BC}(u) &\leq c_1(\Delta ABC) \Big(|\nabla u|_{L^2(\Delta ABC)}^2  +  |\Delta ABC|\Big) ], , \\
		E_{AC}(u) + E_{CD}(u) & \leq c_1(\Delta ACD) \Big(|\nabla u|_{L^2(\Delta ACD)}^2  +  |\Delta ACD|\Big) \, .
	\end{align*}
	Combining these estimates gives
	\begin{align*}
		E_{\text{poly}}(u,P_4) \leq c_{1, \rm max} \Big( |\nabla u|_{L^2(P_4)}^2  +  |P_4| \Big)
	\end{align*}
	where $c_{1, \rm max}$ is the larger of the $c_1$'s of the two triangles.
	
	For the lower bound we must work a bit harder. We start with the observation that there exists a constant $\gamma$
	(depending only on the geometry of triangle $ABC$) such that
	\begin{equation} \label{eqn:desire-upper}
		E_{AC}(u) \leq 3 \Big[ \gamma + E_{AB}(u) + E_{BC}(u) \Big] \, .
	\end{equation}
	The importance of this inequality is that we can bound the energy of one spring in a triangle by the energies of the
	other two springs. The proof of \eqref{eqn:desire-upper} is a straightforward calculation which we briefly postpone.
	Writing \eqref{eqn:desire-upper} in the form
	$\frac{1}{6}E_{AC} -\frac{\gamma}{2} - \frac{1}{2}(E_{AB} + E_{BC}) \leq 0 $, we see that
	\begin{align} \label{eqn:key-step-for-quad-lb}
		E_{\text{poly}}(u,P_4) &\geq E_{AB}(u) + E_{BC}(u) + E_{CD}(u) +
		\frac{1}{6}E_{AC}(u) -\frac{\gamma}{2} - \frac{1}{2} \big( E_{AB}(u) + E_{BC}(u) \big) \nonumber \\
		& \geq \frac{1}{2} \big( E_{AB}(u) + E_{BC}(u) \big) + \frac{1}{6} \big( E_{AC}(u) + E_{CD}(u) \big) - \frac{\gamma}{2} \, .
	\end{align}
	Our lower bound for triangles gives
	\begin{align*}
		E_{AB}(u) + E_{BC}(u) &\geq c_2(\Delta ABC) |\nabla u|_{L^2(\Delta ABC)}^2 - c_3(\Delta ABC) |\Delta ABC| \, , \\
		E_{AC}(u) + E_{CD}(u) &\geq c_2(\Delta ACD) |\nabla u|_{L^2(\Delta ACD)}^2 - c_3(\Delta ACD) |\Delta ACD| \, .
	\end{align*}
	Combining these estimates leads easily to a result of the desired form
	$$
	E_{\text{poly}}(u,P_4) \geq \tilde{c}_2 |\nabla u|_{L^2(P_4)}^2 - \tilde{c}_3 |P_4| \, .
	$$
	
	To finish our discussion of quadrilaterals we now demonstrate \eqref{eqn:desire-upper}. With the notation
	\begin{align*}
		|u(A) - u(C)| &= a_1 \qquad  |A - C| = a_2 \\
		|u(A) - u(B)| &= b_1 \qquad  |A-B| = b_2 \\
		|u(B) - u(C)| &= c_1 \qquad  |B-C| = c_2
	\end{align*}
	the estimate \eqref{eqn:desire-upper} is equivalent to
	\begin{equation} \label{eqn:desired-upper-equivalent-form}
		(a_1- a_2)^2 \leq \eta \big[ \gamma + (b_1-b_2)^2+(c_1 - c_2)^2 \big]
	\end{equation}
	with $\eta = 3$. The triangle inequality implies
	\begin{equation} \label{eqn:triangle-inequality}
		|b_1 - c_1| \leq a_1 < b_1 + c_1, \qquad \qquad |b_2 - c_2| \leq a_2 < b_2 + c_2 \, .
	\end{equation}
	To get \eqref{eqn:desired-upper-equivalent-form}, we begin with the observation that
	\begin{equation} \label{eqn:desired-ineq-part1}
		(a_1 - a_2)^2 = a_1^2 - 2a_1 a_2 + a_2^2 \leq a_1^2 + a_2^2 \leq 2(b_1^2 + c_1^2 + b_2^2 + c_2^2) \, .
	\end{equation}
	The last inequality comes from the triangle inequality \eqref{eqn:triangle-inequality}, since for $i=1,2$ we have
	$$
	\frac{a_i}{2} \leq \frac{b_i + c_i}{2}\leq \sqrt{\frac{b_i^2 + c_i^2}{2}} \qquad \Rightarrow \qquad
	a_i^2 \leq 2(b_i^2 + c_i^2) \, .
	$$
	In view of \eqref{eqn:desired-ineq-part1}, \eqref{eqn:desired-upper-equivalent-form} will hold if
	$$
	2(b_1^2 + c_1^2 + b_2^2 + c_2^2) \leq \eta \big[ \gamma + (b_1-b_2)^2+(c_1 - c_2)^2 \big] \, .
	$$
	It is clear that such an estimate should hold for some $\eta$ and $\gamma$, since $b_2 = |A-B|$ and $c_2 = |B-C|$
	are constants (independent of $u$) and both sides are quadratic in $b_1 = |u(A)-u(B)|$ and $ c_1 = |u(B)-u(C)|$.
	In fact the estimate holds with
	$\eta = 3$ and $\gamma = 6M^2$, if we set $M = \max\{b_2, c_2\}$. To see this, we use the fact that
	$$
	0 \leq \frac{1}{2} (b_1+ c_1)^2 - 6M(b_1 + c_1) + 18M^2 \leq b_1^2 + c_1^2 + b_2^2 + c_2^2 - 6M(b_1 + c_1) +  18M^2 \, .
	$$
	Adding $2(b_1^2 + c_1^2 + b_2^2 + c_2^2)$ to both sides gives
	\begin{multline*}
		2(b_1^2 + c_1^2 + b_2^2 + c_2^2) \leq 3(b_1^2 + c_1^2 + b_2^2 + c_2^2) - 6M(b_1 + c_1) +  18M^2\\
		\leq  3(b_1^2 + c_1^2 + b_2^2 + c_2^2) - 6(b_1b_2 + c_1c_2) +  18M^2 = 3 \big[ 6 M^2 + (b_1 - b_2)^2 + (c_1 - c_2)^2 \big] \, ,
	\end{multline*}
	as claimed. This completes our treatment of quadrilaterals.
	
	We now indicate how the same ideas can be used to obtain the upper and lower bounds for
	convex polygons with $n$ sides, once they are known for polygons with $n-1$ sides. Let $P_n$ have vertices $A_1, \ldots, A_n$
	as shown in \cref{fig:polygon}, and recall that
	$$
	E_\text{poly}(u,P_n) = E_{A_1 A_2}(u) + \ldots E_{A_{n-1}A_n}(u) \, .
	$$
	The upper bound is obtained by observing that
	$$
	E_\text{poly}(u,P_n) \leq E_\text{poly}(u,P')(u) + E_\text{poly}(u,P'')
	$$
	where $P'$ is the triangle $A_1 A_2 A_3$ and $P''$ is the $n-1$-sided polygon $A_1 A_3 A_4 \ldots A_n$. This is
	the analogue of \eqref{eqn:inductive-step-quad-upper}, and by arguing as we did there one obtains the upper bound for
	$E_\text{poly}(u,P_n)$ from the upper bounds for $E_\text{poly}(u,P')$ and $E_\text{poly}(u,P'')$.
	The lower bound is obtained by using \eqref{eqn:desire-upper} for the triangle $P'$:
	\begin{align*}
		E_\text{poly}(u,P_n) & \geq E_\text{poly}(u,P_n) + \frac{1}{6}E_{A_1 A_3}(u) -\frac{\gamma}{2} -
		\frac{1}{2} \big( E_{A_1 A_2}(u) + E_{A_2 A_3}(u) \big) \\
		& \geq \frac{1}{2}E_\text{poly}(u,P') + \frac{1}{6} E_\text{poly}(u,P'') - \frac{\gamma}{2} \, .
	\end{align*}
	This is the analogue of \eqref{eqn:key-step-for-quad-lb}, and arguing as we did there gives the lower bound
	for $E_\text{poly}(u,P_n)$ by combining those for $E_\text{poly}(u,P')$ and $E_\text{poly}(u,P'')$.

	\bibliography{ref}

\begin{thebibliography}{10}

\bibitem{alicandro2004general}
Roberto Alicandro and Marco Cicalese.
\newblock A general integral representation result for continuum limits of
  discrete energies with superlinear growth.
\newblock {\em {SIAM} Journal on Mathematical Analysis}, 36(1):1--37, 2004.

\bibitem{alicandro2011integral}
Roberto Alicandro, Marco Cicalese, and Antoine Gloria.
\newblock Integral representation results for energies defined on stochastic
  lattices and application to nonlinear elasticity.
\newblock {\em Archive for Rational Mechanics and Analysis}, 200(3):881--943,
  2011.

\bibitem{attouch1984variational}
Hedy Attouch.
\newblock {\em Variational Convergence for Functions and Operators}.
\newblock Pitman {A}dvanced {P}ublishing {P}rogram, 1984.

\bibitem{bertoldi2017flexible}
Katia Bertoldi, Vincenzo Vitelli, Johan Christensen, and Martin Van~Hecke.
\newblock Flexible mechanical metamaterials.
\newblock {\em Nature Reviews Materials}, 2(11):1--11, 2017.

\bibitem{borcea2010periodic}
Ciprian~S Borcea and Ileana Streinu.
\newblock Periodic frameworks and flexibility.
\newblock {\em Proceedings of the Royal Society of London A: Mathematical,
  Physical and Engineering Sciences}, 466(2121):2633--2649, 2010.

\bibitem{bossart2021oligomodal}
Aleksi Bossart, David~MJ Dykstra, Jop Van~der Laan, and Corentin Coulais.
\newblock Oligomodal metamaterials with multifunctional mechanics.
\newblock {\em Proceedings of the National Academy of Sciences},
  118(21):e2018610118, 2021.

\bibitem{braides1985homogenization}
Andrea Braides.
\newblock Homogenization of some almost periodic coercive functional.
\newblock {\em Rendiconti, Accademia Nazionale delle Scienze detta dei XL,
  Serie V, Memorie di Matematica, Parte I}, 103:313--322, 1985.

\bibitem{braides-book-1998}
Andrea Braides and Anneliese Defranceschi.
\newblock {\em Homogenization of Multiple Integrals}.
\newblock The Clarendon Press, Oxford University Press, New York, 1998.

\bibitem{braides2008homogenization}
Andrea Braides, Mikhail Maslennikov, and Laura Sigalotti.
\newblock Homogenization by blow-up.
\newblock {\em Applicable Analysis}, 87(12):1341--1356, 2008.

\bibitem{cioranescu2012homogenization}
Doina Cioranescu and Jeannine Saint~Jean Paulin.
\newblock {\em Homogenization of Reticulated Structures}.
\newblock Springer Science \& Business Media, 2012.

\bibitem{conti2015theory}
Sergio Conti and Georg Dolzmann.
\newblock On the theory of relaxation in nonlinear elasticity with constraints
  on the determinant.
\newblock {\em Archive for Rational Mechanics and Analysis}, 217:413--437,
  2015.

\bibitem{czajkowski2022conformal}
Michael Czajkowski, Corentin Coulais, Martin van Hecke, and D~Zeb Rocklin.
\newblock Conformal elasticity of mechanism-based metamaterials.
\newblock {\em Nature Communications}, 13(1):1--9, 2022.

\bibitem{czajkowski2024duality}
Michael Czajkowski and D.~Zeb Rocklin.
\newblock Duality and sheared analytic response in mechanism-based
  metamaterials.
\newblock {\em Physical Review Letters}, 132:068201, Feb 2024.

\bibitem{dacorogna2007direct}
Bernard Dacorogna.
\newblock {\em Direct Methods in the Calculus of Variations}.
\newblock Springer Science \& Business Media, 2007.

\bibitem{degiorgi-1975}
Ennio De~Giorgi.
\newblock Sulla convergenza di alcune successioni d'integrali del tipo
  dell'area.
\newblock {\em Rendiconti di Matematica, Serie VI}, 8:277--294, 1975.

\bibitem{deng2020characterization}
Bolei Deng, Siqin Yu, Antonio~E Forte, Vincent Tournat, and Katia Bertoldi.
\newblock Characterization, stability, and application of domain walls in
  flexible mechanical metamaterials.
\newblock {\em Proceedings of the National Academy of Sciences},
  117(49):31002--31009, 2020.

\bibitem{hutchinson2006structural}
Robert~G Hutchinson and Norman~A Fleck.
\newblock The structural performance of the periodic truss.
\newblock {\em Journal of the Mechanics and Physics of Solids}, 54(4):756--782,
  2006.

\bibitem{iwaniec2011diffeomorphic}
Tadeusz Iwaniec, Leonid~V Kovalev, and Jani Onninen.
\newblock Diffeomorphic approximation of {S}obolev homeomorphisms.
\newblock {\em Archive for Rational Mechanics and Analysis}, 201(3):1047--1067,
  2011.

\bibitem{iwaniec2012hopf}
Tadeusz Iwaniec, Leonid~V Kovalev, and Jani Onninen.
\newblock Hopf differentials and smoothing {S}obolev homeomorphisms.
\newblock {\em International Mathematics Research Notices},
  2012(14):3256--3277, 2012.

\bibitem{kapko2009collapse}
Vitaliy Kapko, Michael~MJ Treacy, Michael~F Thorpe, and SD~Guest.
\newblock On the collapse of locally isostatic networks.
\newblock {\em Proceedings of the Royal Society of London A: Mathematical,
  Physical and Engineering Sciences}, 465(2111):3517--3530, 2009.

\bibitem{li2023some}
Xuenan Li and Robert~V Kohn.
\newblock Some results on the {G}uest-{H}utchinson modes and periodic
  mechanisms of the {K}agome lattice metamaterial.
\newblock {\em Journal of the Mechanics and Physics of Solids}, 178:105311,
  2023.

\bibitem{liforthcoming}
Xuenan Li and Robert V~Kohn.
\newblock A nonlinear homogenization-based perspective on the soft modes and
  effective energies of some conformal metamaterials.
\newblock \url{https://www.arxiv.org/abs/2509.16907}, 2025.

\bibitem{milton2013complete}
Graeme~W Milton.
\newblock Complete characterization of the macroscopic deformations of periodic
  unimode metamaterials of rigid bars and pivots.
\newblock {\em Journal of the Mechanics and Physics of Solids},
  61(7):1543--1560, 2013.

\bibitem{muller1987homogenization}
Stefan M{\"u}ller.
\newblock Homogenization of nonconvex integral functionals and cellular elastic
  materials.
\newblock {\em Archive for {R}ational {M}echanics and {A}nalysis}, 99:189--212,
  1987.

\bibitem{nassar2020microtwist}
Hussein Nassar, Hui Chen, and Guoliang Huang.
\newblock Microtwist elasticity: A continuum approach to zero modes and
  topological polarization in {K}agome lattices.
\newblock {\em Journal of the Mechanics and Physics of Solids}, 144:104107,
  2020.

\bibitem{zheng2022continuum}
Yue Zheng, Imtiar Niloy, Paolo Celli, Ian Tobasco, and Paul Plucinsky.
\newblock Continuum field theory for the deformations of planar kirigami.
\newblock {\em Physical Review Letters}, 128(20):208003, 2022.

\bibitem{zheng2022modeling}
Yue Zheng, Ian Tobasco, Paolo Celli, and Paul Plucinsky.
\newblock Modeling planar {k}irigami metamaterials as generalized elastic
  continua.
\newblock {\em Proceedings of the Royal Society A}, 479(2272):20220665, 2023.

\bibitem{ziemer2012weakly}
William~P Ziemer.
\newblock {\em Weakly differentiable functions: Sobolev spaces and functions of
  bounded variation}.
\newblock Springer Science \& Business Media, 2012.

\end{thebibliography}
	\bibliographystyle{plain}
	
\end{document}